\PassOptionsToPackage{hypertexnames=false}{hyperref}
\documentclass[fleqn,usenatbib]{mnras}

\usepackage{newtxtext,newtxmath}
\usepackage{graphicx}
\usepackage{color,soul}
\usepackage{amsmath}
\usepackage{comment}
\usepackage{longtable}
\usepackage{pdflscape}
\usepackage{enumitem}
\usepackage{mathtools}
\usepackage{booktabs}
\usepackage{xcolor}
\usepackage[export]{adjustbox}
\usepackage{lastpage}
\usepackage{xspace}
\newcommand{\orcid}[1]{}

\newcommand{\ContinuedFloat}{\addtocounter{figure}{-1}}
\newcommand{\phantomcaption}{\stepcounter{figure}}

\def\la{\mathrel{\hbox{\rlap{\hbox{\lower4pt\hbox{$\sim$}}}\hbox{$<$}}}}
\def\ga{\mathrel{\hbox{\rlap{\hbox{\lower4pt\hbox{$\sim$}}}\hbox{$>$}}}}

\def\arcmin{\hbox{$^\prime$}}
\def\arcsec{\hbox{$^{\prime\prime}$}}

\newcommand{\msun}{\,{\rm M}_\odot}

\newcommand{\MSUN}{${\rm M}_\odot$}

\newcommand{\kms}{{\,km\,s$^{-1}$}}

\newcommand{\HI}{\mbox{\normalsize H\thinspace\footnotesize I}\xspace}

\newcommand{\VirgoIII}{Virgo\,\textsc{III}\xspace}

\def\apjl   {{ApJL }}
\def\apjs   {{ApJS }}

\DeclareRobustCommand{\HI}{\texorpdfstring{\mbox{\normalsize H\thinspace\footnotesize I}}{HI}\xspace}
\DeclareRobustCommand{\VirgoIII}{\texorpdfstring{Virgo\,\textsc{III}}{Virgo III}\xspace}

\title[MeerKAT HI-imaging of the Virgo~III filament]{Virgo Filaments VII: MeerKAT HI-imaging of the \VirgoIII filament}

\author[M. Ramatsoku et al.]{
M.~Ramatsoku\orcid{0000-0003-0231-3249},$^{1,2}$\thanks{These authors contributed equally to this work. Emails: m.ramatsoku@ru.ac.za, gautam.nagaraj@epfl.ch}
G.~Nagaraj\orcid{0000-0002-0905-342X},$^{3}$\footnotemark[1]
P.~Jablonka\orcid{0000-0002-9655-1063},$^{3,4}$
R.~A.~Finn\orcid{0000-0001-5334-5166},$^{5}$
B.~Vulcani\orcid{0000-0003-0980-1499},$^{6}$
G.~Rudnick\orcid{0000-0001-5851-1856},$^{7}$
\newauthor
~F.~Combes\orcid{0000-0003-2658-7893},$^{8}$
G.~De~Lucia\orcid{0000-0002-6220-9104},$^{9}$
G.~Castignani\orcid{0000-0003-0444-6897},$^{10}$
D.~Zaritsky\orcid{0000-0002-5177-727X},$^{11}$
Y.~Bah\'e\orcid{0000-0002-3196-5126},$^{12}$
R.~A.~Koopmann\orcid{0000-0002-1340-0543},$^{13}$
\newauthor
~D.~Zakharova\orcid{0009-0001-1809-4821},$^{9}$
and K.~Conger\orcid{0009-0005-0303-0330},$^{7}$\\
$^{1}$Department of Physics and Electronics, Rhodes University, PO Box 94, Makhanda, 6140, South Africa\\
$^{2}$INAF -- Osservatorio Astronomico di Cagliari, Via della Scienza 5, I-09047 Selargius (CA), Italy\\
$^{3}$Laboratoire d'Astrophysique, \'Ecole Polytechnique F\'ed\'erale de Lausanne (EPFL), Route de la Sorge, 1015 Lausanne, Switzerland\\
$^{4}$LIRA, Observatoire de Paris, Universit\'e PSL, CNRS, 5 Place Jules Janssen, 92190 Meudon, France\\
$^{5}$Department of Physics and Astronomy, Siena College, 515 Loudon Road, Loudonville, NY 12211, USA\\
$^{6}$INAF -- Osservatorio Astronomico di Padova, Vicolo Osservatorio 5, I-35122 Padova, Italy\\
$^{7}$Department of Physics and Astronomy, University of Kansas, 1251 Wescoe Hall Drive, Room 1082, Lawrence, KS 66049, USA\\
$^{8}$Observatoire de Paris, LUX, Coll\`ege de France, CNRS, PSL University, Sorbonne University, 75014 Paris, France\\
$^{9}$INAF -- Osservatorio Astronomico di Trieste, Via Tiepolo 11, I-34131 Trieste, Italy\\
$^{10}$INAF -- Osservatorio di Astrofisica e Scienza dello Spazio di Bologna, Via Gobetti 93/3, I-40129 Bologna, Italy\\
$^{11}$Steward Observatory, University of Arizona, 933 North Cherry Avenue, Tucson, AZ 85721-0065, USA\\
$^{12}$School of Physics and Astronomy, University of Nottingham, University Park, Nottingham NG7 2RD, UK\\
$^{13}$Department of Physics \& Astronomy, Union College, Schenectady, NY 12308, USA
}

\date{Accepted 2026 August 11. Received 2026 August 11; in original form 2026 March 06}
\pubyear{2026}

\begin{document}
\raggedbottom
\label{firstpage}
\pagerange{\pageref{firstpage}--\pageref{LastPage}}
\maketitle

\begin{abstract}
We present MeerKAT \HI\ observations of galaxies in the nearby \VirgoIII\ filament. At a median distance of $\sim30$~Mpc, the filament extends over $\approx16$~Mpc across $\sim56$~deg$^{2}$. Our survey comprises 15 targeted MeerKAT pointings centred on \HI\ deficient galaxies. Multi-resolution \HI\ imaging at $8\arcsec$--$90\arcsec$ with an rms of $\sim0.3$ mJy beam$^{-1}$ reaches column densities of $\sim4\times10^{20}$ to $\sim4\times10^{18}$ atoms cm$^{-2}$. This probes \HI\ in galaxy discs at $\sim$1.2~kpc resolution and their outskirts at $\sim$13.5~kpc. We detect \HI\ emission from 80 sources within the $500$--$3000$\,\kms\ range; most are optically detected galaxies, while 10 per cent lack confirmed counterparts. We also identify three \HI\ clouds without optical counterparts near an interacting system, possibly displaced during the interaction. Galaxies in the filament lie below the field \HI\ gas-fraction scaling relations at fixed stellar mass. Crucially, $\sim40$ per cent of galaxies not selected for \HI\ deficiency fall more than $1\sigma$ below these relations, with an even greater fraction lying under the mean relation itself, indicating that the low \HI\ gas fractions are widespread in the filament rather than confined to the pre-selected deficient population. Quantifying \HI\ morphologies using an asymmetry parameter, we find a wide range of disturbances without a clear trend with projected distance to the filament spine or local density, suggesting a combination of local interactions and the filament itself. This paper presents the survey description, catalogue, and \HI\ atlas enabling investigations of gas removal mechanisms in \VirgoIII\ galaxies.
\end{abstract}


%
\section{Introduction}

The dependence of the properties of galaxies on their environment was first noted nearly a century ago \citep{Hubble1936}. Since then, many studies have striven to better understand this connection. In the local Universe, given the relative ease of observation, the question of environment has been a highly researched topic \citep[e.g,][and references therein]{DresslerGalClust1984,DeLucia2007,BlantonMoustakas2009,BoselliGavazziRedSeqEnv2014}. Some of the most robustly established trends include the transition from late-type-dominated to early-type-dominated populations with increasing environmental density \citep[e.g.,][]{DresslerMorphClust1980,PostmanGellerMorphDen1984,DresslerMorphDen1997} as well as higher levels of quenching relative to low-density environments \citep[e.g.,][]{DresslerGalClust1984,KauffmannEnvObs2004,PengEnvGalEv2010}. 

The physical explanation for these observations remains incomplete. While galaxy mass, halo mass, and environment are known to be correlated, it remains challenging to disentangle their relative roles in shaping galaxy evolution \citep[e.g.,][]{DeLucia2007, WinkelCosmicWeb2021, DonnanCosmicWeb2022}. In dense environments, galaxies are known to be subjected to a range of physical processes, including gaseous interactions with the intracluster medium (ICM), such as ram-pressure stripping \citep[RPS;][]{GunnGottClusterEvo1972} and starvation \citep{LarsonS01980,BaloghSFGrad2000}; gravitational interactions such as mergers or harassment \citep[e.g.,][]{MooreGalHaras1996,MooreGalHaras1998}; or an enhancement of internal quenching processes \citep[e.g.,][]{BoselliGavazziEnvEffLateGal2006,VogelsbergerGalPropSim2014,BaloghSatQuench2016}.

Observational studies of environmental effects on galaxy evolution have historically focused on galaxy clusters and their immediate surroundings. Nevertheless, there is ample evidence for continuation of effects out to multiple times the virial radius \citep[e.g.,][]{Balogh2009,PoggiantiSFHCluster1999,LewisEnvSF2002,CorteseFarEnv2006,Haines2009,Hansen2009,WetzelGalEvoGroupClust2012,BaheFarEnv2013, Vulcani2021}. At such large distances from the cluster centre, environmental processes are generally expected to be less efficient than in cluster cores, consistent with the observed weakening of the quenching--density relation \citep[e.g.,][]{BaloghSFGrad2000,LewisEnvSF2002,TreuMorphDist2003,BowerBaloghClusterGroup2004,DresslerSFGalClust2004,NicholClust2004}. In these regions, galaxies are frequently accreted as members of groups rather than in isolation, raising the question of how much processing  of the observed transformation occurs before cluster infall \citep[e.g.,][]{McGeeAccrGroupClust2009,OdekonHIGroupClust2016,VulcaniRPS2018,Raj2020,AnithaShaji2025}.

At larger distances from clusters, most galaxies are increasingly found along filaments delineating the cosmic web \citep[e.g.,][]{York2000,Colless2001}.
The processes active in cosmic filaments, long structures that are part of the cosmic web \citep[e.g.,][]{KitauraCosmWeb2009,DarvishCosmWeb2014}, are even less understood. Hydrodynamical processes are expected to be less prevalent and gravitational interactions more effective in filaments than in cluster cores, due to the generally lower galaxy velocities and lower densities of the intergalactic medium \citep[e.g.,][]{Das2023}. However, this picture is complicated by simulations showing that filaments connected to groups have higher pressure than field galaxies driven primarily by higher galaxy velocities \citep{BaheFarEnv2013}. Moreover, observations of galaxies in filaments report conflicting results regarding their cold gas content. Some works find elevated cold gas reservoirs in filament galaxies \citep[e.g.,][]{KleinerCosmicWeb2017,KotechaFilaments2022}, consistent with scenarios in which filaments facilitate the accretion of cold gas from the cosmic web \citep{Vulcani2018b}. In contrast, other studies report reduced gas content in filament environments \citep[e.g.,][]{Poudel2017,OdekonFilaments2018, HoosainFilaments2024,Zakharova2024}, suggesting that gas supply may instead be suppressed, potentially through tidal effects associated with the filamentary potential or hydrodynamical removal mechanisms such as cosmic web stripping. This apparent tension is further complicated by the fact that filaments are not uniform environments \citep[e.g.,][]{Bahe2025}: they host both galaxy groups and more isolated systems, and the combined effects of these environments may drive a range of evolutionary pathways \citep[e.g.,][]{HoosainFilaments2024,Zakharova2024}. For example in some low-mass groups along filaments, complex disruptions of the neutral and ionised gas are seen, implying that these relatively low-density environments may still be conducive to significant hydrodynamic and gravitational processes \citep{Finn2025NGC5364}.

To understand the physical processes shaping galaxies in filaments, it is essential to examine the neutral atomic gas (\HI). It is the dominant component of the cold interstellar medium and the primary reservoir from which molecular gas and ultimately stars form. It therefore provides an essential diagnostic tool for understanding the drivers of galaxy evolution. Moreover, \HI\ typically extends to large radii at the interface between galaxies and their surroundings, making it one of the first components to experience external interactions. In these outer regions, the gravitational binding to the host galaxy is weaker, and the gas is more diffuse, allowing even relatively mild environmental effects to leave a clear imprint on the \HI\ distribution, as expected in filaments. 

\subsection{The Virgo~\textsc{iii} filament}
In this study, we use data from the MeerKAT radio telescope \citep{Jonas2016,Mauch2020} to study \HI\ properties of galaxies in or close to the nearby \VirgoIII filament. This well-known structure outside the Virgo cluster lies at a median distance of 30~Mpc and extends over a length $\approx 16$ Mpc, spanning right ascensions of 207-225 degrees and declinations of 2.3-5.4 degrees \citep[e.g.,][]{KimVirgoFil2016}. The filament hosts approximately 200 known galaxies \citep{CastignaniII2022}. 

As part of a large project to better understand the atomic and molecular gas in filament environments, \cite{CastignaniI2022} measured integrated CO (1\(\rightarrow\)0) and CO (2\(\rightarrow\)1) with the IRAM 30m telescope and \HI\ fluxes with the Nan\c{c}ay Radio Telescope, complemented by ancillary data from the literature. This effort forms part of the Virgo Filament Survey (VFS), a series of papers characterising galaxies within ${\sim}12$ projected virial radii of the Virgo cluster. Following the gas census and catalogue construction of the first two papers \citep{CastignaniII2022,CastignaniI2022}, the series has examined the predicted gas content of filament galaxies in the GAEA semi-analytic model \citep{Zakharova2024}, the sizes of star-forming disks measured with WISE \citep{Conger2025Virgo}, and resolved narrowband H$\alpha$ imaging, both of the NGC~5364 group \citep{Finn2025NGC5364} and of H$\alpha$ clump statistics across the filaments \citep{Nagaraj2026}. The present work extends the survey to resolved \HI\ imaging.

In the \VirgoIII filament, 36 galaxies have CO measurements, and with the inclusion of the ALFALFA catalogue \citep{HaynesAlfalfa2018}, a total of 109 galaxies have integrated \HI\  measurements \citep{CastignaniII2022}. These integrated measurements provide an important baseline for understanding the global \HI\ gas content of galaxies in this environment.

The deep and spatially resolved \HI\ maps from MeerKAT observations (described in \S \ref{sec:obs_datared}) provide a detailed view of the spatial distribution and kinematics of \HI\ in galaxies. These maps make it possible to identify and characterise signatures of ram pressure stripping, faint tidal features, and other subtle gas perturbations that cannot be studied with integrated \HI\ fluxes alone. Combined with the extensive multi-wavelength ancillary data and measurement of quantities such as stellar masses and star formation rates \citep{CastignaniI2022,CastignaniII2022}, this dataset enables a quantitative assessment of gas asymmetry, truncation, and displacement as a function of galaxy properties and location within the filament.

~\\
This is the first paper in a subseries regarding these MeerKAT observations within the broader VFS series. It is intended to serve as an overview of the \HI\ detections and to illustrate the quality, sensitivity, and scientific potential of the dataset. In subsequent papers, we will explore the physical processes at play in shaping the properties of the \HI\ detected in galaxies in and around galaxy filaments. In Section \ref{sec:sample}, we explain the selection of the sample of galaxies used in this work. Then, in Section \ref{sec:obs_datared}, we detail the MeerKAT observations and the data reduction process, followed by the source detection algorithm in Section \ref{sec:det}. We present an overview of the \HI data products in our catalogue in Section \ref{sec:res} and properties of the \VirgoIII filament in Section \ref{sec:properties} before summarising in Section \ref{sec:conc}. In Appendix\,\ref{sec:full_sample}, we provide the entire catalogue of \HI detections as well as information on galaxy properties and environment. Finally, we present detailed \HI imaging of all galaxies in Appendix\,\ref{sec:allhiim}. 

In this work, we use a $\Lambda$CDM cosmology with $\Omega_M=0.3$ and $\Omega_\Lambda=0.7$ and $H_0=74$ \kms Mpc$^{-1}$. At a distance of \VirgoIII, 1.0\arcsec\ corresponds to $\sim$0.15 kpc.

\section{Sample selection} \label{sec:sample}
The scientific motivation behind the MeerKAT \VirgoIII survey is to better understand gas removal and pre-processing in filament galaxies using resolved \HI\ imaging. As such, we decided on two main categories for target galaxies: \HI deficient galaxies and galaxies with stellar masses of M$_{\star}$ $\geq$ 10$^9$ so that their molecular gas can be easily detected in resolved CO observations for comparisons between atomic and molecular gas. To select the targets, we used the integrated gas measurements and stellar masses compiled by \citet{CastignaniII2022} as part of a large public catalogue of 6780 galaxies in the filaments around the Virgo cluster. For \HI\ deficiency, we used the parameter as defined by \cite{HaynesGiovanelli1984} and catalogued by \citet{CastignaniI2022}. We considered galaxies with def$_{\rm HI}\geq 0.4$ to be deficient. This cut-off corresponds to a $\sim$1--1.5$\sigma$ deviation from the intrinsic scatter of the reference relation, and galaxies with deficiency parameters above this value are likely undergoing environmental processing (e.g., \citealp{Boselli2004, Boselli2014}).

A total of 15 $1.5~\rm{deg}\times 1.5$~deg MeerKAT pointings were placed along the filament (with one being somewhat displaced from the filament but closer to the Virgo cluster). Every pointing includes at least one galaxy of one of the two categories described above and maximizes the number of relevant galaxies---i.e., within the velocity range related to the Virgo cluster of 500-3000 \kms\ and in the field of view. The pointings probe a wide range of local environments. Our full sample contains 72 galaxies, of which 29 have stellar masses M$_{\star}$ $\geq$ 10$^9$ M$_{\odot}$. A total of 66 galaxies have published integrated \HI\ emission measurements, while 6 are \HI\ non-detections. Among the galaxies with available \HI\ measurements, 10 exhibit weak to strong \HI\ deficiencies, ranging between $0.4 <$ def$_{\rm HI}$ $< 1.5$, while the remainder are \HI-normal systems, thereby allowing a direct comparison between \HI-deficient and gas-normal galaxies within a common large-scale environment.

\begin{figure*}
   \centering
 \includegraphics[width=160mm, height=60mm]{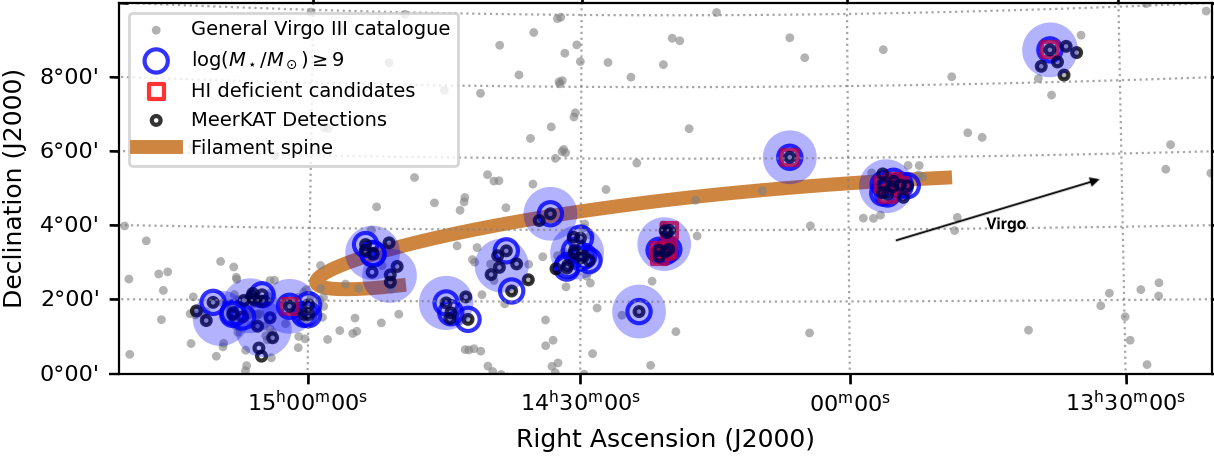}
    \caption{Spatial distribution of galaxies in the \VirgoIII\ filament. The 15 MeerKAT pointings used in this study are shown in blue. Grey points represent all known galaxies within the filament from \citet{CastignaniII2022}. Blue circles indicate galaxies with stellar masses $\log(M_\star/M_\odot)\geq 9$ within the observed pointings. Red squares mark \HI-deficient galaxies. MeerKAT detections, discussed in Sect.\,\ref{sec:det}, are shown as black circles.}\label{fig:spatial_dist_pointing}
 \end{figure*}

The full sample of galaxies, together with their fundamental properties, integrated gas measurements, and environmental parameters, is presented in Table~\ref{sample:tab2}, and their spatial distribution is shown in Fig.\,\ref{fig:spatial_dist_pointing}.

\section{\VirgoIII\ Observations and data reduction} \label{sec:obs_datared}
All \HI\ observations were obtained and reduced using a uniform calibration and imaging strategy. Below, we describe the observational setup and the data reduction procedure applied to all pointings.
 
\subsection{MeerKAT Observations} \label{subsec:obs}
We observed 15 selected pointings containing the sample described in Sect.\,\ref{sec:sample} with MeerKAT \citep{Jonas2016, Mauch2020} in October 2021 (project ID SCI-20210212-PJ-01),  September 2022 (SCI-20220822-PJ-01) and July 2024 (SCI SCI-20230907-MR-01) with at least 58 antennas in the L-band; 856 -- 1712 MHz. We used the 32k SKARAB correlator mode, which samples the observed band with 32768 channels, each with a channel width of 26.123 kHz ($\sim$5.5 \kms\ for \HI\ at redshift $z$ = 0). This mode delivers four linear correlations (HH,VV, HV, VH) with a dump time of 8 seconds. Each pointing was observed for 5 hours of integration on-source.
We also observed the standard bandpass and flux calibrator, J1939-6342, for 10 mins at the beginning, middle, and end of the scan. For each pointing the nearest secondary gain calibrator was observed for 2 mins for every 30 mins scan. Our MeerKAT observations are predominantly taken at night to mitigate against solar radio-frequency interference (RFI).

\subsection{Data reduction} \label{subsec:datared}
We reduced the 32k data on the ILIFU cloud facility\footnote{\url{https://docs.ilifu.ac.za}}, processing only the HH and VV correlations. We selected channels covering the frequency range of $\sim$1373.99 - 1419.69 MHz, which corresponds to recessional velocities in the range of roughly 200 - 9800 \kms. This frequency range, which is broader than the frequency of \VirgoIII, was necessary to obtain a reliable model for the continuum subtraction procedure discussed below.
The visibilities were reduced using the Containerised Automated Radio Astronomical Calibration (\textsc{CARAcal}; \citealp{Josza2022})\footnote{\url{https://caracal.readthedocs.io}}, which uses the Python-based \textsc{Stimela} framework \citep{Makhathini2018}\footnote{\url{https://github.com/SpheMakh/Stimela}}. \textsc{Stimela} provides a containerised environment that is compatible with \textsc{Singularity}, \textsc{Docker}, and \textsc{Podman} and integrates a range of open-source radio interferometry software packages into a single, automated workflow which enables a fully streamlined data calibration and imaging process. 
The calibrators' visibilities were flagged with \textsc{AOFlagger} \citep{Offringa2010} based on the stokes Q.  We used \textsc{casa} \citep{McMullin2007} tasks \textit{bandpass} and \textit{gaincal} to determine the antenna-based time-independent complex flux, bandpass and frequency-independent gains. The task \textit{fluxscale} was then used to bootstrap flux scale and scale gain amplitudes. The resulting complex gain and bandpass solutions were applied on-the-fly to the target visibilities with the \textsc{casa} task \textit{applycal} during the splitting of the target visibilities with \textit{mstransform}. 

The high sensitivity of MeerKAT allows the detection of a large fraction of the continuum emission from extended radio sources, which can resemble faint \HI\ signals in the spectral line cube \citep{Grobler2014}. This, therefore, necessitated a particularly careful continuum subtraction procedure, which we conducted in two steps. 
First, a continuum model was produced through iterative imaging and self-calibration of the target emission. Imaging was carried out with \textsc{wsclean} in Stokes I using multi-scale cleaning (\citealp{Offringa2014, OffringaSmirnov2017}) and a Briggs \textit{robust} parameter of $r = 0$, cleaning to 0.5$\sigma$ within a \textsc{SoFiA-2}-generated mask \citep{Serra2015}. The gain phase was self-calibrated using \textsc{cubical} \citep{Kenyon2018} with a solution interval of 128 seconds, after which the final continuum model was removed from the calibrated visibilities with the \textit{msutils}. Secondly, any remaining continuum emission was subtracted by fitting a second-order polynomial to the visibility spectra with \textsc{uvlin} while excluding channels known to host \HI\ from the fit.
The continuum-subtracted \emph{uv}-data were then imaged with \textsc{wsclean} to produce a final \HI\ cube, using a Briggs \textit{robust} weighting of 0.5 as a compromise between the natural and uniform weight. The cubes generated have a pixel size of 3\arcsec\ and a field of view (FOV) of 1.5 deg$^{2}$. The resulting rms noises per pointing and restoring Gaussian PSF FWHM are listed in Table \ref{noise-per-cube}. 

\setlength{\tabcolsep}{4pt} 

\begin{table}
\caption{Observed \VirgoIII pointings and their noise properties}\label{noise-per-cube}
\resizebox{\columnwidth}{!}{%
\begin{tabular}{lllccc}
\toprule
Pointing & $\alpha_{J2000}$ & $\delta_{J2000}$ & rms & Beam~size & Beam~PA \\
    & (hms) & (dms) & $\left(\mathrm{\frac{mJy}{bm}}\right)$  & (\arcsec$\times$\arcsec) & ($^\circ$) \\
  (1)    &  (2)  & (3)   & (4) & (5) & (6) \\
\midrule
J1337+0853 & 13:37:32 & 08:53:06 & 0.27 & 18.5$\times$09.8 & 147.2 \\
J1355+0512 & 13:55:48 & 05:12:00 & 0.29 & 18.2$\times$08.3 & 155.6 \\
J1406+0601 & 14:06:32 & 06:01:44 & 0.27 & 16.9$\times$11.1 & 155.5 \\
J1420+0336 & 14:20:39 & 03:36:36 & 0.27 & 19.9$\times$10.6 & 146.7 \\
J1423+0143 & 14:23:00 & 01:43:00 & 0.26 & 15.5$\times$10.7 & 128.6 \\
J1430+0322 & 14:30:24 & 03:22:12 & 0.26 & 17.4$\times$09.4 & 152.8 \\
J1433+0427 & 14:33:24 & 04:27:00 & 0.26 & 17.6$\times$09.3 & 154.3 \\
J1438+0300 & 14:38:48 & 03:00:00 & 0.27 & 21.7$\times$09.6 & 155.8 \\
J1444+0157 & 14:44:56 & 01:57:18 & 0.27 & 22.1$\times$08.8 & 157.9 \\
J1451+0242 & 14:51:12 & 02:42:00 & 0.27 & 22.2$\times$09.6 & 155.5 \\
J1453+0319 & 14:53:05 & 03:19:54 & 0.28 & 16.6$\times$09.5 & 155.3 \\
J1502+0150 & 15:02:04 & 01:50:28 & 0.26 & 16.2$\times$15.2 & 148.3 \\
J1504+0112 & 15:04:48 & 01:12:00 & 0.26 & 21.0$\times$09.9 & 154.5 \\
J1506+0151 & 15:06:20 & 01:51:00 & 0.25 & 16.6$\times$09.7 & 151.7 \\
J1509+0130 & 15:09:24 & 01:30:00 & 0.26 & 21.9$\times$09.9 & 155.2 \\
\bottomrule
\multicolumn{6}{p{\columnwidth}}{\textbf{Notes:} Columns are (1) Designated pointing labels, (2) and (3) right ascension (J2000) and declination (J2000) of the pointings' centres, (4) noise rms,  (5) and (6) synthesised beam sizes and position angles.} \\
\end{tabular}}
\end{table}

\subsubsection{Multi-resolution imaging}
To characterise \HI\ in galaxy discs and outskirts simultaneously, we produced a set of multi-resolution data cubes from our observations. Following established MeerKAT \HI\ imaging strategies (e.g. \citealp{Serra2023, deBlok2024}), we adjusted the Briggs robust weighting combined with \emph{uv}-tapering to produce \HI\ cubes at six angular resolutions, labelled R1 to R6, ranging from 8\arcsec\ to 90\arcsec. At a median distance of 30~Mpc for \VirgoIII \citep{CastignaniI2022}, we can map \HI\ at physical resolutions ranging from 1.2 to 14.2\,kpc. Table~\ref{tab:tapering} summarises the Briggs weighting parameters and \emph{uv}-tapering applied to produce these cubes, together with the resulting angular resolutions, rms noise levels, and \HI\ column density sensitivities calculated at the $3\sigma$ level over a linewidth of $\sim25$\kms. The achieved \HI\ column density sensitivities range from approximately $4 \times 10^{20}$ atoms cm$^{-2}$ in the highest-resolution R1 cubes to $4 \times 10^{18}$ atoms cm$^{-2}$ in the lowest-resolution R6 cubes
Unless stated otherwise, we adopt the R2 data cubes as the default resolution used throughout this paper, as they provide a good compromise between angular resolution and sensitivity. At this resolution, the \HI\ mass sensitivity limit of the survey, assuming a spatially unresolved source, is $2.6 \times 10^{6}$ \MSUN\ as the $3\sigma$ noise level over linewidth 45 \kms.  

\setlength{\tabcolsep}{2pt}
\begin{table}
 \caption{Average \HI\ cube properties at varying angular resolutions}
  \begin{tabular}{lcccccc}
    \hline
     Res. &\emph{r} & \emph{uv}-tapering  & Beam~size & Beam~PA & rms  & N$_{\rm HI}^{3\sigma \sim25\rm{km/s}}$    \\ [0.5ex] 
      &        & \arcsec             & \arcsec $\times$ \arcsec & deg & $\mathrm{\frac{mJy}{bm}}$ & atoms/cm$^{2}$ \\ [0.5ex]
    (1)   & (2)    & (3)                 & (4)                    & (5) & (6) & (7) \\ [0.5ex]
    \hline\hline
    R1&0.0 & 0 & $10.2 \times 07.1$ & $150$ & 0.34 & $3.7 \times 10^{20}$ \\
    R2&0.5 & 0 & $17.4 \times 09.7$ & $152$ & 0.27 & $1.3 \times 10^{20}$  \\
    R3&0.5 & 15 & $25.9 \times 20.7$ & $151$ & 0.33 & $4.8 \times 10^{19}$ \\
    R4&0.5 & 30 & $47.1 \times 40.7$ & $161$ & 0.32 & $1.3 \times 10^{19}$ \\
    R5&0.5 & 60 & $68.7 \times 65.9$ & $90$ & 0.38 & $7.0 \times 10^{18}$ \\
    R6&1.0 & 90 & $102.1 \times 96.2$ & $46$ & 0.46 & $3.9 \times 10^{18}$ \\
   \hline 
   \multicolumn{7}{p{\linewidth}}{\textbf{Notes:} Columns are (1) Angular resolution label, (2) Briggs weighting robustness parameter, (3) \emph{r}, (4) Gaussian \emph{uv} taper applied, (5) and (6) averaged synthesised beam sizes and position angles, and (7) averaged noise rms.} 
  \end{tabular}
  \label{tab:tapering}
\end{table}

\section{\HI\ Detections}\label{sec:det}
We used the source-finding application \textsc{SoFiA-2} \citep{Serra2015, Westmeier2021} to search for \HI\ emission in the data cubes. Noise scaling was enabled to account for spatial and spectral variations in the noise, using a spatial window of 170 pixels in each direction and a spectral window of 15 channels. To remove residual baseline ripples, we applied the ripple filter with spatial and spectral windows of 31 pixels and 15 channels, respectively.
Source detection was performed using the \textit{smooth-and-clip} (S+C) method with a detection threshold of \(4\sigma\). Spatial smoothing kernels of 0, 3, and 6 pixels were applied, together with spectral smoothing kernels of 0, 3, and 7 channels, corresponding to velocity widths of 0, 16.5, and 38.6\,\kms. Detected voxels above the \(4\sigma\) threshold were linked, requiring a minimum extent of three pixels in each spatial direction and at least one spectral channel.
To separate real sources from the noise, we adopted a \textit{reliability} threshold of 0.9 and required a minimum integrated signal-to-noise ratio of 3.0. All detections were subsequently inspected visually, and candidates consistent with residual noise or imaging artefacts were removed. This procedure produced the final verified detection masks. We then reran the \emph{linker}, without repeating the source-finding step to derive accurate source parameters within the final detection masks.

Applying this procedure yielded 80 \HI\ detections. Their spatial distribution along the filament is shown in Fig.\,\ref{fig:spatial_dist_pointing}. Of these, 8 were not part of the original target sample but lie within the spatial and velocity range of \VirgoIII. A total of 70 of the 80 detections are associated with clear optical counterparts, while 2 have more tentative but accepted counterparts based on their projected separation from the corresponding galaxy centres. The remaining 8 detections have no previously known optical counterparts. Of these, 6 show evidence of diffuse optical emission; however, they are not currently identified in existing optical catalogues and lack spectroscopic redshift confirmation. The full catalogue is provided in Table~\ref{sample:tab1}.

\subsection{Comparison of the derived \HI mass}\label{HIfluxcomp}

As discussed in Sect.~\ref{sec:sample}, the \VirgoIII sample was selected from single-dish \HI\ observations. Therefore, previous \HI\ mass measurements are available for the majority of galaxies detected with MeerKAT. These measurements were taken with the Nan\c{c}ay Radio Telescope (NRT; \citealp{Theureau2017}), which has a beam size of $4\arcmin \times 22\arcmin$ at $\delta \leq 25^{\circ}$, and with the Arecibo telescope, which has a beam size of $\sim3.3\arcmin$ \citep[e.g.][]{Rosenberg2000, GiovanelliALFALFA2005}.

\begin{figure}
   \centering
   \resizebox{\hsize}{!}{
 \includegraphics{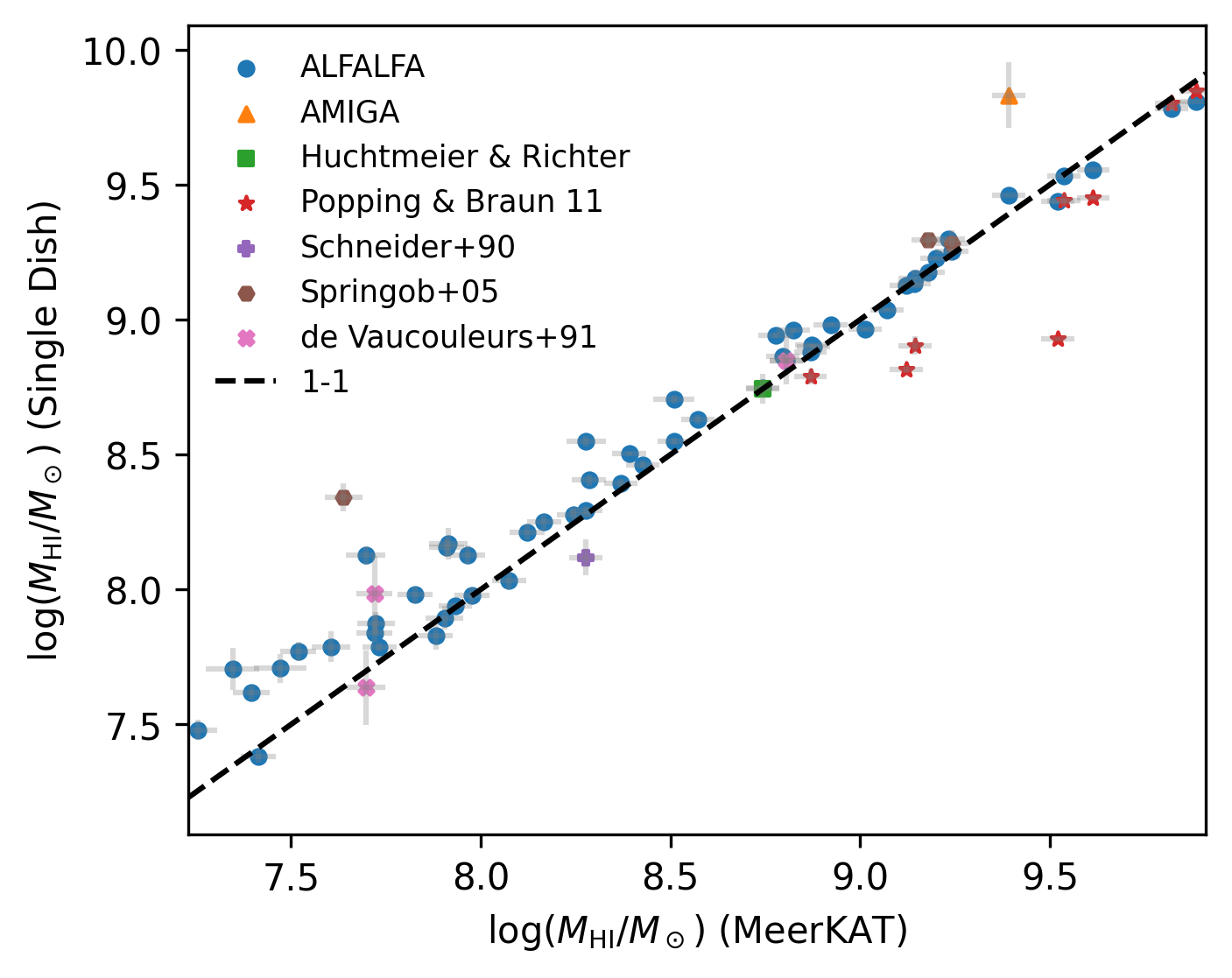}}
    \caption{Comparison of single-dish and MeerKAT \HI\ masses, where the latter measurements are derived from the 1\arcmin-resolution (R5) cubes. The dashed black line represents the 1:1 relation. Different colours and symbols denote the various literature sources of the single-dish \HI\ measurements, with the majority of data points extracted from the ALFALFA survey \citep{HaynesAlfalfa2018}.}\label{fig:compare_fluxes}
 \end{figure}

In Fig.\,\ref{fig:compare_fluxes}, we compare the total \HI\ masses derived from MeerKAT with those from single-dish observations. To best match the beam sizes of the single-dish measurements, we use the MeerKAT \HI\ masses measured from the 1\arcmin-resolution cubes (R5). MeerKAT flux uncertainties are calculated by \textsc{SoFiA-2} as $\sigma_{S_{\rm sum}} = \sqrt{\frac{N_{\rm pix}}{\Omega_{\rm PSF}}} \Delta z~\sigma_{\rm rms}$, where $N_{\rm pix}$ and $\Delta z$ are the total number of pixels in the 3-D source mask and the width of a single spectral channel, respectively, and $\Omega_{\rm PSF}$ is the beam solid angle. This uncertainty is purely statistical. The true uncertainty is difficult to measure given complex effects like gain calibration errors, residual sidelobes because of finite cleaning thresholds, missing short spacings in the interferometer, Eddington bias, source extraction biases, etc. We apply a 10\% uncertainty floor in quadrature to the MeerKAT statistical uncertainties as a simple approximation of these other effects. The single-dish uncertainties are taken as reported. 

Overall, the cross-matched measurements are in decent agreement, with a root mean squared error of $0.19$~dex around the 1:1 relation for galaxies with $\log M_{\rm HI}$/\MSUN\ $> 8.5$. At lower masses ($\log M_{\rm HI}$/\MSUN\ $< 8.5$), however, the single-dish \HI\ masses, which are predominantly from ALFALFA \citep{Haynes2018}, are systematically higher than those measured with MeerKAT. This discrepancy is likely driven by differences in sensitivity. The ALFALFA survey has a typical rms noise of $\sim2.0$\,mJy\,beam$^{-1}$ per channel for a channel width comparable to our data (5\,\kms; \citealp{Haynes2018}), whereas the MeerKAT observations reach an rms noise level that is approximately $5\times$ lower (see Table\,\ref{tab:tapering}). As shown in Fig.~3 of \citet{HaynesALFALFA2011}, at the median distance of the \VirgoIII filament, the ALFALFA \HI\ mass detection limit is approximately $\log(M_{\rm HI}$/\MSUN) $\sim8.5$. Below this limit, we are likely seeing an Eddington bias where, near the detection threshold, faint sources are preferentially included only when noise scatters their measured flux upward, leading to the observed offset toward higher single-dish \HI\ masses \citep{Riseley2016}.

\begin{figure}
    \centering
    \includegraphics[width=\columnwidth]{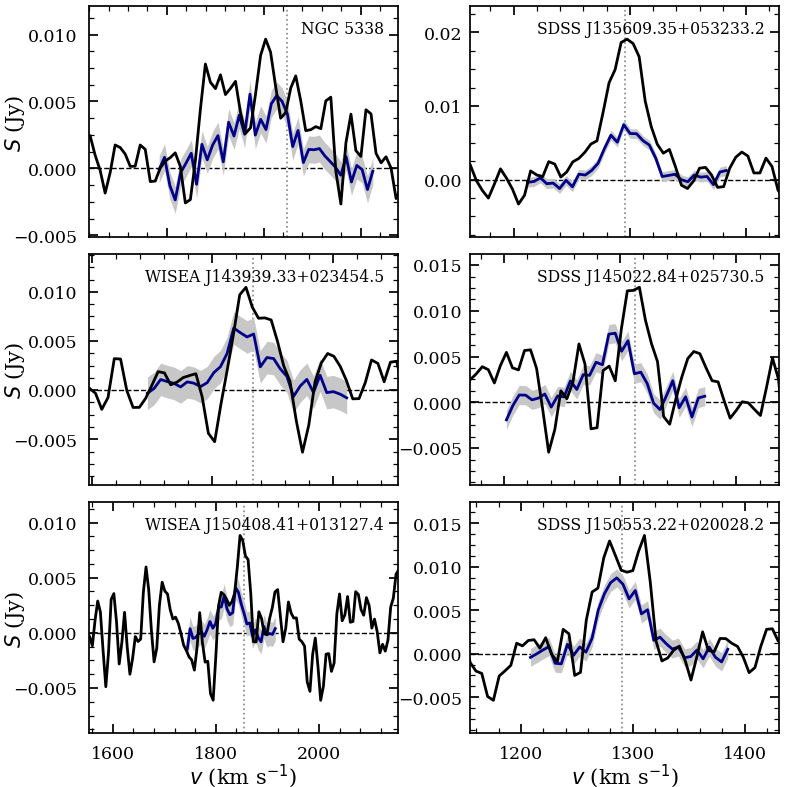}
    \caption{\HI\ spectra of galaxies whose single-dish and MeerKAT \HI\ masses deviate by more than $2.5\sigma$ from the 1:1 relation in Fig.\,\ref{fig:compare_fluxes} at lower masses end ($\log (M_{\rm HI}$/\MSUN) $<$ 8.5). MeerKAT spectra extracted from the 1\arcmin-resolution (R5) cubes are shown in blue, with the shaded band indicating the uncertainty per channel at 1$\sigma$. The corresponding ALFALFA spectra \citep{Haynes2018} are shown in black. The dotted vertical line marks the systemic velocity.}\label{HIspec}
\end{figure}

In order to limit detecting too many noise features as real sources, we exclude pixels with very low surface brightnesses, keeping a $4\sigma$ detection threshold. We have considered whether faint \HI\ emission below our $4\sigma$ detection threshold could remain not deconvolved in the residual cubes and lead us to underestimate the MeerKAT masses. To test this, we compare the MeerKAT and ALFALFA spectra in Fig.\,\ref{HIspec} of the galaxies deviating by more than $2.5\sigma$ from the 1:1 relation in Fig.\,\ref{fig:compare_fluxes}, and have a single-dish spectrum available in the literature. The MeerKAT profiles recover the same line shapes and velocity centroids; however, they lie below them by a roughly uniform factor across all channels. Emission left in the residuals would instead produce a deficit concentrated in the extended, low-surface-brightness channels, distorting the profile shape rather than scaling it down uniformly \citep{deBlok2024}. We therefore attribute the low-mass offset to the ALFALFA sensitivity limit and the associated Eddington bias.

We find the detection threshold to play a more noticeable role for faint sources at our high resolutions (R1 and R2) than at the 1\arcmin\ resolution (R5), where our mass comparison is made. At R5, the total flux for several sources increases by only $\lesssim 5\%$ when the detection threshold is lowered to $3\sigma$, confirming that the choice of threshold has little effect on the masses at the resolution used for our comparison.

A particularly strong example of the interplay between the resolution, detection threshold, and total detected mass (an outlier, to be clear) is NGC~5576, an early-type galaxy \citep{deVaucouleurs1991Catalog}, with an ALFALFA \HI\ mass of $\log (M_{\rm HI}$/\MSUN) $= 8.28 \pm 0.11$ but a much lower \cite{deVaucouleurs1991Catalog} \HI\ mass of $\log (M_{\rm HI}$/\MSUN) $= 7.64 \pm 0.14$ (Table~A.2). At MeerKAT R2 resolution, we detect a compact \HI\ feature approximately $131\arcsec$ from the optical centre of NGC~5576, with no clear optical counterpart, as shown in Fig.\,\ref{fig:HIclouds}. This feature is blueshifted by $\sim170$\,\kms\ relative to the systemic velocity of NGC~5576 and may trace \HI\ associated with the galaxy outskirts.

At R2 resolution, this feature has an \HI\ mass of $\log (M_{\rm HI}$/\MSUN) $= 6.7 \pm 0.5$, increasing by $\sim0.4$~dex if the detection threshold is set to $3\sigma$. At lower resolutions, the feature (with a $4\sigma$ detection threshold) becomes larger and more clearly linked to NGC~5576, with its mass increasing to $\log (M_{\rm HI}$/\MSUN) $= 7.70 \pm 0.15$ at R5, in good agreement with \citet{deVaucouleurs1991Catalog}. The increase in mass by changing the detection threshold to $3\sigma$ becomes negligible (under $0.05$~dex) at the low resolution end.

\begin{figure}
    \centering
    \includegraphics[width=\columnwidth]{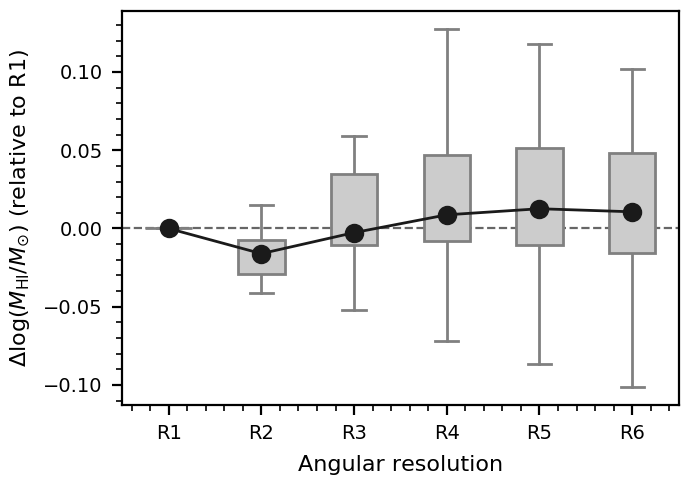}
    \caption{Difference in \HI\ mass measured at each angular resolution relative to the highest resolution cube (R1), shown as a function of angular resolution. Black points indicate the median value at each resolution, while the grey boxes show the interquartile range. The error bars indicate the full range/spread of the distribution.}\label{HImassvsres}
\end{figure}

This trend of increasing masses at lower resolutions is present in our full MeerKAT sample, though with significant scatter and typically much smaller changes than seen for NGC~5576. In Fig.\,\ref{HImassvsres}, we show the median difference in \HI\ mass measured at each angular resolution relative to the highest-resolution cube (R1). The median \HI\ mass actually decreases between R1 and R2 by $\sim 0.016$~dex, and then increases monotonically towards lower angular resolution. Between R1 and R2, the decrease is caused by the slightly larger impact of diffusing the observed flux over a wider area and thus going under the $4\sigma$ threshold than the impact of increasing the sensitivity. The later increasing trend converges by the 1\arcmin-resolution (R5), suggesting that at this resolution the majority of the \HI\ emission associated with each galaxy in our survey is recovered by MeerKAT. Further \emph{uv}-tapering to lower angular resolutions is limited by the rapidly increasing rms noise in the most heavily tapered cubes (R6), as it down-weights a large fraction of the measured visibilities and reduces the effective sensitivity of the interferometer as expected for MeerKAT (e.g. \citealt{Briggs1995, Booth2009}).

\section{\HI\ data products overview} \label{sec:res}
This section provides an overview of the \HI\ data products derived from our MeerKAT observations and describes how they are presented in the atlas included in this paper. For all detections, a uniform set of \HI\ products is shown at a single reference angular resolution (R2), enabling direct and consistent comparison across the sample. Additionally, a couple of representative galaxies are used to illustrate how imaging the data at different angular resolutions affects the recovery of \HI\ emission.

\subsection{Spatial \HI products}
The primary \HI\ diagnostic used in this work is the zeroth-moment (moment-0) map, which traces the spatial distribution of the integrated \HI\ emission and provides a direct view of its extent and morphology. As this product is often the first indicator of spatial structure on different angular scales, we examine the moment-0 maps at multiple angular resolutions. Additionally, spatially integrated \HI\ spectra are extracted using the same source masks, to produce global profiles that are directly comparable to those obtained from single-dish observations and that quantify the total \HI\ emission recovered at each resolution.

\subsection{\HI kinematics}
To characterise the \HI\ kinematics of the detected galaxies, we also derive first- and second-moment maps (moment-1 and moment-2) from the primary beam-corrected data cubes, using the final \textsc{SoFiA-2} masks to isolate the \HI\ emission. The moment-1 maps represent the intensity-weighted mean velocity along each line of sight and provide a two-dimensional velocity field describing the large-scale kinematics of the atomic gas. These maps form the basis for quantitative kinematic analyses, such as rotation-curve modelling with tilted-ring techniques. The moment-2 maps represent the intensity-weighted velocity dispersion of the \HI\ emission along each line of sight and reflect the measured line-of-sight velocity width of the detected emission. These maps were derived from the masked, primary beam-corrected data cubes. Pixels with signal below a $3\sigma$ threshold in the corresponding moment-0 maps were blanked before computing the moment-2 maps. Additionally, position--velocity diagrams (PVDs) were generated for each detection to provide a complementary representation of the \HI\ kinematics. The PVDs show the \HI\ emission projected along a defined spatial axis as a function of line-of-sight velocity. For each galaxy, the slice orientation was determined from the kinematic major axis measured from the velocity field maps. As with the other products, the PVDs were extracted from the masked, primary beam-corrected cubes using \textsc{SoFiA-2}. The slice width is often of a few synthesised beams at the adopted angular resolution.

\subsection{Atlas presentation}
\begin{figure*}
  \centering
  \includegraphics[width=0.72\textwidth]{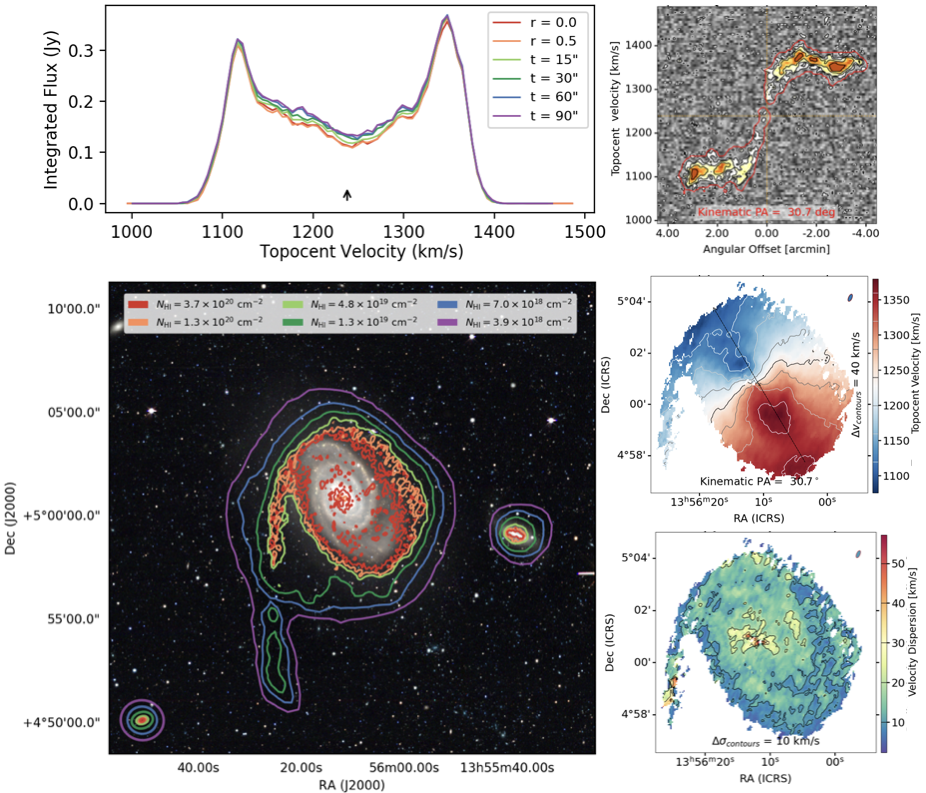}\hfill
  \includegraphics[width=0.72\textwidth]{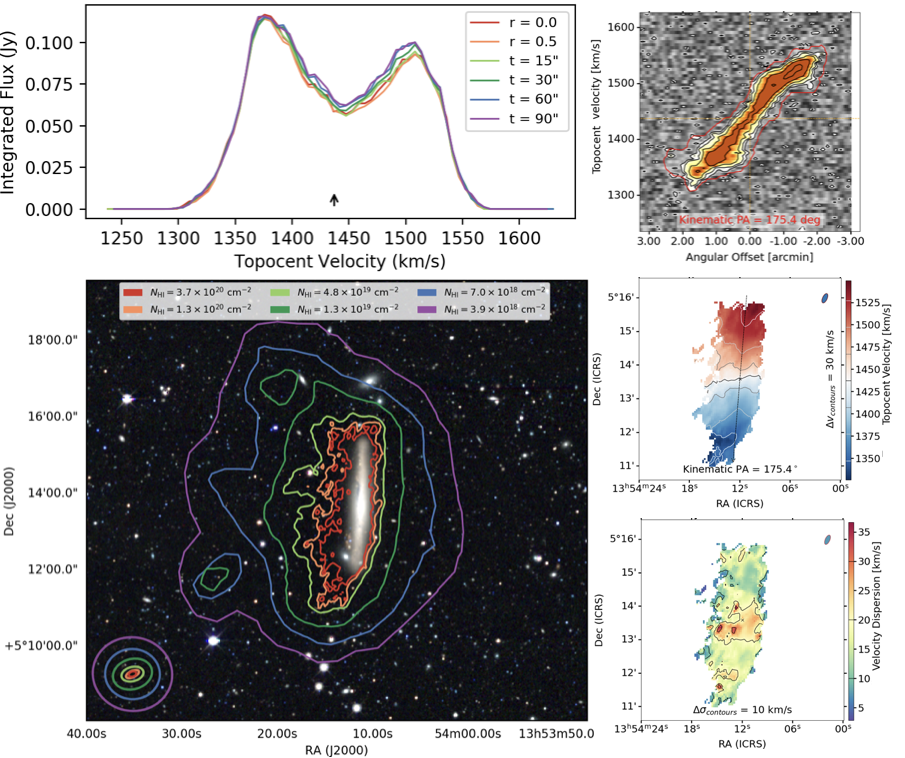}
    \caption{Examples of the MeerKAT \HI\ data products and atlas layout. The top composite panel shows NGC\,5364 and the bottom shows NGC\,5348. In each composite panel, the bottom left shows the \HI column density contours overlaid on the DECaLS \textit{grz} optical image. Contours are shown at the $3\sigma$ column density sensitivity for each angular resolution: R1 (red), R2 (orange), R3 (light green), R4 (dark green), R5 (blue), R6 (purple). The top left panel shows the integrated \HI\ spectra at each resolution. The top right panel is the position velocity diagram extracted at the default, R2 angular resolution along the kinematic major axis, and the red contour is the \textsc{SoFiA-2} detection mask. The middle and lower right panels are the moment-1 (velocity field) maps and  moment-2 (velocity dispersion ) maps, respectively.} \label{multiresIm}
\end{figure*}

Figure~\ref{multiresIm} presents the layout and full set of \HI\ data products for two representative galaxies, NGC~5364 and NGC~5348. For each system, moment-0 maps and spatially integrated \HI\ spectra are shown at all six angular resolutions (R1--R6). The moment-1 and moment-2 maps, as well as the position--velocity diagrams, are shown at the default R2 angular resolution, consistent with the presentation adopted for the full atlas. All products shown in Fig.~\ref{multiresIm} were generated using the SoFiA Imaging Pipeline (SIP)\footnote{\url{https://github.com/kmhess/SoFiA-image-pipeline/}}.
The multi-resolution presentation in Fig.~\ref{multiresIm} is intended to demonstrate the effect of angular resolution on the recovered \HI\ emission. A more detailed analysis of these galaxies is provided by \citet{Finn2025NGC5364}. For consistency across the sample, the atlas figures presented in Appendix~\ref{sec:allhiim} are shown at default (R2) resolution for direct comparison of \HI\ morphology and kinematics among all detected galaxies.

\section{\HI properties in the Virgo~\textsc{iii} filament}\label{sec:properties}
\subsection{\HI deficiency}
The \HI\ deficiency is one of the most direct observational indicators of environmental gas removal in galaxies. Quantifying the degree of \HI\ deficiency therefore provides a first-order assessment of the impact of the environment on the cold gas reservoirs of galaxies.

\begin{figure}
    \centering
    \includegraphics[width=85mm, height=65mm]{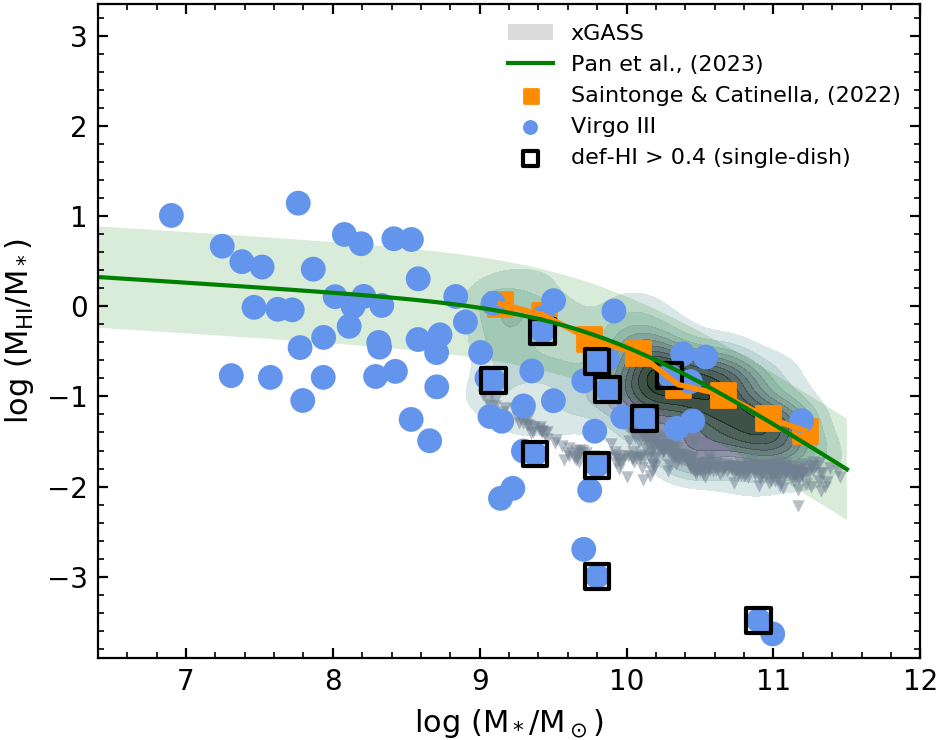}
    \caption{\HI\ gas as a function of stellar mass for galaxies in the \VirgoIII filament. The blue points are our detected galaxies. Those outlined by black open squares were classified as \HI-deficient (def$_{\rm HI} \geq 0.4$) from single-dish measurements in the literature \citep{CastignaniI2022}, which formed part of our target selection criteria. The grey shading shows the xGASS sample, with triangles indicating upper limits for xGASS non-detections. The orange points show the \HI\ scaling relation from xGASS extracted from \citet{Saintonge2022}. The green solid line shows the Bayesian \HI\ gas scaling model of \citet{Pan2023} with its intrinsic scatter shown as green shading.}\label{HIdefsc}
\end{figure}

Using integrated \HI\ measurements, \citet{CastignaniI2022} demonstrated that galaxies residing in nearby Virgo filaments, including Virgo~\textsc{iii}, are on average \HI-deficient relative to isolated systems of the same morphological type from the AMIGA sample \citep{Jones2018}. The galaxies targeted by our MeerKAT survey were selected in part based on this evidence for reduced \HI\ content. However, as mentioned in Section~\ref{sec:sample}, our selection was also optimised to detect other galaxies within the pointings, resulting in a broader sample that includes galaxies not selected on the basis of \HI\ deficiency. This provides an opportunity to assess whether \HI\ depletion in Virgo~\textsc{iii} is a widespread environmental effect, or whether it is confined to the pre-selected population. With our MeerKAT data, we revisit this result using our \HI\ mass measurements. We compute \HI\ deficiency by comparing the observed \HI\ gas fractions, $\log(M_{\rm HI}/M_\star)$, to reference gas-fraction scaling relations as a function of stellar mass. As a first reference, we adopt the xGASS scaling relations \citep{Catinella2018, Saintonge2022}, which provide a well-characterised baseline for nearby field galaxies. However, the xGASS sample is limited to stellar masses $\log(M_\star/M_\odot) \gtrsim 9$, whereas our MeerKAT detections extend to significantly lower stellar masses (Fig.~\ref{HIdefsc}). As a result, the xGASS \HI\ scaling relation is not constrained over the full stellar mass range of our sample. To extend the comparison to lower stellar masses, we additionally make use of the \HI\ gas scaling relation presented by \citet{Pan2023}. This relation is derived using a Bayesian technique as shown in their Equation~5 and Table~4 (Model B), which models the dependence of the \HI\ mass on stellar mass while accounting for intrinsic scatter and measurement uncertainties. The intrinsic scatter $\sigma_{\HI}$ is the Gaussian standard deviation of $\log M_{\HI}$ at fixed stellar mass. The relation is constrained down to $\log(M_\star/M_\odot)\sim8$, significantly below the stellar mass limit of the xGASS sample.

To distinguish between galaxies selected on the basis of \HI\ deficiency and others detected within our MeerKAT pointings, we show the two subsamples separately in Fig.~\ref{HIdefsc}. As expected by construction, the def$_{ \rm HI} \geq 0.4$ subsample lies systematically below the field gas-fraction scaling relations, with the majority of galaxies offset by a median of $\sim0.45$\,dex, and substantial scatter ($\sim 0.3-0.5$ dex). Importantly, a large fraction ($\sim40$\%) of the non-\HI-deficient-selected galaxies also lie below the scatter of the \citet{Pan2023} relation over the stellar mass range $7.5 \lesssim \log(M_\star/M_\odot) \lesssim 10.5$. Relative to the mean relation the fraction is larger still, with $\sim60$\% lying below it. This suggests that environmental \HI\ depletion in Virgo~\textsc{iii} is not only confined to the pre-selected \HI-deficient population, and that a significant fraction of galaxies across the full stellar mass range of our survey show suppressed \HI\ gas fractions relative to field expectations. The remaining $\sim$60\% of non-\HI-deficient-selected galaxies are broadly consistent with the Pan et al.~scaling relation, indicating that Virgo~\textsc{iii} also retains a population of galaxies with field-like \HI\ content.

\begin{figure}
    \centering
    \includegraphics[width=\columnwidth]{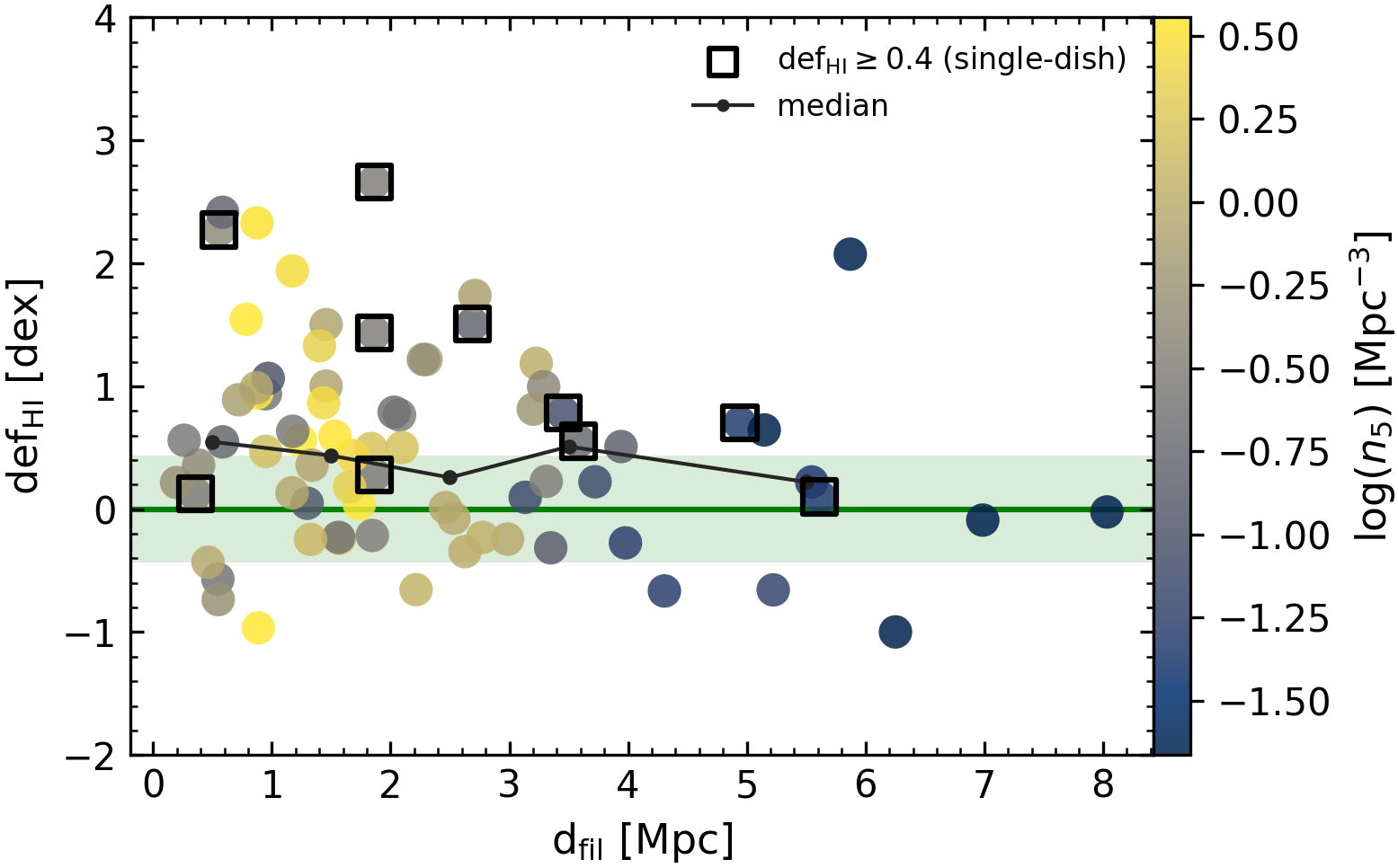}
    \caption{\HI\ deficiency as a function of distance to the filament spine. The deficiency is computed as the difference between the expected gas fraction from the \citet{Pan2023} scaling relation and the observed \HI-gas fraction, with positive values indicating gas-poor galaxies. The green line and shading mark this scaling relation and its intrinsic scatter. Points are colour-coded by the local density, $\log(n_5)$. The black line shows the running median in 1\,Mpc bins of $d_{\rm fil}$ (with at least five galaxies). Black open squares mark galaxies classified as \HI-deficient from single-dish measurements.}\label{HIdefdfil}
\end{figure}

To investigate whether the \HI\ deficiencies measured above are related to the large-scale filamentary environment, we show Fig.~\ref{HIdefdfil}, which is the \HI-deficiency as a function of distance to the filament spine, $d_{\rm fil}$, as well as local density, $n_5$. We find no clear monotonic dependence of \HI\ deficiency on $d_{\rm fil}$. In other words, gas-poor galaxies are found both close to and several Mpc from the spine. However, \HI-deficient galaxies show a mild, though not statistically significant, preference for denser local environments.  We note, though, that $n_5$ and $d_{\rm fil}$ are not independent quantities because the high-density regions in \VirgoIII\ are dominated by groups embedded within the filament. As a result, the two effects are difficult to separate. Indeed, a partial correlation analysis\footnote{Spearman rank correlation of def$_{\rm HI}$ with $d_{\rm fil}$ at fixed $n_5$, computed on rank residuals ($\rho = -0.04$, $p = 0.73$).} shows that, once the dependence on local density is taken into account, no residual trend of \HI\ deficiency with $d_{\rm fil}$ remains. Our MeerKAT data are therefore consistent with \HI\ removal being driven primarily by local density, and we find no evidence for an additional dependence on proximity to the large-scale filament.

\subsection{\HI Asymmetries}\label{subsec:asymmetries}

In this study, we use the asymmetry index as a simple and robust tracer of large-scale disturbances in the \HI\ distribution since the extended \HI\ component is particularly sensitive to external processes such as tidal interactions and ram-pressure stripping. For each detected galaxy, the \HI\ asymmetry was computed using the \texttt{statmorph} package \citep{RodriguezGomezSM2019}. The asymmetry index compares the original image to a version rotated by $180^\circ$, using the flux-weighted centroid of the \HI\ distribution as the centre of rotation. The asymmetry index is defined as follows. 

\begin{equation}
 A = \frac{\sum_{i,j} \left| I_{ij} - I^{180}_{ij} \right|}{\sum_{i,j} \left| I_{ij} \right|} - A_{\rm bgr},
 \end{equation}
where $I_{ij}$ and $I^{180}_{ij}$ denote the pixel intensities in the original and rotated images, respectively and $A_{\rm bgr}$ accounts for the contribution from background noise. In this work, asymmetries are measured on masked \HI\ maps, with the calculation restricted to pixels within the source detection mask. As a result, we set $A_{\rm bgr}$ to zero. The detection masks are defined at a $4\sigma$ detection threshold (see Sect.\,\ref{sec:det}), therefore all the pixels used in the asymmetry calculation have a signal-to-noise ratio of at least 4, which minimises the contribution of the noise within the mask. Restricting the calculation to the source mask reduces background noise; however, because the asymmetry parameter is positive definite, it remains subject to some noise bias \citep{RudnickRix1998}.

\begin{figure}
    \centering
    \includegraphics[width=85mm, height=50mm]{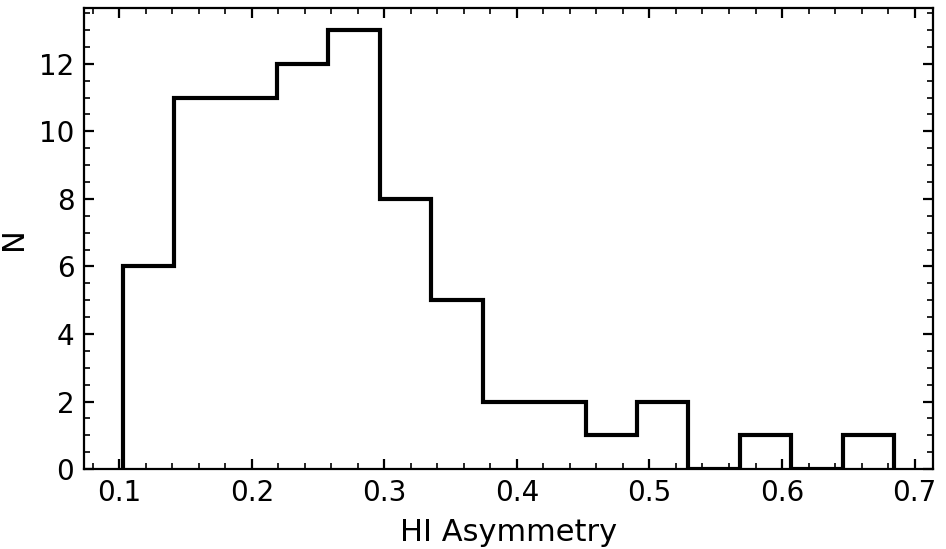}
    \caption{Distribution of \HI\ asymmetry indices measured from the \VirgoIII \HI intensity maps.}
    \label{fig:asymmetries}
\end{figure}

The asymmetry index was measured for all detected galaxies.  Interacting pairs (NGC~5774/NGC~5775 and IC~1067/IC~1066; see Appendix A) are excluded from the analysis, as their \HI\ emission is blended both spatially and in velocity, therefore cannot be considered as individual systems. The index ranges from values close to 0 for highly symmetric systems to values exceeding 1 for strongly disturbed gas distributions. In our sample shown in Fig.~\ref{fig:asymmetries}, the measured \HI\ asymmetries ranges from $A \simeq 0.1$ to $A \simeq 0.7$, with a median value of $A \simeq 0.25$   and a mean of $A \simeq 0.27$. The median absolute deviation of the distribution is 0.06, indicating that it extends towards higher asymmetry indices. Therefore, in general, our \HI distributions are considerably more disturbed/asymmetric than stellar light in galaxies. For example, \cite{RodriguezGomezSM2019} create a sample of $\sim 27,000$ galaxies from Pan-STARRS \citep{KaiserPanstarrs2002} and measure the asymmetries of the $i-$band images and find a typical asymmetry of $A \sim 0.01$.

\begin{figure}
    \centering
    \includegraphics[width=\columnwidth]{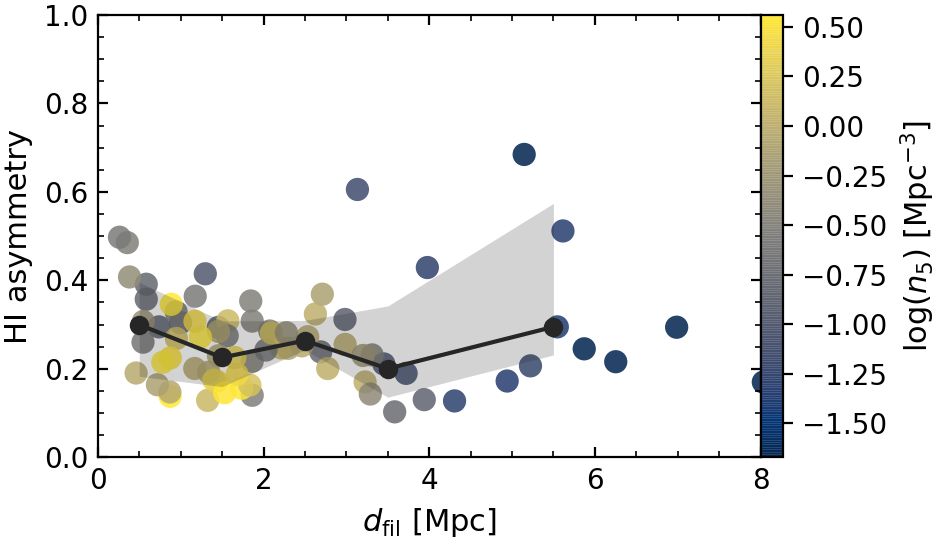}
    \caption{\HI asymmetry as a function of the distance to the filament spine. Points show individual galaxies, coloured by local density log($n_5$), where $n_5$ is the inverse of the volume occupied by the five nearest neighbours. The black points and solid line mark the median \HI asymmetry in 1 Mpc d$_{\rm fil}$ bins, while the shaded region indicates the 16th - 84th percentile range of the distribution in each bin, shown only for bins containing at least five galaxies.}
    \label{fig:asy_vs_dfil}
\end{figure}

Figure~\ref{fig:asy_vs_dfil} shows the \HI\ asymmetry index as a function of projected distance to the \VirgoIII filament spine, $d_{\rm fil}$, and the local density $n_5$ (inverse of the volume encapsulating the nearest five neighbours, shown as the colour of the points). The filament distance ($d_{\rm fil}$) is defined as the 3-D separation between each galaxy and the filament spine, as described in \citet{CastignaniI2022}. All galaxies in our sample found in pockets of higher local density (e.g., $n_5 \gtrsim 1$~Mpc$^{-3}$) are within 3-4 Mpc of the filament spine. 

To better understand the impact of the environment on the asymmetry of the \HI distribution in our galaxies, we compute the mean asymmetry in bins of $d_{\rm fil}$, with uncertainties given by the standard error on the mean. The binned measurements show that there is no strong trend between asymmetry and $d_{\rm fil}$ or $n_5$. The mean asymmetry remains between 0.2 and 0.4 over the full range of $d_{\rm fil}$. 

Nevertheless, there are some interesting subtle trends that could hint at environmental effects, which we will study in depth in the next paper of the subseries. Directly next to the spine (within $\sim 0.5$~Mpc), there is a spike of more asymmetric \HI distributions, possibly pointing to filament-based effects. There is also a weak anti-correlation between asymmetry and $n_5$ (Spearman $\rho = -0.28$, $p = 0.01$), with the most asymmetric galaxies found at lower densities. Since $n_5$ and $d_{\rm fil}$ are anti-correlated, the two trends partially offset each other. These results suggest that proximity to the filament spine and local density each have an independent, albeit weak, connection to the disturbance of the \HI\ in \VirgoIII\ galaxies, painting a complex picture of how the atomic gas is impacted in galaxies within this environment.

\subsection{\HI clouds with no optical counterparts}

\begin{figure}
   \centering
 \includegraphics[width=\columnwidth]{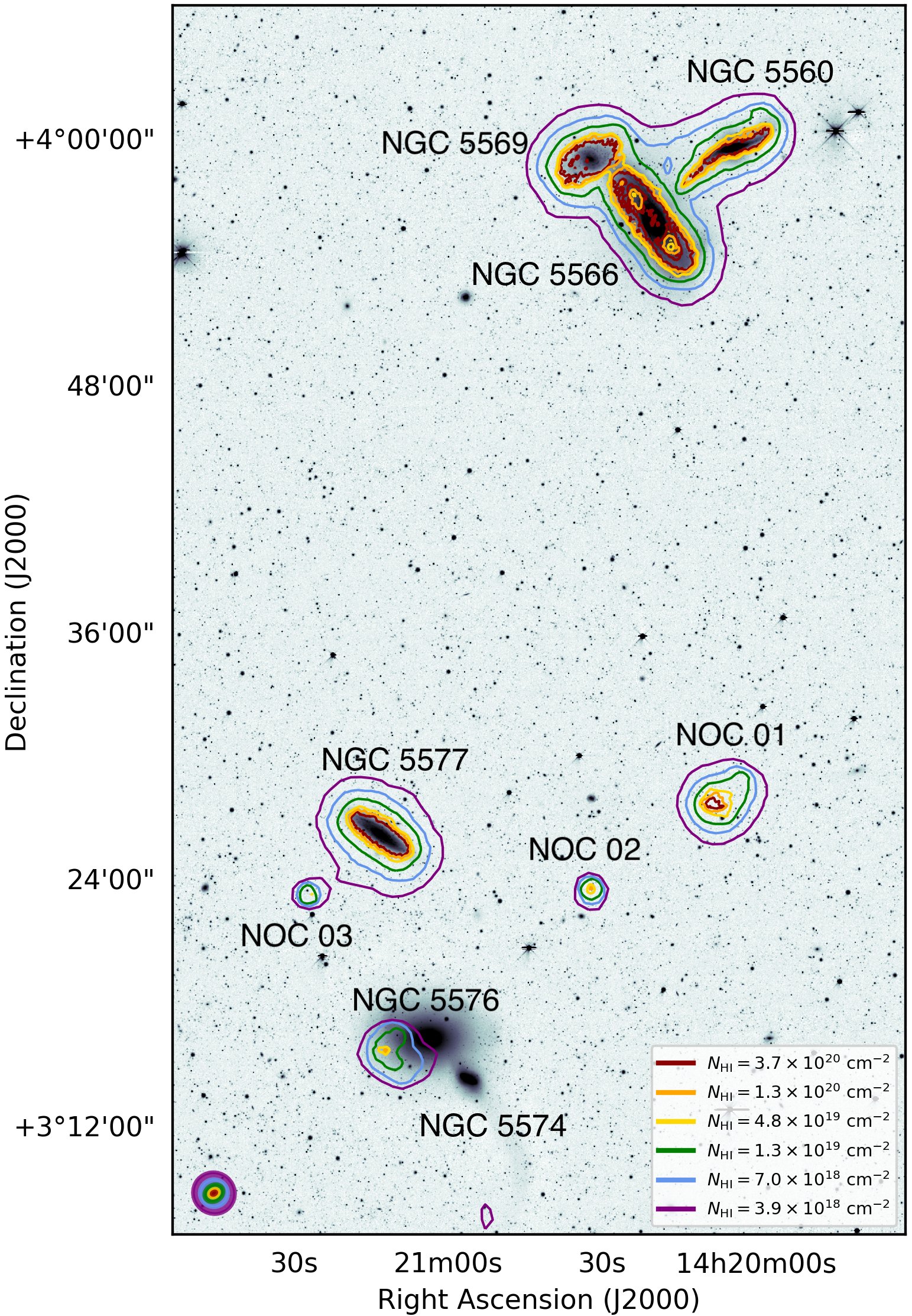}
    \caption{\HI\ contours at multiple angular resolutions, each at its corresponding column density sensitivity (see legend), overlaid on the DECaLS $g$-band image of a portion of the NGC~5566 group. The field includes the close association of NGC~5566, NGC~5569, and NGC~5560 at the top, and the two elliptical galaxies NGC~5576 and NGC~5574 and the spiral NGC~5577 at the bottom. The three \HI\ detections with no known optical counterparts are marked by ``NOC''. The filled ellipses in the bottom-left corner show the synthesised beams at each resolution, with colours matching the contours.}\label{fig:HIclouds}
 \end{figure}

In our 80 HI detections at the default resolution R2, 70 have clear optical counterparts. A further 2 \HI\ detections have more tentative but still accepted optical counterparts, based on their projected distance from the corresponding galaxy centre, while the remaining 8 have no known counterparts (see  Table~\ref{sample:tab1}). Of the 8 detections with no counterparts in the literature, 6 show evidence of a diffuse optical source counterpart, but no information is known about these galaxies as they are not (yet) found in any catalogues, signifying a lack of redshift confirmation. However, 2 \HI sources have no visible counterpart, at least down to the 29 mag arcsec$^{-2}$ $r-$band surface brightness limits of the DESI Legacy Surveys (e.g., \citealp{Delgado2023}), suggesting that they were likely stripped from one or more galaxies in the course of their interactions. These two sources, in addition to a third detected only at lower resolutions, are shown in Figure \ref{fig:HIclouds}, overlaid on the g-band image from DECaLS. 

The three \HI detections with no optical counterparts (hereafter NOCs) are located in the vicinity of the galaxy system containing NGC 5577, NGC 5576, and NGC 5574 and can be seen in the lower half of the image shown in Fig.~\ref{fig:HIclouds}. Both NGC 5576 and NGC 5574 are elliptical galaxies, while NGC 5577 is an \HI-deficient spiral galaxy with a truncated \HI\ disc. 
The close projected separation of NOC~03 ($\log M_{\rm HI}/\msun = 6.79 \pm 0.10$) from NGC~5577, and their similar velocities (see Table~\ref{sample:tab1}), suggests that it may be an extension of gas removed from this galaxy. The origin of NOC~02 ($\log M_{\rm HI}/\msun = 6.79 \pm 0.07$) and the large NOC~01 ($\log M_{\rm HI}/\msun = 8.02 \pm 0.05$) are less clear. In conjunction with the highly off-centre \HI\ emission of NGC 5576, the most plausible scenario could be that those clouds were displaced by NGC~5576 and/or NGC~5574 during a tidal interaction event. 

\begin{figure}
    \centering
    \includegraphics[width=65mm, height=110mm]{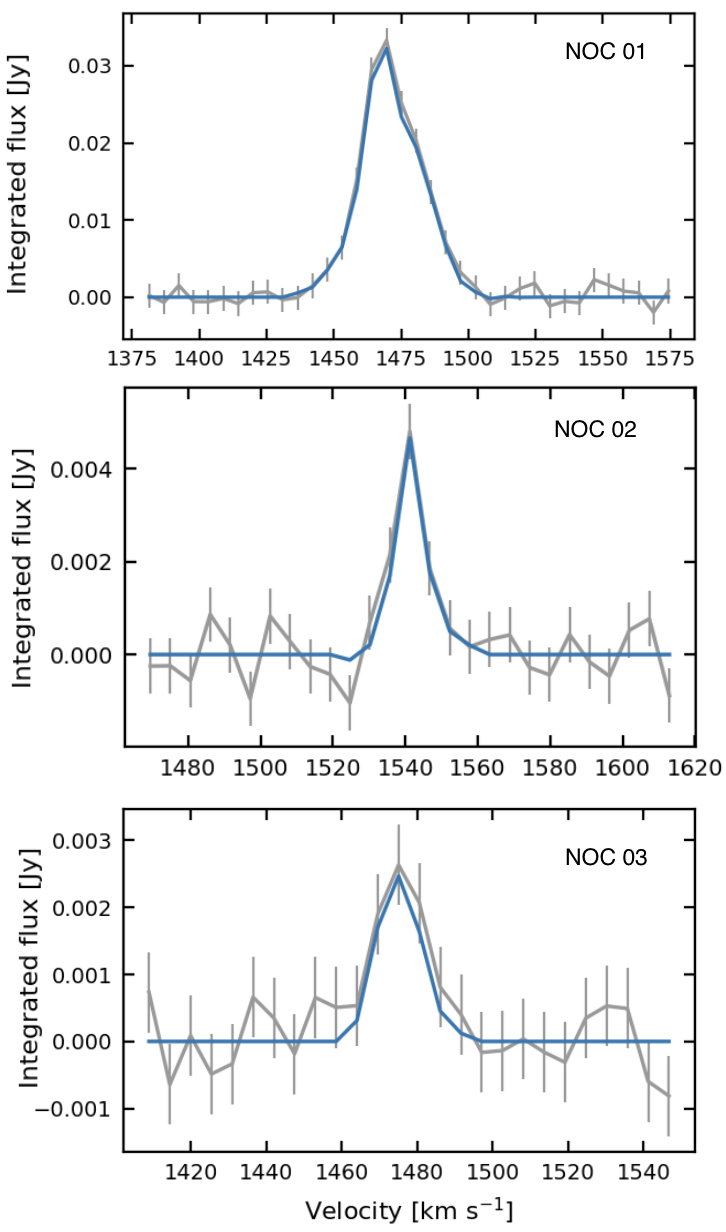}
    \caption{Integrated \HI spectra of the three \HI clouds without optical counterparts, as labelled in Fig.~\ref{fig:HIclouds}. The spectra were extracted from the R4 data cube. The grey lines show the spectra including noise around the detection, while the blue line shows the emission within the detection mask.}
    \label{fig:HIclouds_spectra}
\end{figure}

We present the \HI\ spectra of the three clouds in Fig.~\ref{fig:HIclouds_spectra}, all of which show clear signals. The peak velocities of the clouds fall within the velocity range occupied by NGC~5577 (see Fig.~\ref{fig:HIclouds}), consistent with a physical association with the NGC~5577/5576 system. The spectra are characterised by narrow, single-peaked profiles, consistent with gas that is not strongly rotationally supported, at least at the spatial scales we are probing. Such kinematic properties hint that the gas could have been displaced from a parent system recently. We note, however, that the presence of lower column-density \HI\ connecting the clouds and the nearby galaxies cannot be excluded at the sensitivity limits of the current data. Similar kinematic and morphological properties have been observed for detached \HI\ clouds in nearby galaxy groups, where tidal and hydrodynamical encounters produce narrow-lined, low-rotation \HI\ structures spatially offset from their parent systems (e.g. \citealp{Serra2013, HessWilcots2013, Koribalski2018, Serra2024}).

\begin{figure*}
   \centering
\includegraphics[width=\textwidth]{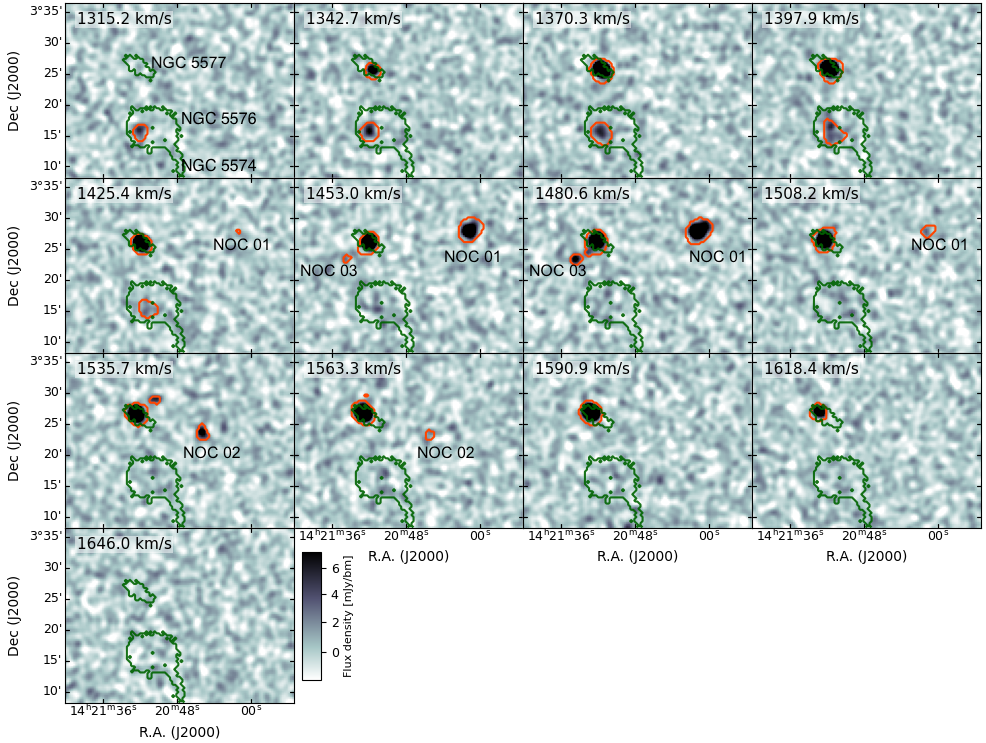}
    \caption{Channel maps of three \HI\ clouds with no optical counterparts extracted from the R4 data cube. The channel width is 27.5 \kms, obtained by binning the native spectral resolution of 5.5 \kms\ by a factor of five. Orange contours outline the detected \HI\ emission in each channel. The \HI\ clouds without counterparts are labelled following Fig.~\ref{fig:HIclouds}, while emission associated with galaxies that have optical counterparts is not labelled. The green contour outlines the 26 mag arcsec$^{-2}$ $r$-band surface-brightness level from the DESI legacy survey.}\label{fig:channel_maps}
 \end{figure*}

Further information on the NOCs is provided by the channel maps around the velocity range of the system as extracted from the R4 cube, as shown in Figure\,\ref{fig:channel_maps}. The 5.5 km/s channels were binned by a factor of 5 for visual clarity resulting in velocity steps of 27.5 km/s per channel. There are five distinct sources in the channel maps. The unlabeled sources correspond to systems with known optical counterparts: NGC~5577 (middle left) and NGC~5576 (directly below NGC~5577). NGC~5577 shows \HI\ emission over the velocity range 1342.7--1618.4 \kms, while NGC~5576 is detected from 1315.2--1425.4 \kms. 

The channel maps show that NOC~03 appears at 1453.0 and 1480.6~\kms, coincident both spatially and in velocity with the \HI\ emission of NGC~5577, supporting an association in which the cloud traces gas displaced from this galaxy. NOC~01 and NOC~02, by contrast, are spatially isolated: NOC~01 is detected between 1453.0 and 1508.2~\kms\ and NOC~02 between 1535.7 and 1563.3~\kms, in both cases with no \HI\ emission from a known galaxy nearby in either position or velocity. All three clouds are confined to a small number of channels ($\lesssim$100~\kms\ in extent), consistent with the narrow, single-peaked profiles seen in their integrated spectra.

\section{Summary} \label{sec:conc}

We present deep, spatially resolved MeerKAT observations of \HI\ in galaxies residing within the nearby \VirgoIII filament. This well-studied structure lies at a median distance of $\sim$30~Mpc, extends over $\approx16$~Mpc, and covers a sky area of roughly 56~deg$^{2}$. Our survey comprises 15 targeted MeerKAT pointings designed to probe environmental processing in this region. We therefore prioritised systems with previously reported evidence for \HI\ deficiency and those with stellar masses $M_{\star} \geq 10^{9}\,M_{\odot}$, while also including other targets within the pointings. Observations were obtained in 2021, 2022, and 2024 in the 32k correlator mode, providing a spectral resolution of 5.5\,\kms.

To characterise both the \HI\ emission in the galaxy disc and outskirts, we generated multi-resolution \HI\ image cubes spanning angular resolutions from 8\arcsec\ to 90\arcsec. Unless otherwise stated, we adopt the R2 cubes with a typical synthesised beam of 17.4\arcsec $\times$ 9.7\arcsec\ as the default resolution for the presented data products. At this resolution, our \HI\ data reach \HI\ column-density sensitivities of order $10^{20}$\,atoms\,cm$^{-2}$ (3$\sigma$ over a $\sim25$\,\kms\ linewidth).

Source finding was performed using \textsc{SoFiA-2}, followed by visual validation. We detect \HI\ emission from 80 sources within the $500$--$3000$\,\kms\ velocity range of the \VirgoIII\ filament. Of these detections, 70 have clear optical counterparts and 2 have tentative but accepted counterparts based on projected offsets, while 8 have no confirmed counterparts; 6 of the latter show diffuse optical emission in deep imaging but lack redshift confirmation. We provide an \HI\ atlas at the default R2 angular resolution, including moment-0 (integrated intensity), moment-1 (velocity field), and moment-2 (velocity dispersion) maps, as well as position--velocity diagrams for each detection. The multi-resolution performance of the data products is demonstrated using representative galaxies. A full catalogue of source properties and the complete imaging atlas are presented in the appendices.

We compare the \HI\ masses derived from MeerKAT with available single-dish measurements, predominantly from ALFALFA. The measurements are broadly consistent at $\log(M_{\rm HI}/M_{\odot}) \gtrsim 8.5$, while at lower masses the single-dish values are systematically higher. We attribute this primarily to the ALFALFA sensitivity limit and the associated selection effect near its detection threshold at the distance of the \VirgoIII\ filament.

Consistent with our survey design and earlier integrated \HI\ studies, \VirgoIII\ galaxies are, on average, offset below the field on the \HI\ gas-fraction scaling relations. Using reference relations from xGASS at $\log(M_{\star}/M_{\odot}) \gtrsim 9$ and an extended low-mass relation from \citet{Pan2023} down to $\log(M_{\star}/M_{\odot}) \sim 8$, we find a typical offset of $\sim0.45$\,dex in $\log(M_{\rm HI}/M_{\star})$ relative to field expectations, with substantial scatter.

We also quantify large-scale \HI\ disturbances using an asymmetry index measured from the moment-0 maps. The \HI\ asymmetries span $A \simeq 0.1$--0.7 in the analysed sample (excluding the blended interacting pairs NGC~5775/NGC~5774 and IC~1067/IC~1066), with a median of $A \approx 0.25$; the interacting system NGC~5775/NGC~5774 reaches $A \simeq 1.1$. We find no strong monotonic trend of \HI\ asymmetry with (3-D) distance to the filament spine, suggesting that the large-scale filament environment plays at most a minor role in directly affecting \HI morphologies. In addition, we identify three \HI\ clouds without optical counterparts in the vicinity of the NGC~5577/NGC~5576/NGC~5574 system. These clouds exhibit narrow line profiles and velocities consistent with the group environment, suggesting a possible origin of gas displaced during galaxy interactions.
This paper provides the survey description, catalogue, and atlas of data products that form the basis for subsequent analyses of the physical mechanisms shaping \HI\ properties in the \VirgoIII filament.

\section*{Acknowledgements}
MR's research is supported by the SARAO HCD programme via the "New Scientific Frontiers with Precision Radio Interferometry II" research group grant.

G.R. gratefully acknowledges support from NSF grants AST-1716690 and AST-2308126 and from a NASA ADAP grant 80NSSC21K0641.

This work was supported from the Swiss National Science Foundation, under the funding program ``Galaxy evolution in the cosmic web'' ($200021\_213076$).

R.A.F. gratefully acknowledges support from NSF grants AST-1716657 and AST-2308127 and from a NASA ADAP grant 80NSSC21K0640.

YMB acknowledges support from UK Research and Innovation through a Future Leaders Fellowship (grant agreement MR/X035166/1

The team acknowledges the support of the International Space Sciences Institute in Bern, Switzerland.  ISSI hosted multiple team meetings that focused on Virgo Filament Survey science.

The MeerKAT telescope is operated by the South African Radio Astronomy Observatory, which is a facility
of the National Research Foundation, an agency of the Department of Science and Innovation.

We acknowledge the use of the ilifu cloud computing facility – www.ilifu.ac.za, a partnership between the University of Cape Town, the University of the Western Cape, Stellenbosch University, Sol Plaatje University and the Cape Peninsula University of Technology. The ilifu facility is supported by contributions from the Inter-University Institute for Data Intensive Astronomy (IDIA – a partnership between the University of Cape Town, the University of Pretoria and the University of the Western Cape), the Computational Biology division at UCT and the Data Intensive Research Initiative of South Africa (DIRISA).
This research has made use of the NASA/IPAC Extragalactic Database, which is funded by the National Aeronautics and Space Administration and operated by the California Institute of Technology. It has also made use of the Astrophysics Data System, funded by NASA under Cooperative Agreement 80NSSC21M0056. 

(Part of) the data published here have been reduced using the CARACal pipeline, partially supported by ERC Starting grant number 679627 “FORNAX”, MAECI Grant Number ZA18GR02, DST-NRF Grant Number 113121 as part of the ISARP Joint Research Scheme, and BMBF project 05A17PC2 for D-MeerKAT. Information about CARACal can be obtained online under the URL: https://caracal.readthedocs.io

This paper uses data that were obtained by The Legacy Surveys: the Dark Energy Camera Legacy Survey (DECaLS; NOAO Proposal ID No. 2014B-0404; PIs: David Schlegel and Arjun Dey), the Beijing-Arizona Sky Survey (BASS; NOAO Proposal ID No. 2015A-0801; PIs: Zhou Xu and Xiaohui Fan), and the Mayall z-band Legacy Survey (MzLS; NOAO Proposal ID No. 2016A-0453; PI: Arjun Dey). DECaLS, BASS, and MzLS together include data obtained, respectively, at the Blanco telescope, Cerro Tololo Inter-American Observatory, National Optical Astronomy Observatory (NOAO); the Bok telescope, Steward Observatory, University of Arizona; and the Mayall telescope, Kitt Peak National Observatory, NOAO. NOAO is operated by the Association of Universities for Research in Astronomy (AURA) under a cooperative agreement with the National Science Foundation. Please see http://legacysurvey.org for details regarding the Legacy Surveys for details regarding the Legacy Surveys. BASS is a key project of the Telescope Access Program (TAP), which has been funded by the National Astronomical Observatories of China, the Chinese Academy of Sciences (the Strategic Priority Research Program `The Emergence of Cosmological Structures' grant no. XDB09000000), and the Special Fund for Astronomy from the Ministry of Finance. The BASS is also supported by the External Cooperation Program of Chinese Academy of Sciences (grant no. 114A11KYSB20160057) and Chinese National Natural Science Foundation (grant no. 11433005). The Legacy Surveys imaging of the DESI footprint is supported by the Director, Office of Science, Office of High Energy Physics of the U.S. Department of Energy under contract no. DE-AC02-05CH1123, and by the National Energy Research Scientific Computing Center, a DOE Office of Science User Facility under the same contract; and by the U.S. National Science Foundation, Division of Astronomical Sciences under Contract No. AST-0950945 to the National Optical Astronomy Observatory.

\section{Data availability}
The MeerKAT raw data presented in this article are available from the SARAO archive (\url{https://archive.sarao.ac.za/}) following the end of the proprietary period. The derived data products presented in Table~\ref{sample:tab1} are available upon via Zenodo at \url{https://doi.org/10.5281/zenodo.22304630}.


\bibliographystyle{mnras}
\bibliography{clean-VirgoFilHI}


\clearpage
\onecolumn
\appendix
\section{The \VirgoIII\ detection catalogue and parent sample}\label{sec:full_sample}

This appendix presents the full data used in this study. Table~\ref{sample:tab1} lists the derived \HI\ properties of all sources detected with MeerKAT in the \VirgoIII\ filament, together with their associated optical-counterparts. Table~\ref{sample:tab2} lists the fundamental and environmental properties of the parent sample selected for observation; galaxies are ordered by group association (delineated by horizontal lines and identified in the final column), with values compiled from \citet{CastignaniII2022} and \citet{CastignaniI2022}. The derived \HI\ source properties in Table~\ref{sample:tab1} are publicly available in machine-readable format via Zenodo (\url{https://doi.org/10.5281/zenodo.22304630}).

{
\begin{landscape}
\setlength{\tabcolsep}{6pt} 
\begin{longtable}{c | c | c |c |c |c |c |c |c | c |c |c |c |c}
\caption{Fundamental properties of all \HI\ sources detected with MeerKAT in \VirgoIII.}
\label{sample:tab1}\\
\hline\hline
VFID & Galaxy  & $\alpha_{J2000}$ & $\delta_{J2000}$  & v$_{\rm rad}$ & D$_{\rm L}$ & $w_{20}$ & $w_{50}$ & S$_{\rm int}$ & log(M$_{\rm HI}$) & Asym. & Opt. & $\Delta d_{\rm opt-HI}$ & Notes \\
     &         & deg              & deg               & km\,s$^{-1}$  & Mpc         & km\,s$^{-1}$ & km\,s$^{-1}$ & Jy\,km\,s$^{-1}$ & $M_\odot$ &       &      & arcsec & \\
(1)  & (2)     & (3)              & (4)               & (5)           & (6)         & (7)      & (8)      & (9)           & (10)              & (11)  & (12) & (13)                        & (14)  \\
\hline
\endfirsthead
\caption{continued.}\\
\hline\hline
VFID & Galaxy  & $\alpha_{J2000}$ & $\delta_{J2000}$  & v$_{\rm rad}$ & D$_{\rm L}$ & $w_{20}$ & $w_{50}$ & S$_{\rm int}$ & log(M$_{\rm HI}$) & Asym. & Opt. & $\Delta d_{\rm opt-HI}$ & Notes \\
     &         & deg              & deg               & km\,s$^{-1}$  & Mpc         & km\,s$^{-1}$ & km\,s$^{-1}$ & Jy\,km\,s$^{-1}$ & $M_\odot$ &       &      & arcsec & \\
(1)  & (2)     & (3)              & (4)               & (5)           & (6)         & (7)      & (8)      & (9)           & (10)              & (11)  & (12) & (13)                        & (14)  \\
\hline
\endhead
\hline
\endfoot
\hline
\multicolumn{14}{p{0.98\linewidth}}{\raggedright\textbf{Notes.} 
Columns are: 
(1) designated Virgo filament survey ID; 
(2) galaxy \HI\ ID and in (3) and (4) right ascension (J2000) and declination (J2000) of the galaxies' \HI\ centroids, respectively; 
(5) radial velocity; 
(6) luminosity distance;
(7) and (8) \ion{H}{i} profile line widths measured at 20\% and 50\% of the \ion{H}{i} profile peak flux, respectively; 
(9) integrated \HI\ flux and associated uncertainty, the latter combines the SoFiA-2 statistical value with a 10\% uncertainty added in quadrature to approximate systematic effects; see Sect.,\ref{HIfluxcomp} for details;
(10) total \ion{H}{i} mass and associated uncertainty; 
(11) \ion{H}{i} asymmetry (Asym).
(12) Optical counterparts, galaxy IDs marked (a)-(u) have been shortened for brevity. Their full designates IDs are (a) SDSS J135609.35+053233.2, (b) WISEA J135504.45+051121.7, (c) GALEXASC J143043.52+032043.4, (d) SDSS J143048.22+034749.6, (e) WISEA J144245.97+020725.9, (f) WISEA J145354.55+032344.4, (g) WISEA J145118.69+033640.1, (h) SDSS J150553.22+020028.2, (i) WISEA J150658.40+015939.9, (j) WISEA J150603.26+021105.1, (k) SDSS J150611.22+020544.2, (l) WISEA J150448.51+015850.9, (m) SDSS J150504.39+015751.7, (n) SDSS J143549.94+023618.4, (o) WISEA J143915.36+031621.3, (p) SDSS J143958.22+024451.9, (q) SDSS J145022.84+025730.5, (r) SDSS J145106.74+023126.9, (s) WISEA J150408.41+013127.4, (t) SDSS J150349.93+005831.7, (u) SDSS J150528.74+011733.0. Galaxies with names beginning with ``NOC'' have no confirmed optical counterparts. Those listed as ``---'' are \HI\ detections with a coincident optical counterpart but no catalogued galaxy ID; these are flagged 2 in column (14), since no redshift is available. $^{*}$NOC 03 from Fig.\,~\ref{fig:HIclouds} is not detected at this resolution, so we present the findings from the next resolution, R3.
(13) The projected separation between the \HI\ centres and the closest optical counterpart. (14) Information about galaxies' optical counterparts. The numbers have the following meanings: 0) Confirmed counterpart. 1) Confirmed lack of a counterpart (down to at least the surface brightness limits of the DESI Legacy Surveys). 2) Likely optical counterpart, but no redshift information.}  \\
\endlastfoot 
5351&J133830.48+082634.8&204.627&8.443&1017.38&23.77&128.38&116.16& 5.01 $\pm$ 0.51 & 8.82 $\pm$ 0.04 &0.21& UGC 08629 &1.06& 0 \\
5292&J133732.16+085309.6&204.384&8.886&1147.52&12.62&283.00&261.63& 85.0 $\pm$ 8.5 & 9.50 $\pm$ 0.04 &0.51& NGC 5248 &3.59& 0 \\
5283&J133545.60+085812.0&203.940&8.970&1159.23&22.44&136.61&119.04& 13.5 $\pm$ 1.3 & 9.20 $\pm$ 0.04 &0.13& UGC 08575 &3.1& 0 \\
5337&J133643.68+083256.4&204.182&8.549&1160.84&12.36&54.76&39.57& 1.33 $\pm$ 0.14 & 7.67 $\pm$ 0.05 &0.29& CGCG 073-051 &7.78& 0 \\
5305&J133437.92+084727.6&203.658&8.791&1226.76&15.57&100.80&79.29& 3.59 $\pm$ 0.39 & 8.31 $\pm$ 0.05 &0.61& CGCG 073-036 &10.78& 0 \\
5381&J133602.40+081106.0&204.010&8.185&1235.49&33.17&125.44&110.40& 3.04 $\pm$ 0.35 & 8.89 $\pm$ 0.05 &0.29& FGC 1642 &4.02& 0 \\
5857&J135326.64+051228.8&208.361&5.208&807.48&12.48&78.30&60.45& 0.27 $\pm$ 0.04 & 7.00 $\pm$ 0.06 &0.25& NGC 5338 &2.3& 0 \\
5889&J135610.56+050046.8&209.044&5.013&1239.40&17.17&291.50&268.32& 54.9 $\pm$ 5.5 & 9.58 $\pm$ 0.04 &0.41& NGC 5364 &20.88& 0 \\
5802&J135609.36+053234.8&209.039&5.543&1096.41&14.65&53.58&38.74& 0.53 $\pm$ 0.06 & 7.42 $\pm$ 0.05 &0.19& J1356+0532$^{a}$ &2.22& 0 \\
5892&J135538.40+045902.4&208.910&4.984&1167.78&22.05&86.28&68.48& 0.45 $\pm$ 0.05 & 7.71 $\pm$ 0.05 &0.24& NGC 5360 &3.98& 0 \\
5842&J135458.56+052009.6&208.744&5.336&1372.41&24.87&286.46&272.96& 4.44 $\pm$ 0.45 & 8.81 $\pm$ 0.04 &0.17& NGC 5356 &6.83& 0 \\
5907&J135353.76+045316.8&208.474&4.888&1275.59&12.08&44.08&29.00& 0.71 $\pm$ 0.08 & 7.38 $\pm$ 0.05 &0.22& AGC 238709 &0.78& 0 \\
5855&J135412.00+051333.6&208.550&5.226&1437.18&18.18&210.36&180.99& 16.0 $\pm$ 1.6 & 9.09 $\pm$ 0.04 &0.49& NGC 5348 &11.14& 0 \\
5859&J135503.60+051113.2&208.765&5.187&1391.84&14.39&57.54&40.40& 1.50 $\pm$ 0.15 & 7.86 $\pm$ 0.04 &0.43& J1355+0511$^{b}$ &18.38& 0 \\
5851&J135607.04+051518.0&209.029&5.255&1167.67&19.41&572.42&511.93& 0.25 $\pm$ 0.03 & 7.35 $\pm$ 0.05 & -- & NGC 5363 &5.95& 0 \\
5709&J140631.92+060144.4&211.633&6.029&1021.41&24.09&262.11&250.07& 5.36 $\pm$ 0.54 & 8.86 $\pm$ 0.04 &0.1& NGC 5470 &1.3& 0 \\
6033&J141020.08+035558.8&215.083&3.933&1495.10&18.53&443.89&409.15& 12.5 $\pm$ 1.3 & 9.00 $\pm$ 0.04 &0.29& NGC 5566 &2.35& 0 \\
6120&J141112.24+031543.2&215.301&3.262&1334.51&21.27&61.84&52.01& 0.04 $\pm$ 0.01 & 6.67 $\pm$ 0.10 &0.29& NGC 5576 &131.36& 0 \\
6091&J141113.20+032613.2&215.305&3.437&1484.91&23.21&255.40&241.48& 10.8 $\pm$ 1.1 & 9.13 $\pm$ 0.04 &0.14& NGC 5577 &3.8& 0 \\
-- &J141007.68+032743.2&215.032&3.462&1471.21& -- &38.87&24.13& 0.83 $\pm$ 0.09 & 8.02 $\pm$ 0.05 &0.31& NOC 01 &982.33& 1 \\
-- &J141032.16+032334.8&215.134&3.393&1542.16& -- &17.50&8.76& 0.05 $\pm$ 0.01 & 6.79 $\pm$ 0.07 &0.22& NOC 02 &633.86& 1 \\
-- &J142127.06+032313.6&215.363&3.387&1475.33& -- &21.16&14.76& 0.05 $\pm$ 0.01 & 6.79 $\pm$ 0.10 &0.18& NOC 03$^*$ &272.9& 2 \\
6018&J141004.08+035931.2&215.017&3.992&1724.15&19.15&236.59&193.58& 5.27 $\pm$ 0.53 & 8.65 $\pm$ 0.04 &0.3& NGC 5560 &4.04& 0 \\
6020&J141032.40+035858.8&215.135&3.983&1759.84&16.43&103.47&82.22& 8.20 $\pm$ 0.82 & 8.71 $\pm$ 0.04 &0.31& NGC 5569 &5.09& 0 \\
6380&J141632.64+014330.0&215.864&1.725&1387.27&25.12&230.20&211.79& 23.1 $\pm$ 2.3 & 9.53 $\pm$ 0.04 &0.26& UGC 09215 &7.1& 0 \\
6157&J142326.88+030032.4&217.862&3.009&1452.54&21.98&236.91&200.43& 6.48 $\pm$ 0.66 & 8.86 $\pm$ 0.04 &0.39& IC 1024 &5.23& 0 \\
6157&J142330.64+025609.6&217.886&2.936&1409.96&21.98&39.11&21.95& 0.08 $\pm$ 0.02 & 6.97 $\pm$ 0.09 &0.36& IC 1024 &273.2& 0 \\
6175&J142745.12+025454.0&218.188&2.915&1523.76&24.22&167.53&144.80& 5.25 $\pm$ 0.55 & 8.86 $\pm$ 0.05 &0.31& CGCG 047-085 &1.56& 0 \\
6108&J142259.44+032045.6&217.681&3.346&1519.76&24.26&90.98&79.42& 0.89 $\pm$ 0.09 & 8.09 $\pm$ 0.04 &0.16& J1430+0320$^{c}$ &2.8& 0 \\
6048&J142201.92+034622.8&217.508&3.773&1704.67&38.83&213.08&201.10& 4.89 $\pm$ 0.49 & 9.23 $\pm$ 0.04 &0.17& IC 1022 &0.38& 0 \\
6121&J142138.88+032640.0&217.413&3.267&1636.29&25.88&115.04&103.92& 0.31 $\pm$ 0.04 & 7.68 $\pm$ 0.05 &0.37& NGC 5636 &2.65& 0 \\
6046&J142248.24+034752.8&217.701&3.798&1666.68&26.22&38.53&24.65& 0.55 $\pm$ 0.06 & 7.95 $\pm$ 0.05 &0.2& J1430+0347$^{d}$ &2.9& 0 \\
6094&J142241.12+032533.6&217.672&3.426&1746.01&27.24&75.94&56.72& 0.39 $\pm$ 0.04 & 7.84 $\pm$ 0.05 &0.17& CGCG 047-069 &2.76& 0 \\
6130&J142200.96+031315.6&217.504&3.221&1836.50&25.60&158.06&144.18& 7.58 $\pm$ 0.76 & 9.06 $\pm$ 0.04 &0.27& UGC 09310 &0.91& 0 \\
6135&J142903.84+030856.4&217.266&3.149&1844.54&24.17&160.12&147.88& 2.03 $\pm$ 0.21 & 8.44 $\pm$ 0.04 &0.16& UGC 09285 &0.76& 0 \\
5964&J142324.72+042703.6&218.353&4.451&1573.63&26.36&112.29&93.81& 44.0 $\pm$ 4.4 & 9.85 $\pm$ 0.04 &0.32& NGC 5668 &4.55& 0 \\
5991&J142314.88+041539.6&218.663&4.261&1674.52&21.08&111.47&94.97& 6.95 $\pm$ 0.7 & 8.86 $\pm$ 0.04 &0.26& UGC 09380 &6.84& 0 \\
6394&J143824.48+014051.6&221.102&1.681&1565.96&28.69&344.24&318.49& 33.0 $\pm$ 3.3 & 9.80 $\pm$ 0.04 &0.35& NGC 5740 &4.57& 0 \\
6340&J143856.16+015716.4&221.233&1.954&1715.13&29.58&646.19&622.47& 31.8 $\pm$ 3.2 & 9.81 $\pm$ 0.04 &0.27& NGC 5746 &2.45& 0 \\
6444&J143830.88+013118.8&221.127&1.522&1457.42&23.30&55.64&42.87& 0.48 $\pm$ 0.06 & 7.78 $\pm$ 0.05 &0.33& CGCG 020-010 &4.84& 0 \\
6301&J143749.92+020906.4&220.692&2.124&1591.76&25.03&91.37&79.77& 0.92 $\pm$ 0.11 & 8.13 $\pm$ 0.05 &0.2& J1442+0207$^{e}$ &1.42& 0 \\
6446&J143734.24+020030.0&220.624&1.501&1889.23&28.87&133.06&120.98& 0.9 $\pm$ 0.1 & 8.25 $\pm$ 0.06 &0.29& CGCG 019-084 &1.18& 0 \\
6081&J144432.56+033328.8&223.469&3.558&1628.30&18.69&387.34&128.61& 101 $\pm$ 10 & 9.91 $\pm$ 0.04 &1.11& NGC 5775/4 &61.61& 0 \\
6110&J144343.92+031850.4&223.267&3.314&1570.28&24.53&219.48&207.06& 12.5 $\pm$ 1.3 & 9.25 $\pm$ 0.04 &0.68& IC 1067/IC 1066 &8.19& 0 \\
6102&J144414.72+032345.6&223.478&3.396&1539.29&24.27&45.65&25.16& 0.55 $\pm$ 0.06 & 7.88 $\pm$ 0.05 &0.41& J1453+0323$^{f}$ &0.83& 0 \\
6072&J144326.56+033610.8&222.826&3.611&1653.60&25.84&43.41&25.89& 0.51 $\pm$ 0.06 & 7.90 $\pm$ 0.05 &0.31& J1451+0336$^{g}$ &3.01& 0 \\
6352&J145200.24+015316.8&225.001&1.888&1359.50&24.89&335.93&314.06& 9.31 $\pm$ 0.94 & 9.13 $\pm$ 0.04 &0.27& NGC 5806 &12.31& 0 \\
6362&J145956.40+015050.4&225.515&1.841&1235.67&19.69&115.70&99.46& 2.00 $\pm$ 0.2 & 8.26 $\pm$ 0.04 &0.21& UGC 09661 &1.71& 0 \\
6411&J145932.40+013723.2&225.115&1.623&1515.34&23.75&87.62&66.93& 1.14 $\pm$ 0.12 & 8.18 $\pm$ 0.05 &0.23& NGC 5811 &7.36& 0 \\
-- &J144000.00+013653.6&224.999&1.615&1984.24& -- &61.72&37.00& 0.36 $\pm$ 0.05 & 7.68 $\pm$ 0.06 &0.28& --- &$\sim 0$& 2 \\
-- &J145759.28+020904.0&226.262&2.151&1231.18& -- &35.08&24.51& 0.13 $\pm$ 0.02 & 7.35 $\pm$ 0.07 &0.22& --- &$\sim 0$& 2 \\
6327&J145748.64+020025.2&226.473&2.007&1291.15&20.46&60.53&39.47& 0.36 $\pm$ 0.04 & 7.55 $\pm$ 0.05 &0.23& J1505+0200$^{h}$ &4.04& 0 \\
6332&J145901.44+015936.0&226.744&1.994&1318.55&20.68&32.18&25.20& 0.03 $\pm$ 0.01 & 6.53 $\pm$ 0.09 &0.14& J1506+0159$^{i}$ &5.24& 0 \\
6290&J145923.36+021105.6&226.514&2.185&1643.19&25.51&122.38&102.66& 0.98 $\pm$ 0.1 & 8.18 $\pm$ 0.05 &0.27& J1506+0211$^{j}$ &1.07& 0 \\
6308&J145924.40+020542.0&226.545&2.095&1801.68&27.65&159.92&142.79& 4.07 $\pm$ 0.41 & 8.86 $\pm$ 0.04 &0.35& J1506+0205$^{k}$ &7.51& 0 \\
6403&J150805.76+013905.6&227.024&1.651&1835.69&27.71&105.84&96.77& 0.09 $\pm$ 0.02 & 7.20 $\pm$ 0.08 &0.31& CGCG 021-013 &1.94& 0 \\
6334&J150821.28+015850.8&226.201&1.981&1939.76&29.38&31.43&22.22& 0.07 $\pm$ 0.01 & 7.13 $\pm$ 0.09 &0.14& J1504+0158$^{l}$ &3.26& 0 \\
6420&J150852.96+013636.0&227.039&1.610&2107.25&31.86&46.85&32.19& 0.3 $\pm$ 0 & 7.85 $\pm$ 0.06 &0.25& CGCG 021-015 &5.56& 0 \\
6338&J150926.88+015751.2&226.269&1.964&2368.91&34.74&60.80&40.40& 0.28 $\pm$ 0.03 & 7.89 $\pm$ 0.05 &0.13& J1505+0157$^{m}$ &1.25& 0 \\
6437&J151030.00+023639.6&226.775&1.543&2529.50&16.82&209.08&190.83& 14.6 $\pm$ 1.5 & 8.98 $\pm$ 0.04 &0.68& NGC 5850 &26.48& 0 \\
6098&J144218.24+032439.6&219.576&3.411&1590.12&24.99&181.46&162.39& 1.90 $\pm$ 0.19 & 8.44 $\pm$ 0.04 &0.17& NGC 5692 &2.75& 0 \\
6170&J144327.60+025700.0&219.765&2.950&1565.10&27.02&118.28&99.53& 8.06 $\pm$ 0.81 & 9.14 $\pm$ 0.04 &0.25& UGC 09432 &11.42& 0 \\
6230&J143750.16+023621.6&218.959&2.606&1549.05&20.60&77.68&57.58& 0.19 $\pm$ 0.02 & 7.28 $\pm$ 0.05 &0.36& J1435+0236$^{n}$ &2.51& 0 \\
6119&J144515.36+031622.8&219.814&3.273&1588.43&24.95&117.89&103.06& 1.23 $\pm$ 0.13 & 8.26 $\pm$ 0.04 &0.13& J1439+0316$^{o}$ &2.05& 0 \\
6272&J144340.96+021735.6&219.420&2.293&1753.96&19.39&314.61&292.28& 5.36 $\pm$ 0.54 & 8.67 $\pm$ 0.04 &0.24& NGC 5690 &9.35& 0 \\
6149&J144910.56+030252.0&219.287&3.048&1747.12&27.12&84.40&65.48& 1.17 $\pm$ 0.12 & 8.30 $\pm$ 0.04 &0.25& SHOC 474 &3.86& 0 \\
6194&J145558.56+024456.0&219.994&2.749&1728.70&26.96&49.16&36.83& 0.22 $\pm$ 0.03 & 7.58 $\pm$ 0.05 &0.28& J1439+0244$^{p}$ &4.51& 0 \\
6115&J144432.56+031742.8&223.261&3.295&1572.48&32.19&202.91&183.78& 0.94 $\pm$ 0.1 & 8.35 $\pm$ 0.05 &0.35& IC 1066 &4.52& 0 \\
6197&J143707.68+024326.4&222.782&2.724&1512.22&23.93&78.15&60.38& 0.58 $\pm$ 0.06 & 7.89 $\pm$ 0.05 &0.19& SDSSCGB 38563 &3.26& 0 \\
-- &J144432.00+024745.6&223.282&2.796&1619.46& -- &25.52&19.59& 0.03 $\pm$ 0.01 & 6.76 $\pm$ 0.10 &0.15& --- &$\sim 0$ & 2 \\
6169&J143815.04+025732.8&222.596&2.959&1696.25&26.23&55.97&34.45& 0.19 $\pm$ 0.02 & 7.49 $\pm$ 0.05 &0.16& J1450+0257$^{q}$ &1.09& 0 \\
6242&J143706.72+023121.6&222.777&2.523&2049.11&30.85&43.24&29.63& 0.32 $\pm$ 0.04 & 7.85 $\pm$ 0.05 &0.5& J1451+0231$^{r}$ &7.85& 0 \\
6443&J144408.64+015450.0&226.036&1.524&1834.58&27.82&64.52&48.67& 0.08 $\pm$ 0.01 & 7.16 $\pm$ 0.07 &0.21& J1504+0131$^{s}$ &3.32& 0 \\
-- &J144459.84+002826.4&226.248&0.474&1989.53& -- &60.91&33.17& 0.04 $\pm$ 0.01 & 6.70 $\pm$ 0.08 &0.28& --- &$\sim 0$& 2 \\
6519&J144411.12+005840.0&225.957&0.978&2052.81&30.60&95.01&76.60& 1.12 $\pm$ 0.11 & 8.39 $\pm$ 0.04 &0.19& J1503+0058$^{t}$ &2.72& 0 \\
6476&J144456.72+011731.0&226.368&1.292&2279.92&33.71&35.72&21.97& 0.18 $\pm$ 0.02 & 7.68 $\pm$ 0.05 &0.23& J1505+0117$^{u}$ &4.66& 0 \\
-- &J144525.92+004143.2&226.322&0.695&2349.54& -- &89.04&72.05& 0.12 $\pm$ 0.02 & 7.25 $\pm$ 0.06 &0.17& --- &$\sim 0$& 2 \\
6461&J144902.88+012618.0&227.742&1.439&1539.92&26.30&140.74&127.15& 0.87 $\pm$ 0.09 & 8.15 $\pm$ 0.04 &0.19& UGC 09751 &5.15& 0 \\
6342&J151016.56+015602.4&227.569&1.934&1730.33&29.30&209.28&194.93& 1.88 $\pm$ 0.19 & 8.57 $\pm$ 0.04 &0.15& UGC 09746 &1.3& 0 \\
6390&J151202.16+014156.4&228.009&1.699&2005.60&27.04&160.87&143.22& 2.64 $\pm$ 0.27 & 8.65 $\pm$ 0.04 &0.23& UGC 09760 &3.44& 0 \\
\hline 
\end{longtable}
\end{landscape}
}
\begin{landscape}
\setlength{\tabcolsep}{4pt} 

\begin{longtable}{ccccccccccccc}
\caption{Fundamental properties of the galaxies selected for observations in the \VirgoIII filament and their environments. }\label{sample:tab2}\\
\hline\hline
VFID & Galaxy  & $\alpha_{J2000}$ & $\delta_{J2000}$  & log(M$_{\rm HI}$) &log(M$_{\rm H_{2}}$) & log(M$_\star$) & def$_{\rm HI}$ & $t_{\rm hub}$ & d$_{\rm cl}$ & d$_{\rm fil}$ & $n_5$ & Group ID \\
&  & deg & deg & \MSUN\ & \MSUN\ & \MSUN &  &  & Mpc & Mpc &  &     \\
(1) & (2) & (3) & (4) & (5) & (6) & (7) & (8) & (9) & (10) & (11) & (12) & (13) \\
\hline
\endfirsthead
\caption{continued.}\\
\hline
VFID & Galaxy  & $\alpha_{J2000}$ & $\delta_{J2000}$  & log(M$_{\rm HI}$) &log(M$_{\rm H_{2}}$) & log(M$_\star$) & def$_{\rm HI}$ & $t_{\rm hub}$ & d$_{\rm cl}$ & d$_{\rm fil}$ & $n_5$ & Group ID \\
&  & deg & deg & \MSUN\ & \MSUN\ & \MSUN &  &  & Mpc & Mpc &  &     \\
(1) & (2) & (3) & (4) & (5) & (6) & (7) & (8) & (9) & (10) & (11) & (12) & (13) \\
\hline
\endhead
\hline
\endfoot
\hline
\multicolumn{13}{p{0.5\linewidth}}{\raggedright\textbf{Notes.} 
Values in this table come from catalogue of \cite{CastignaniII2022} in all cases except for \HI deficiencies from ALFALFA, which are recalculated based on the catalogue's ALFALFA \HI masses and Hyperleda Hubble types using Equations~10~and~11 from \cite{CastignaniI2022}. Columns are: 
(1) designated Virgo filament survey ID; 
(2) galaxy ID, (a)–(c) full SDSS names have been shortened for brevity: 
(a) SDSS J135609.37+053233.9, 
(b) SDSS J143048.23+034749.4, 
(c) WISEA J143915.36+031621.3,
(d) SDSS J143958.22+024451.9,
(e) SDSS J143549.94+023618.4,
(f) SDSS J145022.84+025730.5,
(g) SDSS J145106.74+023126.9,
(h) WISEA J150658.40+015939.9,
(i) SDSS J150448.49+015851.3,
(j) SDSS J150528.74+011733.0,
(k) WISEA J150408.41+013127.4,
(l) SDSS J150349.93+005831.7;
(3) and (4) right ascension (J2000) and declination (J2000) of the galaxies, respectively; 
(5) total \HI mass \citep[][if starred and ALFALFA if not]{CastignaniI2022}; 
(6) H$_{2}$ mass; 
(7) stellar mass; 
(8) \HI deficiency \citep[][if starred and ALFALFA if not]{CastignaniI2022}; 
(9) Hubble type $t_{\rm hub}$ (quantitative measure between -5 and 10, ranging from elliptical to irregular); 
(10) and (11) distance to the Virgo cluster and Virgo\,\textsc{iii} filament, respectively; 
(12) local density $n_5$; 
(13) galaxy group ID.} \\
\endlastfoot

5283 & UGC08575 & 203.940 & 8.969 & 9.30 & -- & 8.25 & 0.08 & 9.5 & 10.03 & 4.31 & 0.05 & NGC5248	\\
5305 & PGC047826 & 203.658 & 8.794 & 8.54 & -- & 8.25 & -0.33 & 10.0 & 4.48 & 3.13 & 0.07 & 	\\
5337 & PGC048046 & 204.183 & 8.547 & 7.80 & -- & 8.05 & 0.45 & 10.0 & 4.01 & 5.55 & 0.04 & 	\\
5351 & UGC08629 & 204.628 & 8.442 & 8.92 & -- & 7.89 & -0.08 & 9.8 & 11.38 & 5.22 & 0.06 & 	\\
5381 & PGC091299 & 204.009 & 8.186 & 8.76 & -- & 7.94 & -0.13 & 6.6 & 20.39 & 6.99 & 0.01 & 	\\
5292 & NGC5248 & 204.383 & 8.885 & 8.93$^*$ & 9.61 & 10.5 & 0.45$^*$ & 4.0 & 3.99 & 5.62 & 0.05 & 	\\
\hline																													
5857 & NGC5338 & 208.361 & 5.208 & 7.49 & -- & 9.09 & 0.94 & -2.0 & 5.07 & 5.87 & 0.02 & --	\\
\hline																													
5907 & AGC238709 & 208.474 & 4.888 & 7.40 & -- & 6.58 & 0.95 & -- & 5.15 & 6.25 & 0.02 & NGC5363	\\
5892 & NGC5360 & 208.911 & 4.985 & 7.98$^*$ & 7.57 & 9.09 & 0.43$^*$ & 0.1 & 10.68 & 2.69 & 0.16 & 	\\
5889 & NGC5364 & 209.050 & 5.015 & 9.45$^*$ & 7.99 & 10.42 & 0.05$^*$ & 4.0 & 6.87 & 1.30 & 0.11 & 	\\
5859 & PGC1279452 & 208.769 & 5.189 & 7.95 & -- & 7.68 & -0.17 & 10.0 & 5.40 & 3.98 & 0.05 & 	\\
5855 & NGC5348 & 208.547 & 5.227 & 8.82$^*$ & 8.02 & 9.52 & 0.73$^*$ & 3.8 & 7.45 & 0.36 & 0.31 & 	\\
5851 & NGC5363 & 209.030 & 5.255 & 8.19$^*$ & 9.63 & 10.84 & 1.16$^*$ & 0.1 & 8.48 & 0.55 & 0.41 & 	\\
5802 & J135609$^{a}$ & 209.039 & 5.543 & 7.64 & -- & 7.51 & 0.95 & 10.0 & 5.52 & 3.72 & 0.07 & 	\\
5842 & NGC5356 & 208.744 & 5.334 & 8.85$^*$ & 8.95 & 9.96 & 0.69$^*$ & 3.8 & 13.17 & 4.94 & 0.05 & 	\\
\hline																													
5709 & NGC5470 & 211.633 & 6.029 & 8.79$^*$ & 8.39 & 9.48 & 0.69$^*$ & 3.3 & 12.91 & 3.58 & 0.18 & --	\\
\hline																													
6018 & NGC5560 & 215.018 & 3.993 & 8.99$^*$ & 8.78 & -- & 0.52$^*$ & 3.1 & 9.60 & 1.00 & 0.13 & NGC5566	\\
6020 & NGC5569 & 215.134 & 3.983 & 8.78 & -- & -- & -0.20 & 6.0 & 7.85 & 2.98 & 0.13 & 	\\
6033 & NGC5566 & 215.083 & 3.934 & 9.19$^*$ & 9.26 & -- & 0.39$^*$ & 1.6 & 9.17 & 1.45 & 0.12 & 	\\
6091 & NGC5577 & 215.305 & 3.436 & 8.90$^*$ & 8.54 & 9.77 & 0.61$^*$ & 3.8 & 12.97 & 1.86 & 0.30 & 	\\
6120 & NGC5576 & 215.265 & 3.271 & 7.64$^*$ & 7.49 & -- & 1.49$^*$ & -4.9 & 11.35 & 0.74 & 0.12 & 	\\
\hline																													
6380 & UGC09215 & 215.863 & 1.726 & 9.44$^*$ & 7.77 & 9.17 & -0.25$^*$ & 6.4 & 14.87 & 2.99 & 0.91 & --	\\
\hline																													
6121 & NGC5636 & 217.413 & 3.266 & 8.34$^*$ & 7.84 & 9.77 & 0.37$^*$ & -0.3 & 15.73 & 2.71 & 0.75 & NGC5638	\\
6135 & UGC09285 & 217.266 & 3.149 & 8.52 & -- & 9.14 & 0.38 & 6.0 & 14.21 & 1.84 & 1.66 & 	\\
6048 & IC1022 & 217.508 & 3.773 & 9.28$^*$ & 7.46 & 8.95 & -0.08$^*$ & 3.4 & 27.92 & 8.03 & 0.02 & 	\\
6094 & PGC051845 & 217.673 & 3.426 & 8.14 & -- & 8.97 & 0.02 & -2.4 & 17.00 & 3.23 & 1.03 & 	\\
6108 & AGC249448 & 217.681 & 3.345 & 8.06 & -- & 7.81 & 0.83 & -- & 14.36 & 1.74 & 51.88 & 	\\
6046 & J143048$^{b}$ & 217.701 & 3.797 & 7.98 & -- & 7.43 & 0.34 & 10.0 & 16.06 & 2.78 & 1.09 & 	\\
6157 & IC1024 & 217.863 & 3.008 & 9.05$^*$ & 8.68 & 9.63 & -0.63$^*$ & -1.8 & 12.48 & 0.58 & 0.16 & 	\\
6175 & PGC051971 & 218.188 & 2.915 & 9.01 & -- & 8.26 & -0.62 & 4.6 & 14.45 & 1.57 & 1.38 & 	\\
6130 & UGC09310 & 217.505 & 3.220 & 9.10 & -- & 8.75 & 0.10 & 8.1 & 15.51 & 2.53 & 0.91 & 	\\
\hline																													
5964 & NGC5668 & 218.351 & 4.450 & 9.85$^*$ & 8.20 & 9.61 & -0.69$^*$ & 6.9 & 16.27 & 2.63 & 1.03 & NGC5668	\\
5991 & UGC09380 & 218.663 & 4.263 & 8.93 & -- & 8.45 & 0.03 & 9.9 & 11.84 & 0.55 & 0.23 & 	\\
\hline		

6272 & NGC5690 & 219.421 & 2.291 & 9.83$^*$ & 9.10 & 10.4 & -0.45$^*$ & 5.2 & 10.86 & 2.03 & 0.22 & NGC5690 \\	
\hline
6098 & NGC5692 & 219.575 & 3.410 & 8.74$^*$ & 8.12 & 9.31 & 0.22$^*$ & 4.0 & 15.37 & 1.44 & 2.77 & NGC5692 \\
6119 & J143915$^{c}$ & 219.814 & 3.273 & 8.41 & -- & 7.91 & -0.28 & 7.0 & 15.38 & 1.33 & 1.33 & \\
\hline
6170 & UGC09432 & 219.769 & 2.950 & 9.16 & -- & 8.06 & -0.29 & 9.6 & 17.23 & 2.21 & 1.20 & -- \\	\hline
6230 & J143549$^{d}$ & 218.959 & 2.605 & 8.40 & -- & 7.78 & 0.53 & 9.5 & 11.64 & 1.18 & 0.26 & -- \\	\hline
6149 & SHOC474 & 219.287 & 3.047 & 8.56 & -- & 7.92 & -0.26 & 10.0 & 17.22 & 2.46 & 0.92 & -- \\	\hline
6194 & J143958$^{e}$ & 219.993 & 2.748 & 7.89 & -- & 7.70 & 0.15 & 10.0 & 17.24 & 2.10 & 1.63 & -- \\	\hline

6446 & PGC052534 & 220.623 & 1.500 & 8.55 & -- & 8.91 & -0.24 & 0.0 & 19.17 & 2.08 & 0.30 & NGC5746	\\
6301 & PGC1216689 & 220.692 & 2.124 & 8.22 & -- & 8.09 & 0.10 & 7.0 & 15.71 & 1.17 & 0.89 & 	\\
6444 & PGC052652 & 221.129 & 1.522 & -- & -- & 8.60 & -- & 4.0 & 14.37 & 0.95 & 0.24 & 	\\
6394 & NGC5740 & 221.102 & 1.680 & 9.84$^*$ & 9.40 & 10.17 & -0.25$^*$ & 3.0 & 19.09 & 1.85 & 0.32 & 	\\
6340 & NGC5746 & 221.233 & 1.955 & 9.80$^*$ & 9.11 & 10.78 & 0.34$^*$ & 3.0 & 19.91 & 1.56 & 0.24 & 	\\
\hline		

6197 & SDSSCGB38563 & 222.781 & 2.724 & 7.85 & -- & 7.12 & 0.54 & -- & 15.15 & 0.46 & 0.92 & -- \\	\hline
6169 & J145022$^{f}$ & 222.595 & 2.958 & 7.81 & -- & 8.28 & 0.36 & 10.0 & 17.09 & 0.72 & 0.75 & -- \\	\hline
6242 & J145106$^g$ & 222.778 & 2.524 & 8.17 & -- & 8.27 & -0.00 & 10.0 & 21.35 & 0.26 & 0.26 & -- \\	\hline					

6110 & IC1067 & 223.272 & 3.332 & 9.36$^*$ & 8.01 & 9.79 & -0.06$^*$ & 3.0 & 15.74 & 0.20 & 0.51 & NGC5775	\\
6115 & IC1066 & 223.262 & 3.296 & 9.27$^*$ & 8.46 & 9.50 & -0.04$^*$ & 3.2 & 22.64 & 0.97 & 0.10 & 	\\
6102 & AGC238948 & 223.477 & 3.396 & 7.99 & -- & 7.99 & 0.26 & -- & 15.55 & 0.38 & 0.44 & 	\\
6081 & NGC5775 & 223.490 & 3.544 & 10.01 & -- & 10.50 & -0.49 & 5.1 & 11.19 & 3.35 & 0.13 & 	\\
6077 & NGC5774 & 223.427 & 3.583 & 9.86 & -- & 9.54 & -0.75 & 6.9 & 18.04 & 0.70 & 0.51 & 	\\
6072 & AGC238947 & 222.827 & 3.612 & 7.90 & -- & 6.79 & 0.36 & -- & 16.76 & 0.55 & 0.51 & 	\\
\hline																													
6352 & NGC5806 & 225.002 & 1.891 & 9.30$^*$ & 9.36 & 10.29 & 0.29$^*$ & 3.2 & 16.50 & 0.95 & 1.51 & NGC5846	\\
6411 & NGC5811 & 225.113 & 1.624 & 8.29 & -- & 9.15 & 0.44 & 9.1 & 15.58 & 1.46 & 0.81 & 	\\
6362 & UGC09661 & 225.515 & 1.841 & 8.12$^*$ & 7.91 & 8.94 & 0.50$^*$ & 8.0 & 12.46 & 3.45 & 0.09 & 	\\
6327 & PGC1213020 & 226.472 & 2.008 & 7.81 & -- & 7.48 & 0.21 & 10.0 & 13.24 & 3.31 & 0.31 & 	\\
6332 & J150658$^{h}$ & 226.743 & 1.995 & -- & -- & 7.57 & -- & 10.0 & 13.46 & 3.29 & 0.40 & 	\\
6290 & PGC1218738 & 226.514 & 2.185 & 8.26 & -- & 8.56 & 0.27 & 9.0 & 17.34 & 1.24 & 3.56 & 	\\
6308 & PGC1215798 & 226.547 & 2.096 & 8.94 & -- & 7.83 & -0.20 & 8.0 & 19.23 & 0.89 & 3.80 & 	\\
6403 & PGC054037 & 227.023 & 1.652 & -- & 7.83 & 9.27 & -- & 10.0 & 19.42 & 1.17 & 2.87 & 	\\
6334 & J150448$^{i}$ & 226.202 & 1.981 & -- & -- & 7.99 & -- & 10.0 & 20.72 & 0.87 & 5.20 & 	\\
6420 & PGC054045 & 227.039 & 1.609 & -- & -- & 9.17 & -- & 10.0 & 23.19 & 2.30 & 0.75 & 	\\
6338 & PGC1211621 & 226.268 & 1.964 & 8.19 & -- & 8.28 & -0.00 & 10.0 & 25.69 & 3.94 & 0.14 & 	\\
6437 & NGC5850 & 226.782 & 1.545 & 8.97 & -- & 10.71 & 0.46 & 3.1 & 10.89 & 5.15 & 0.02 & 	\\
6342 & UGC09746 & 227.570 & 1.934 & 9.02 & -- & 9.23 & 0.31 & 4.1 & 20.93 & 1.53 & 3.49 & \\	
6461 & UGC09751 & 227.743 & 1.438 & 8.47 & -- & 8.47 & 0.49 & 6.1 & 18.34 & 1.68 & 2.72 & 	\\
6476 & J150528$^j$ & 226.369 & 1.292 & -- & -- & 8.40 & -- & 10.0 & 24.79 & 3.20 & 0.58 & \\	
6390 & UGC09760 & 228.010 & 1.698 & 9.37 & -- & 8.53 & -0.17 & 6.6 & 19.03 & 1.65 & 2.28 & \\	
6443 & J150408$^k$ & 226.035 & 1.525 & 7.71 & -- & 8.63 & 0.95 & 0.5 & 19.31 & 0.78 & 4.14 & \\
6519 & J150309$^l$ & 225.958 & 0.977 & 8.65 & -- & 8.47 & 0.07 & 10.0 & 21.85 & 1.34 & 0.84 & \\	
\hline																									
\hline													

\hline	
\end{longtable}
\end{landscape}
\section{Atlas of \HI\ data products}\label{sec:allhiim}
This appendix presents the \HI\ imaging atlas for all detections with confirmed optical counterparts, shown at the default R2 angular resolution. Each galaxy is displayed on a single page comprising five panels: the spatially integrated \HI\ spectrum (top left), with the profile width $W_{20}$ indicated; the position--velocity diagram extracted along the kinematic major axis (top right), with the fitted position angle marked; the \HI\ column-density contours overlaid on the DECaLS optical image (lower left), with contour levels quoted at the $3\sigma$ column-density sensitivity; and the moment-1 (velocity field) and moment-2 (velocity dispersion) maps (middle and lower right). This layout follows that described in Sect.~\ref{sec:res} and illustrated for two representative galaxies across all six angular resolutions in Fig.~\ref{multiresIm}. The complete atlas is presented in Fig.~\ref{fig:AllGalImage}.

\newgeometry{margin=1.5cm}
\begin{figure*}
    \centering
    \resizebox{0.86\hsize}{!}{
    \includegraphics{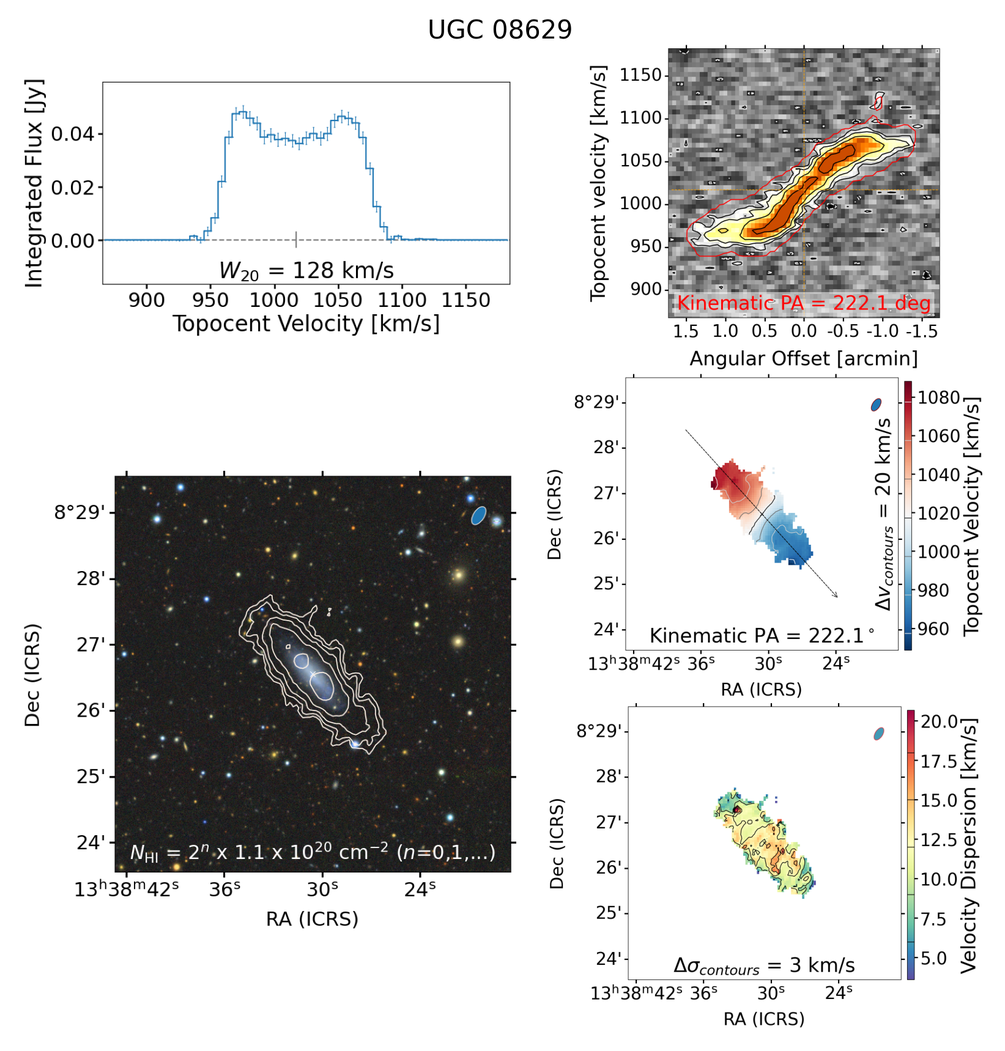}\hspace{0.5cm}
    \includegraphics{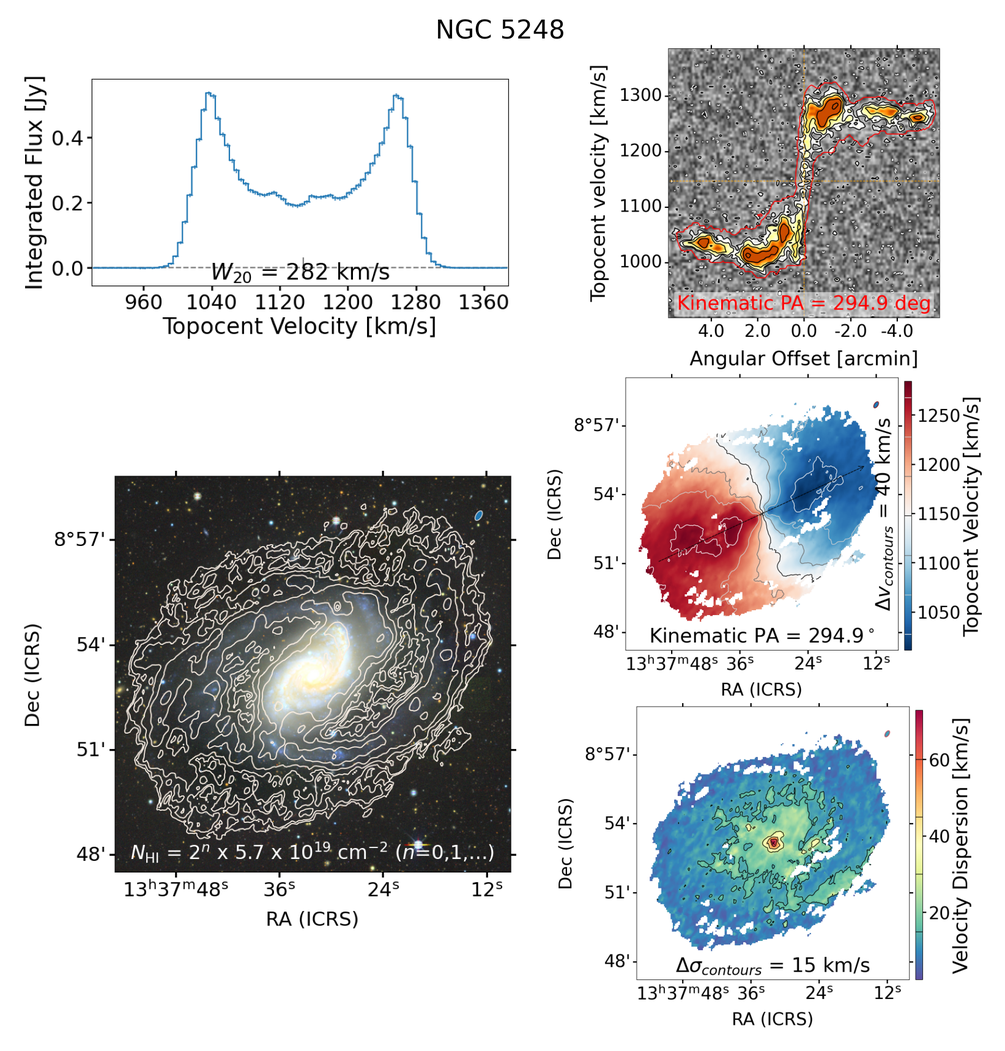}}
    \vspace{0.05cm}
    \resizebox{0.86\hsize}{!}{
    \includegraphics{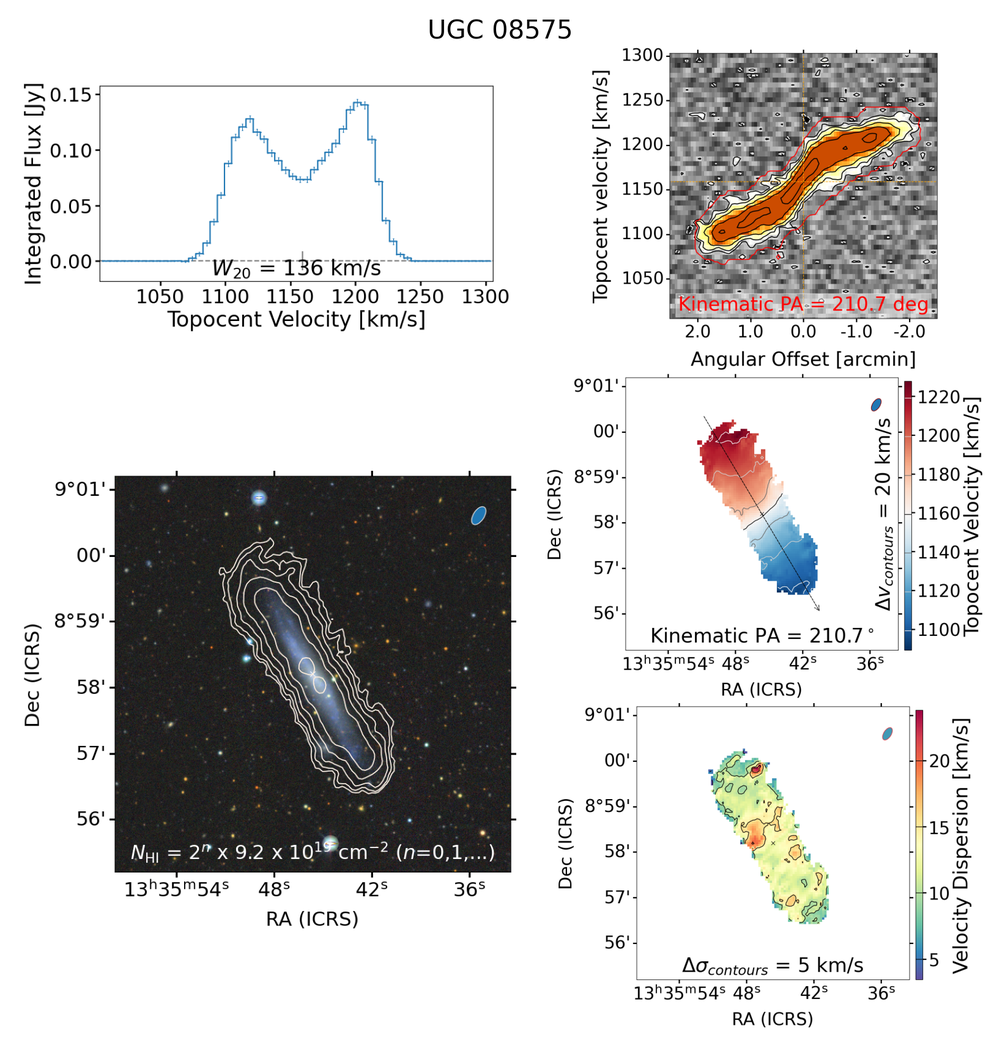}\hspace{0.5cm}
    \includegraphics{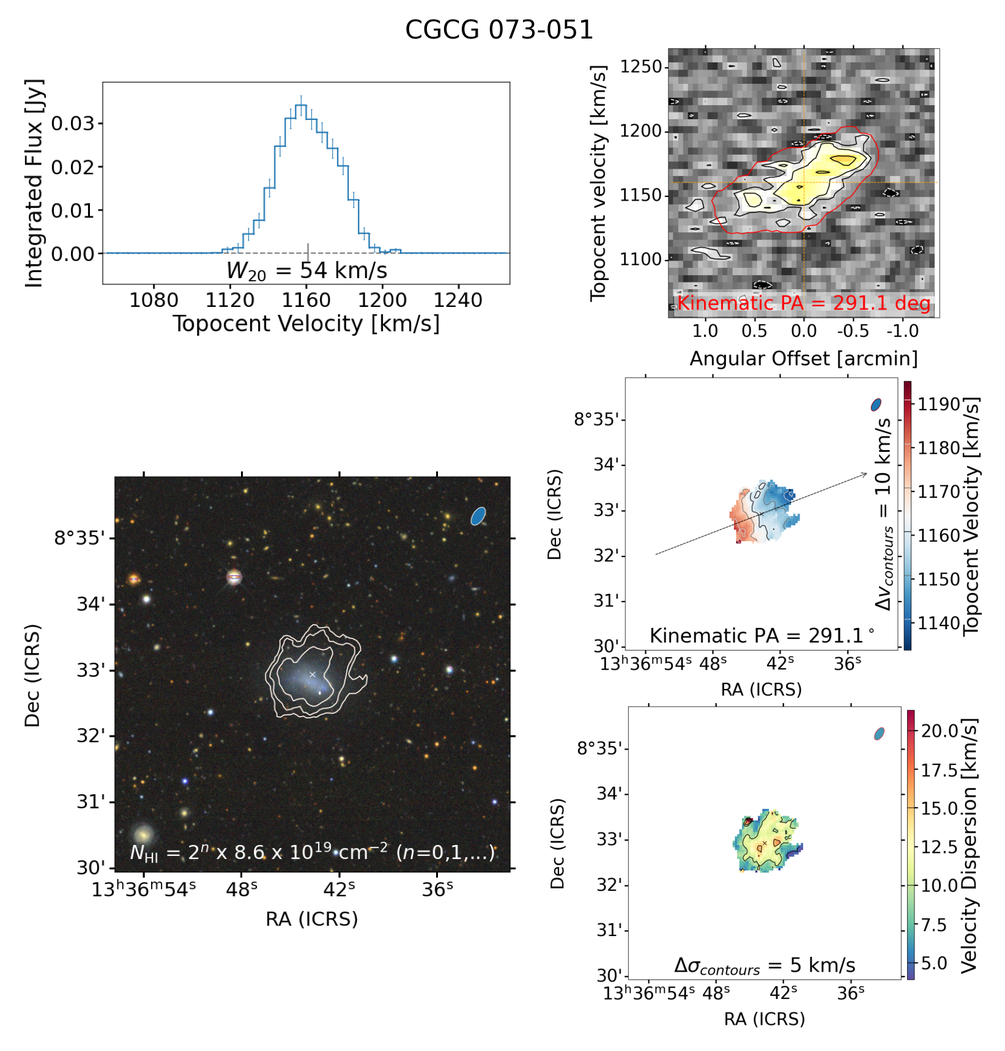}}
    \vspace{0.05cm}
    \resizebox{0.86\hsize}{!}{
    \includegraphics{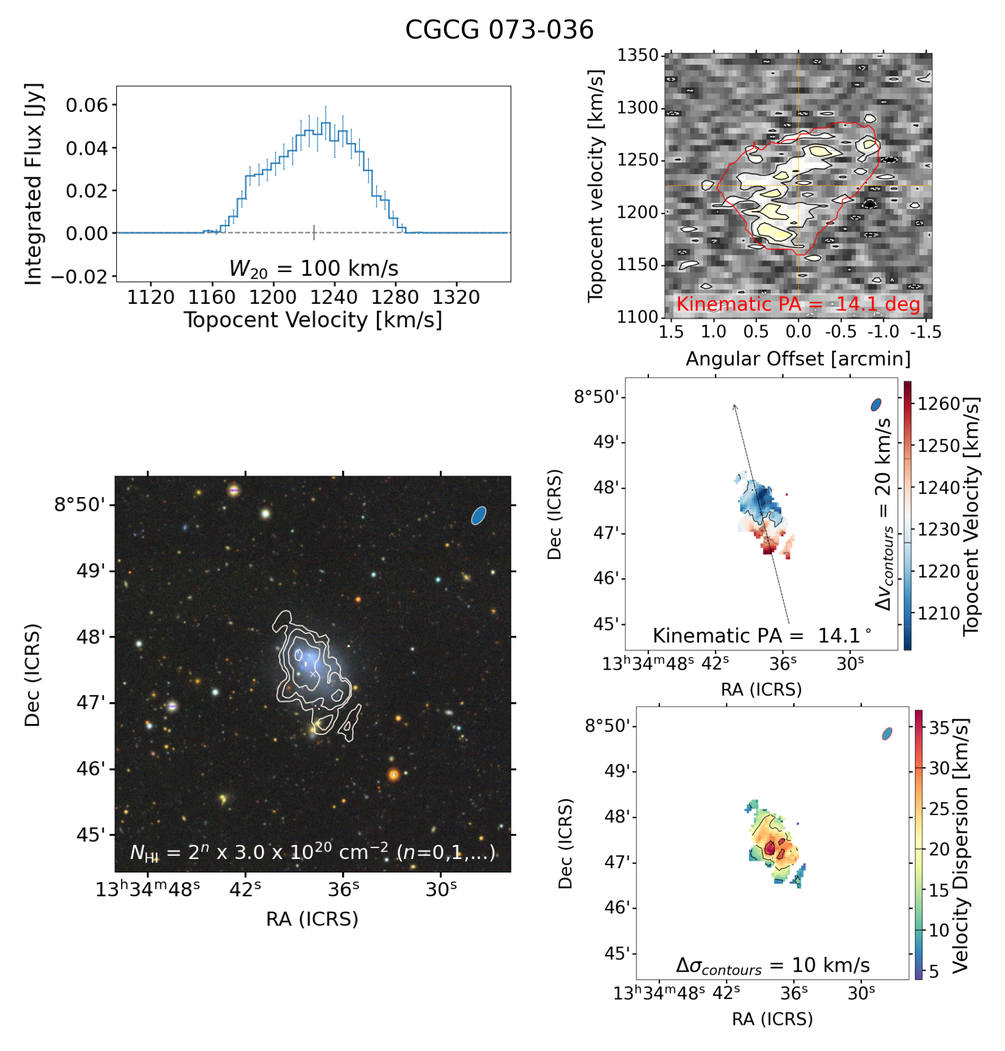}\hspace{0.5cm}
    \includegraphics{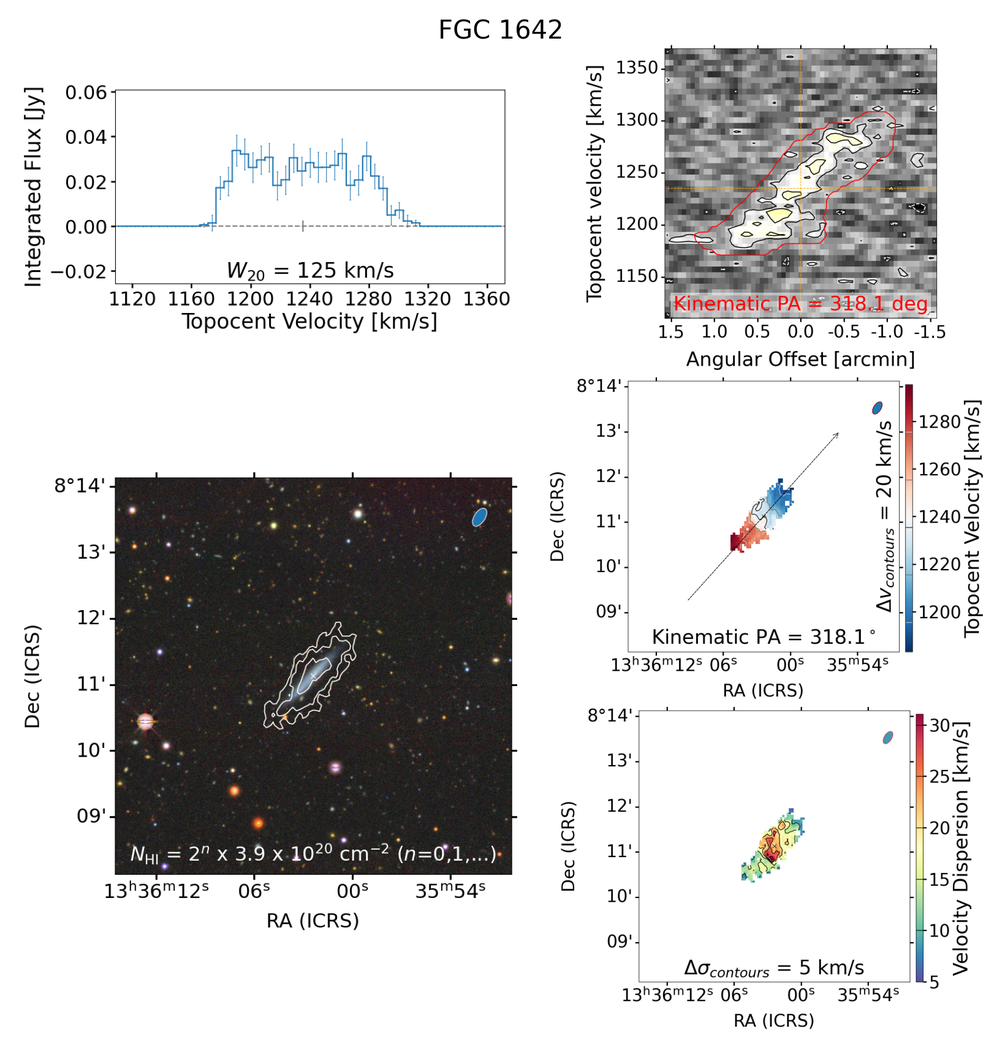}}
    \phantomcaption
\end{figure*}

\clearpage
\begin{figure*}
    \ContinuedFloat
    \centering
    \resizebox{0.86\hsize}{!}{
    \includegraphics{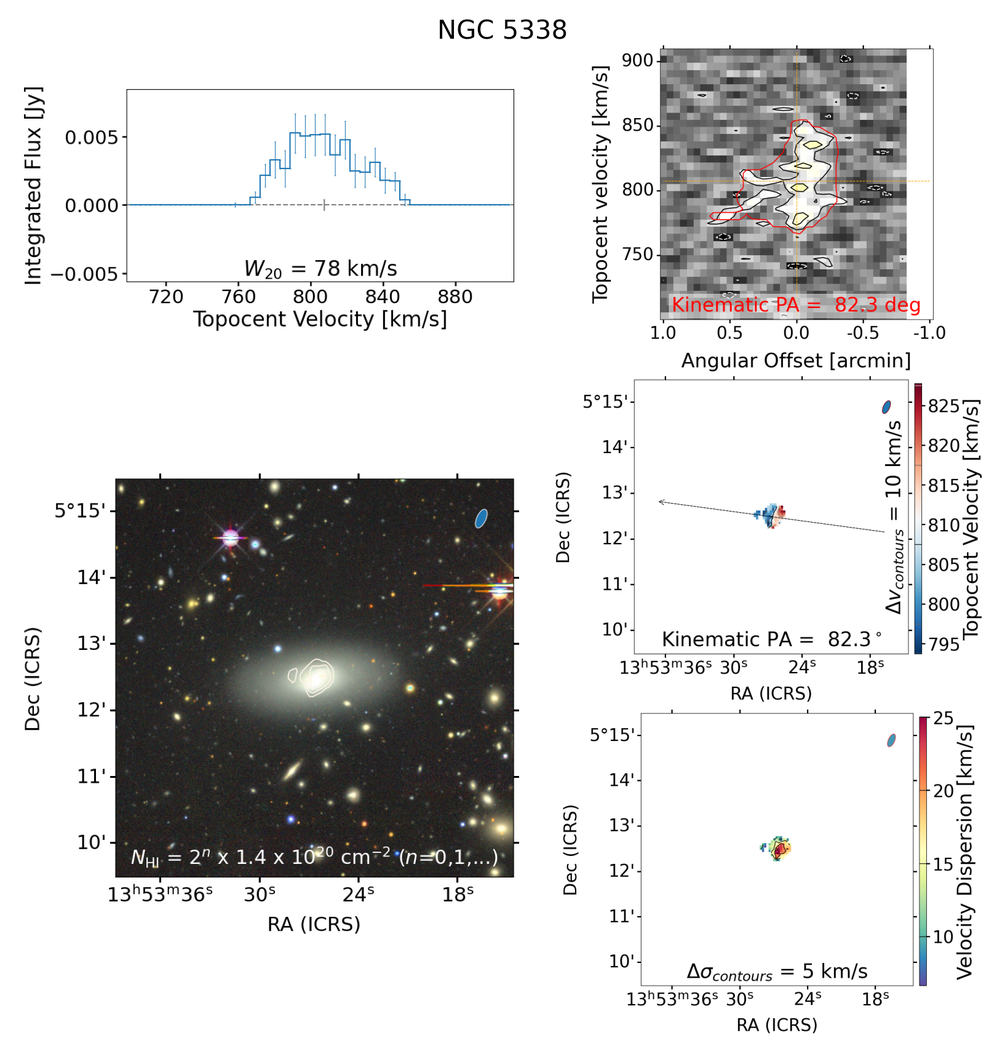}\hspace{0.5cm}
    \includegraphics{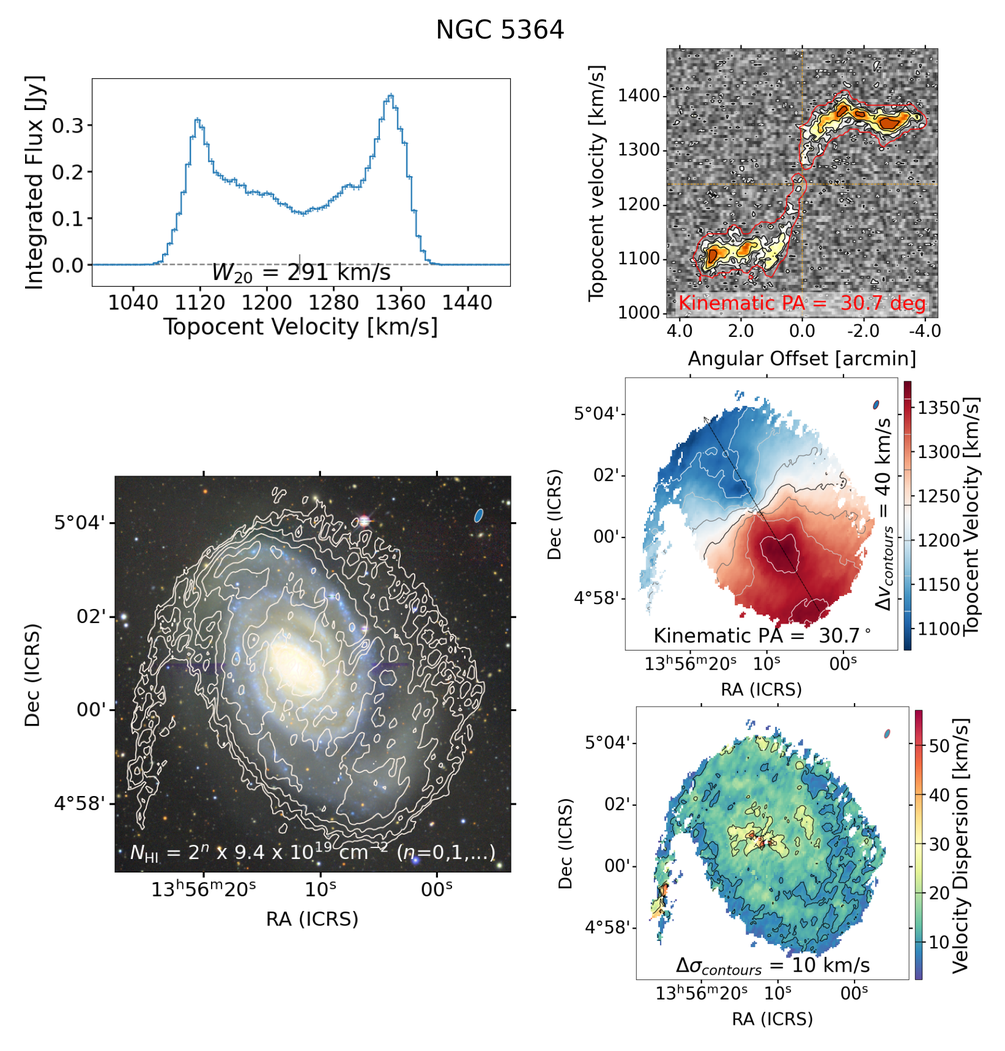}}
    \vspace{0.3cm}
    \resizebox{0.86\hsize}{!}{
    \includegraphics{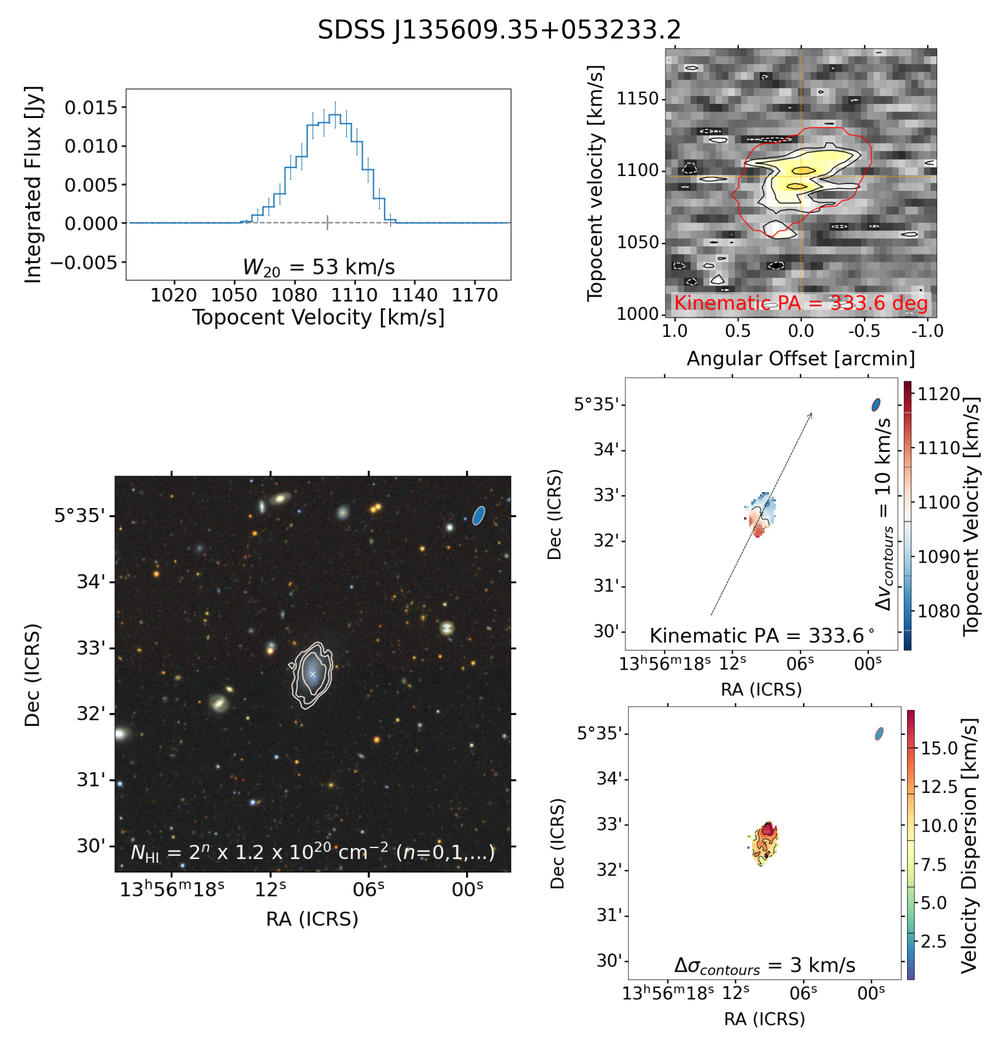}\hspace{0.5cm}
    \includegraphics{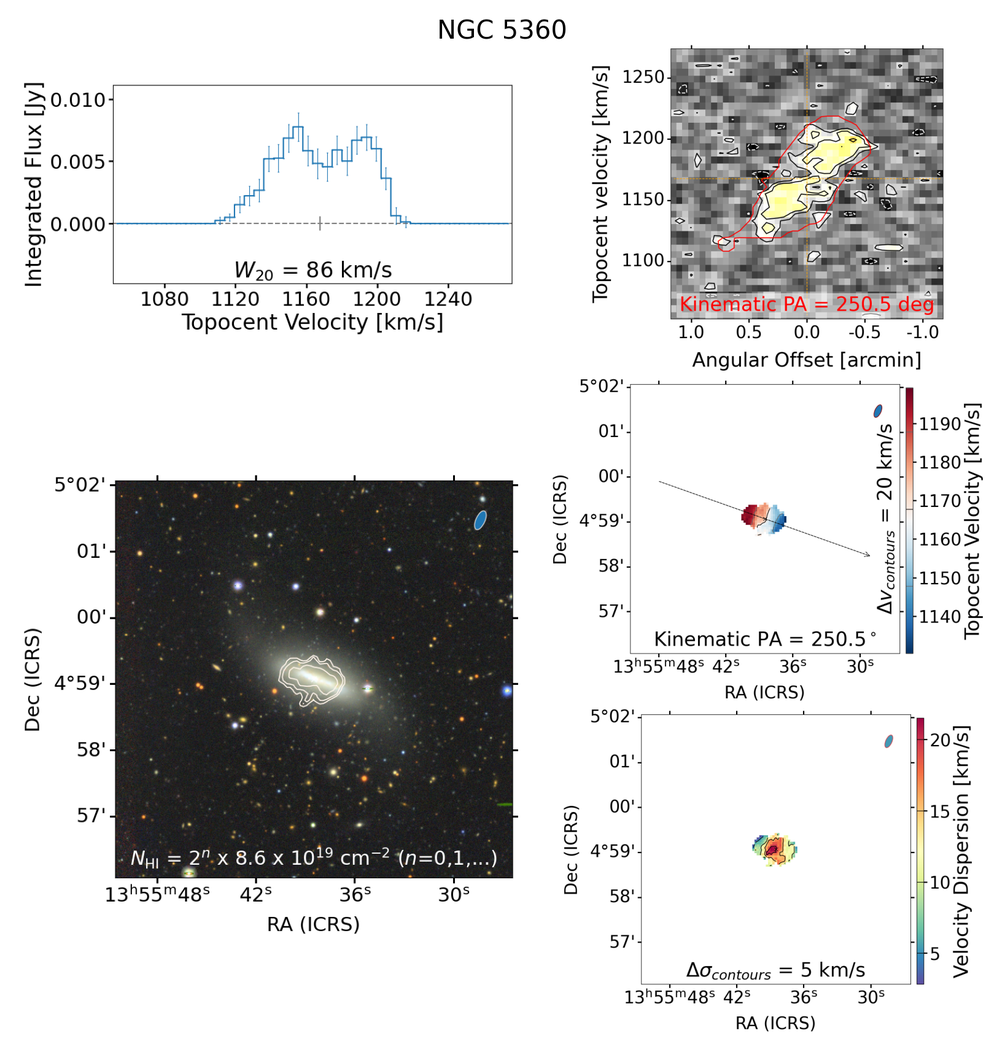}}
    \vspace{0.3cm}
    \resizebox{0.86\hsize}{!}{
    \includegraphics{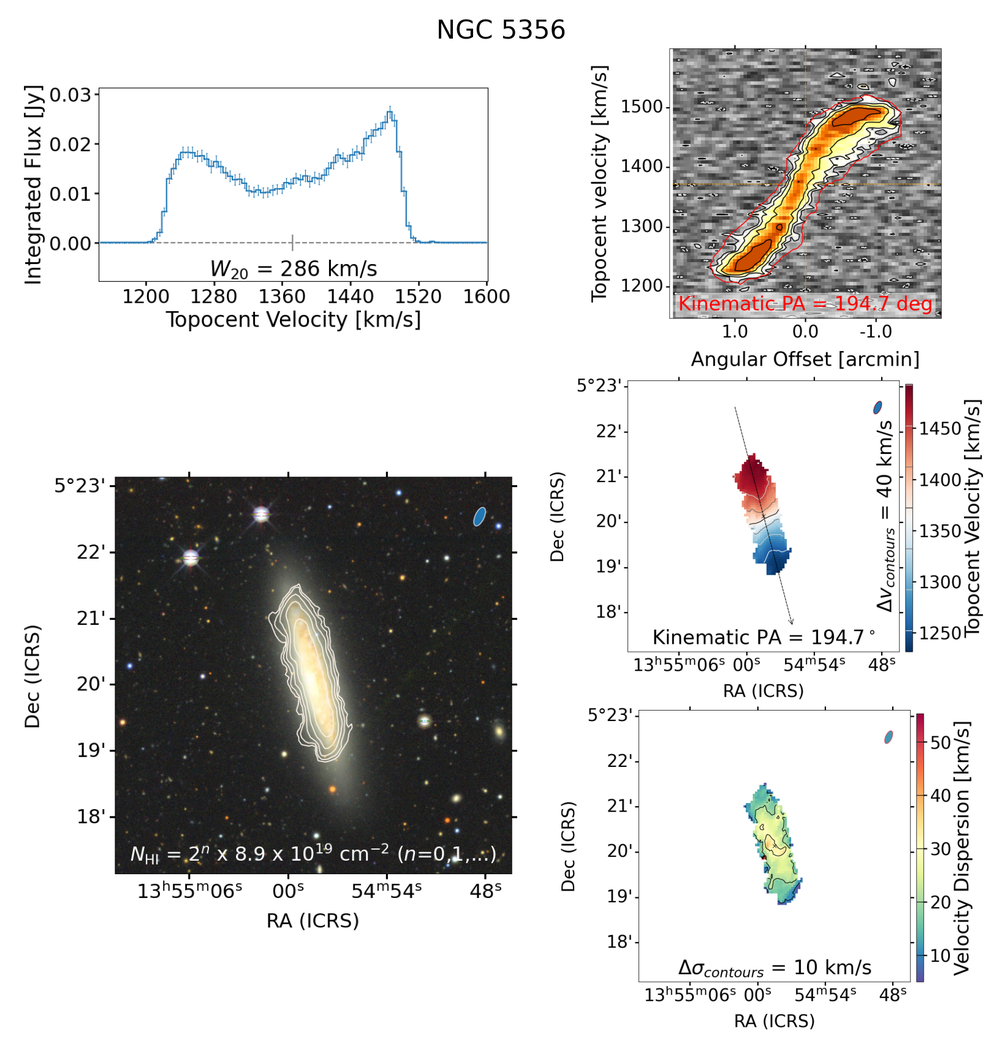}\hspace{0.5cm}
    \includegraphics{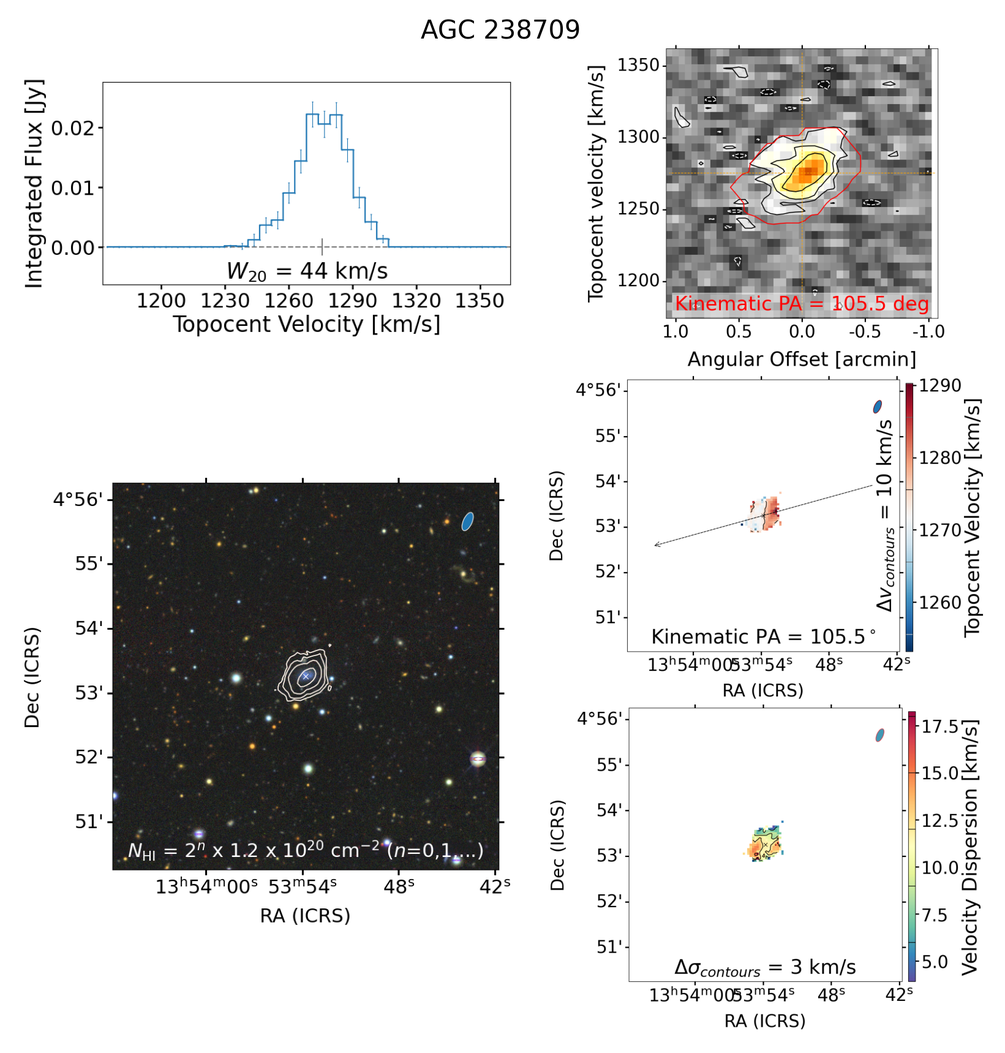}}
    \phantomcaption
\end{figure*}

\clearpage
\begin{figure*}
    \ContinuedFloat
    \centering
    \resizebox{0.86\hsize}{!}{
    \includegraphics{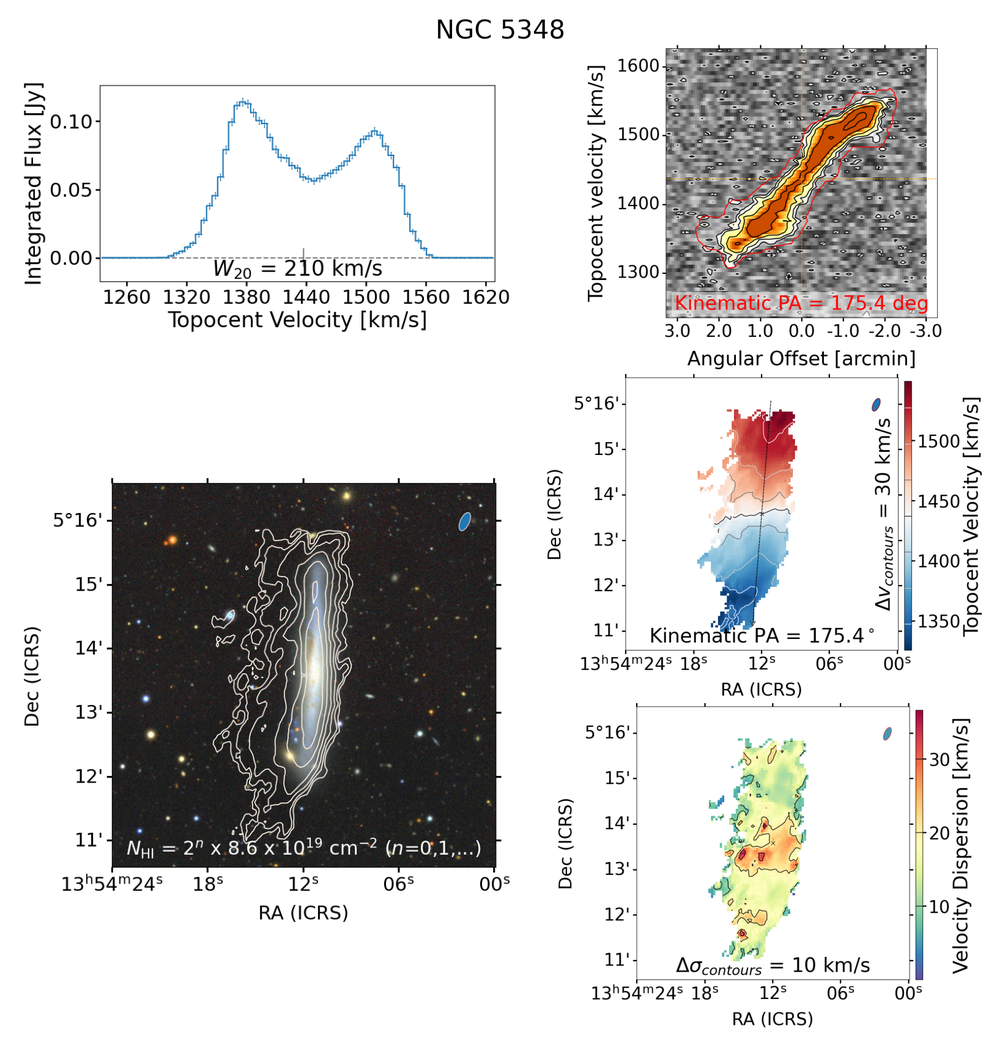}\hspace{0.5cm}
    \includegraphics{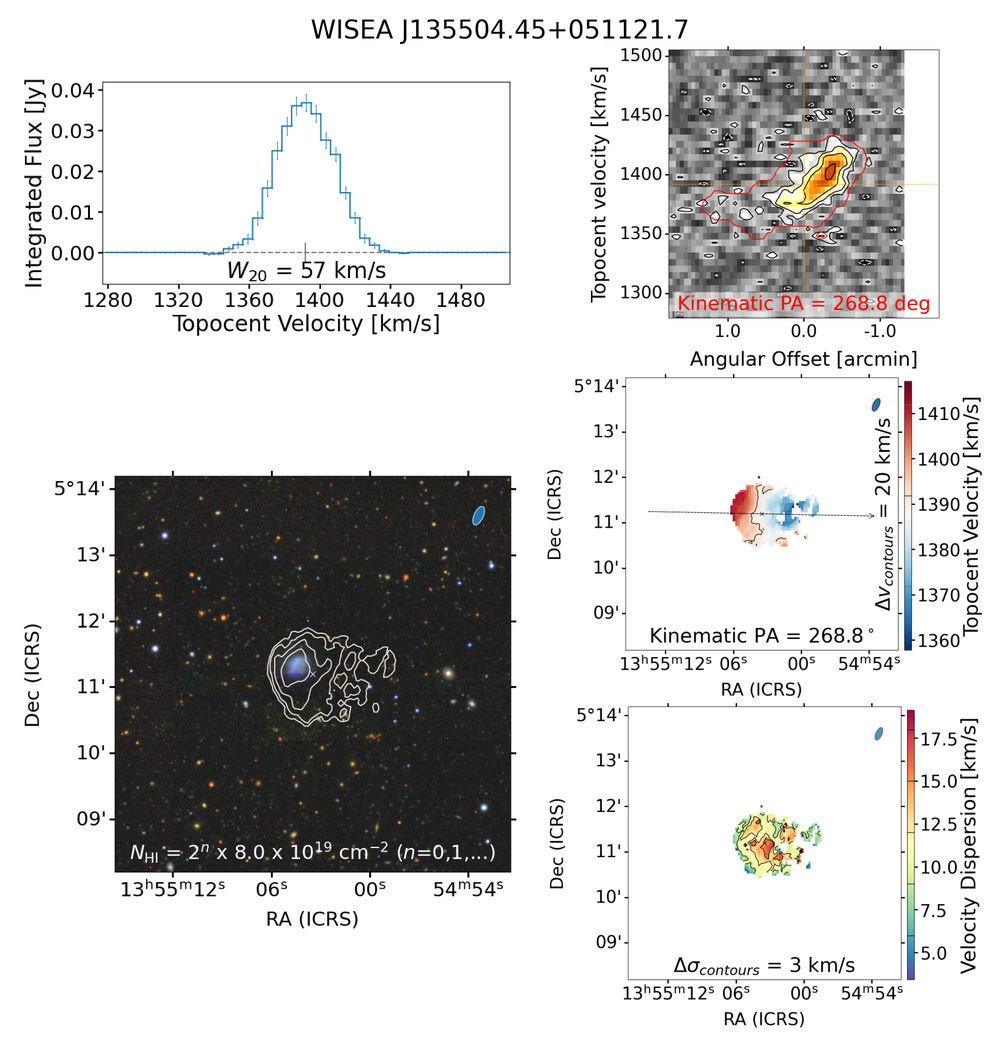}}
    \vspace{0.3cm}
    \resizebox{0.86\hsize}{!}{
    \includegraphics{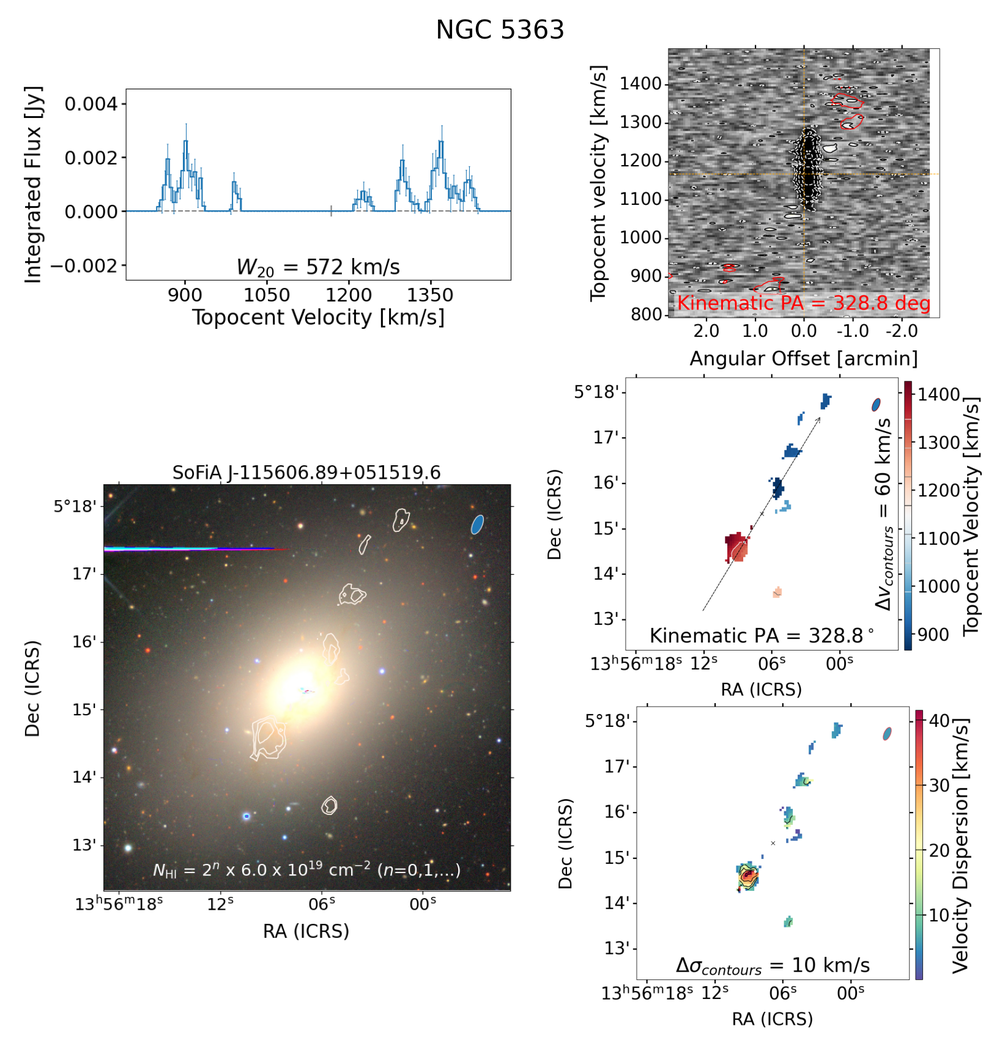}\hspace{0.5cm}
    \includegraphics{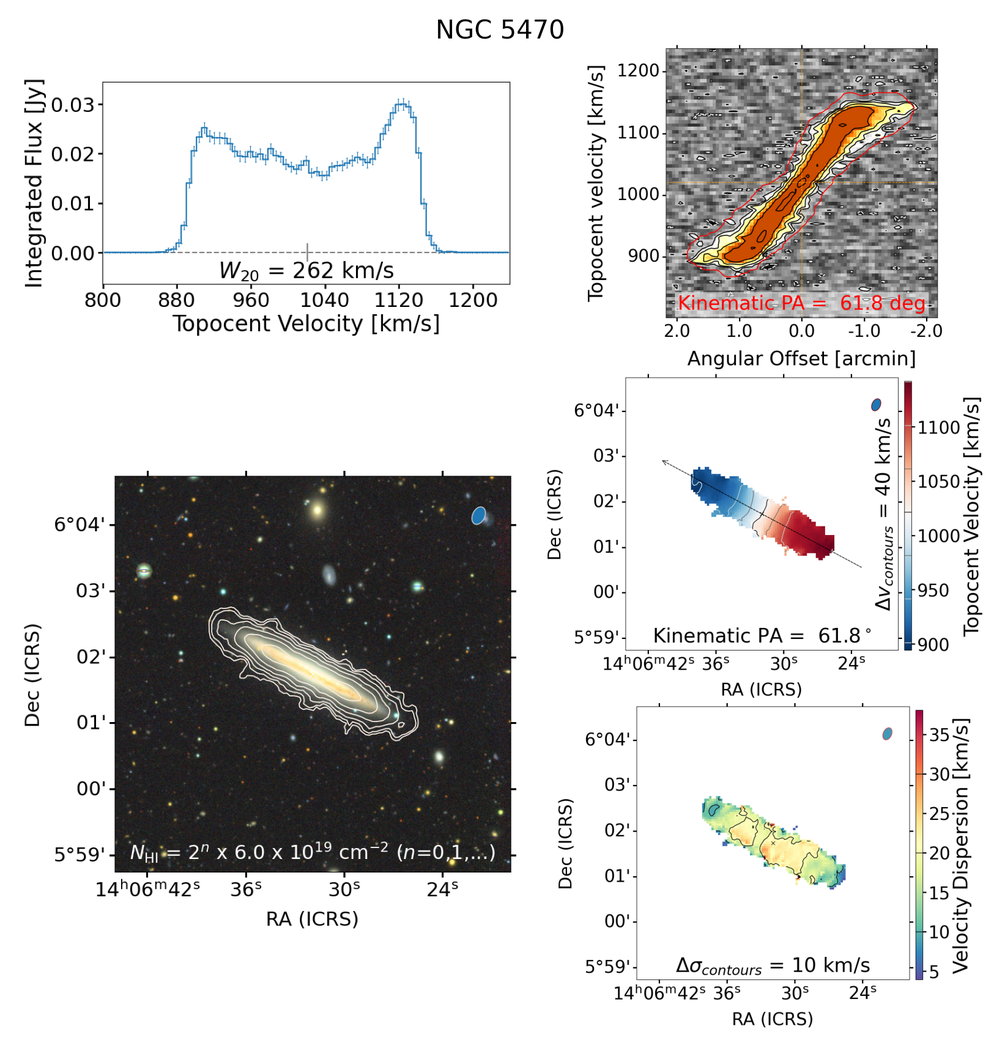}}
    \vspace{0.3cm}
    \resizebox{0.86\hsize}{!}{
    \includegraphics{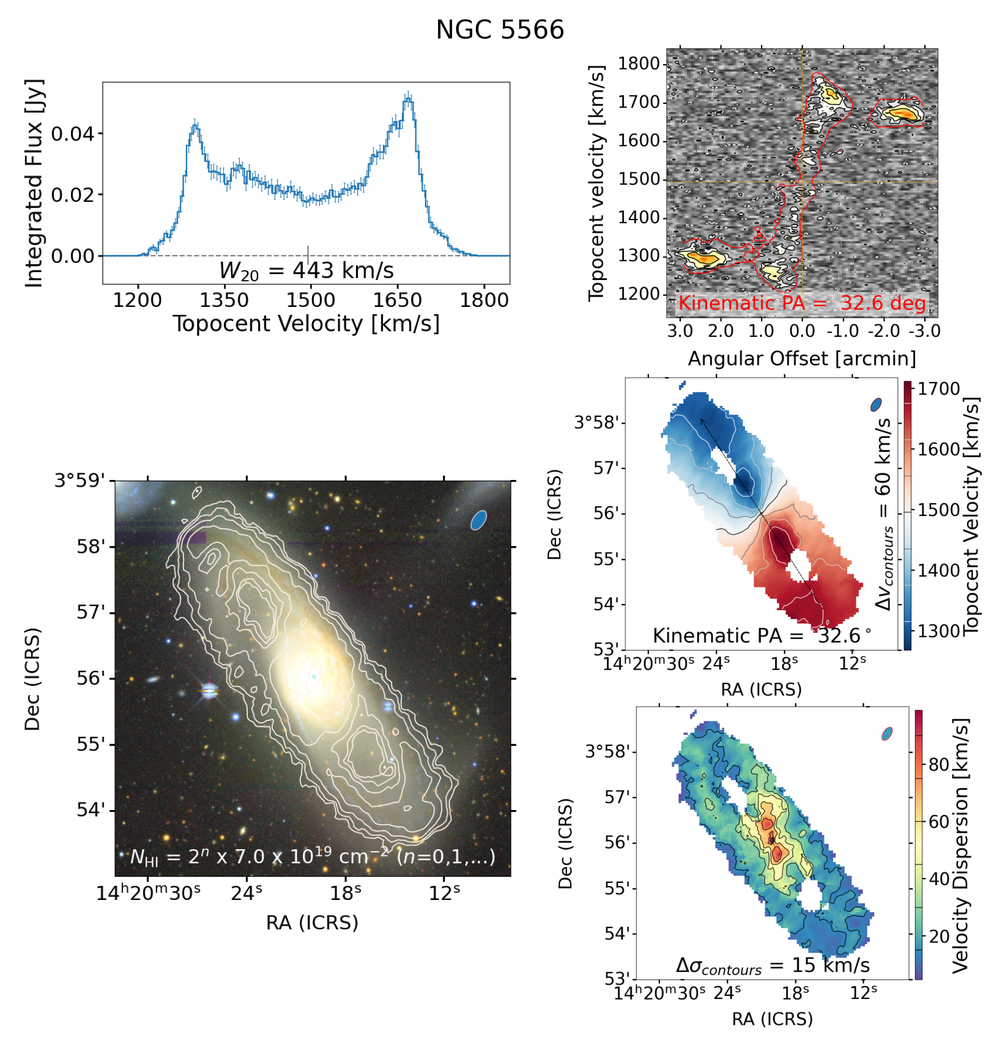}\hspace{0.5cm}
    \includegraphics{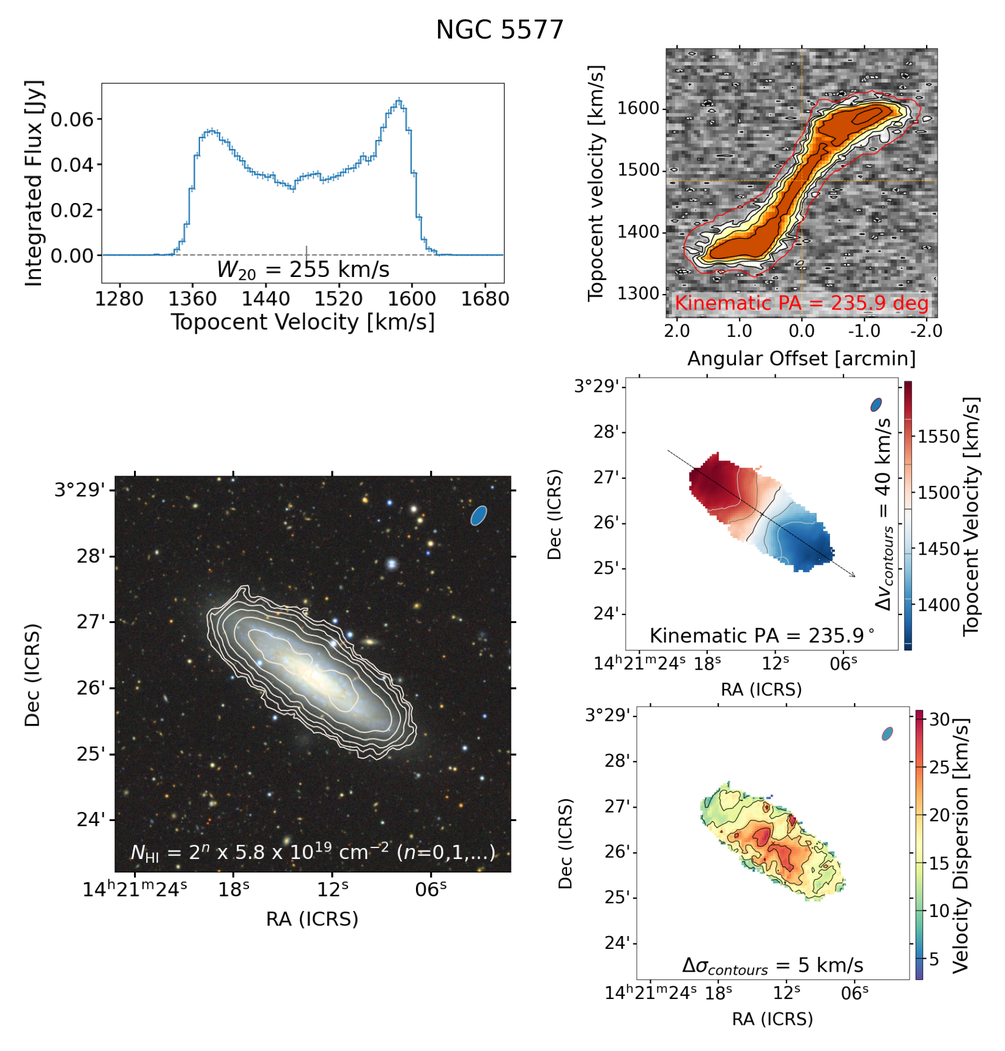}}
    \phantomcaption
\end{figure*}

\clearpage
\begin{figure*}
    \ContinuedFloat
    \centering
    \resizebox{0.86\hsize}{!}{
    \includegraphics{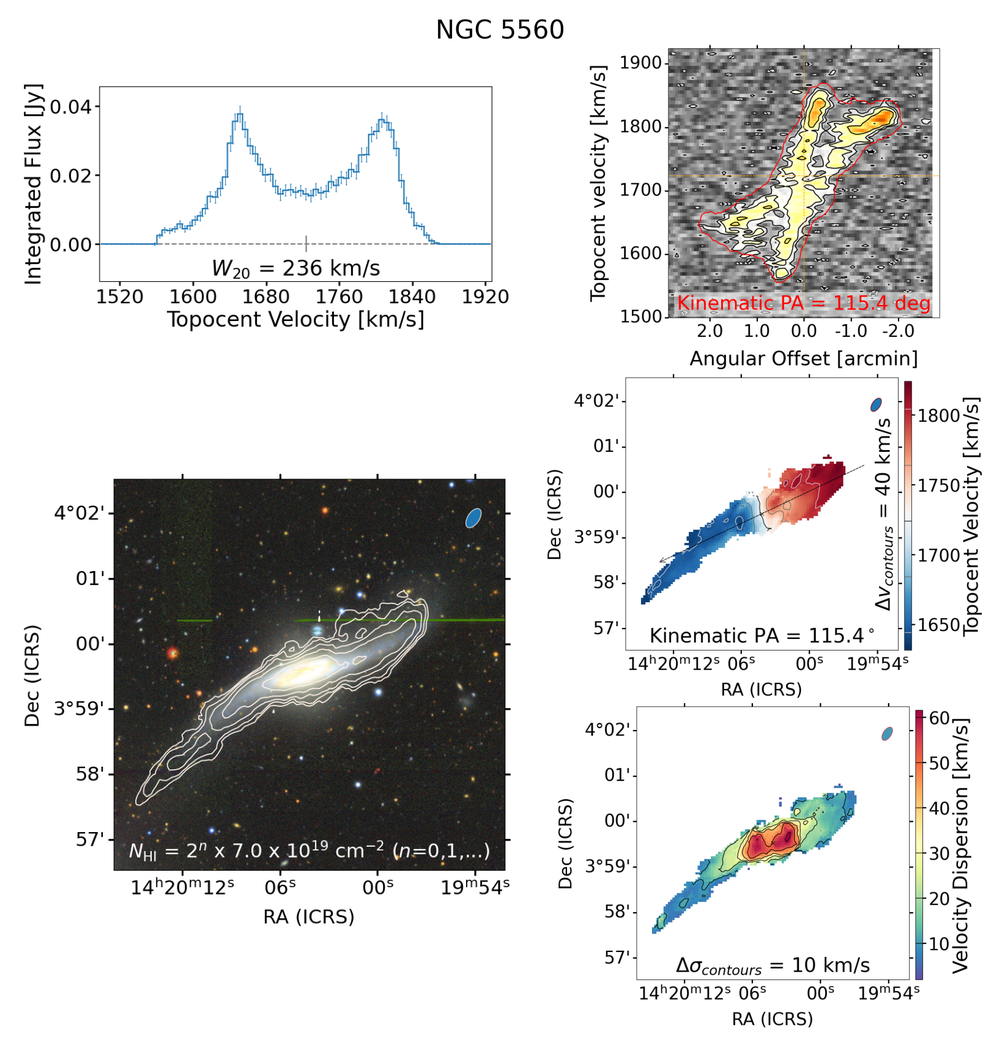}\hspace{0.5cm}
    \includegraphics{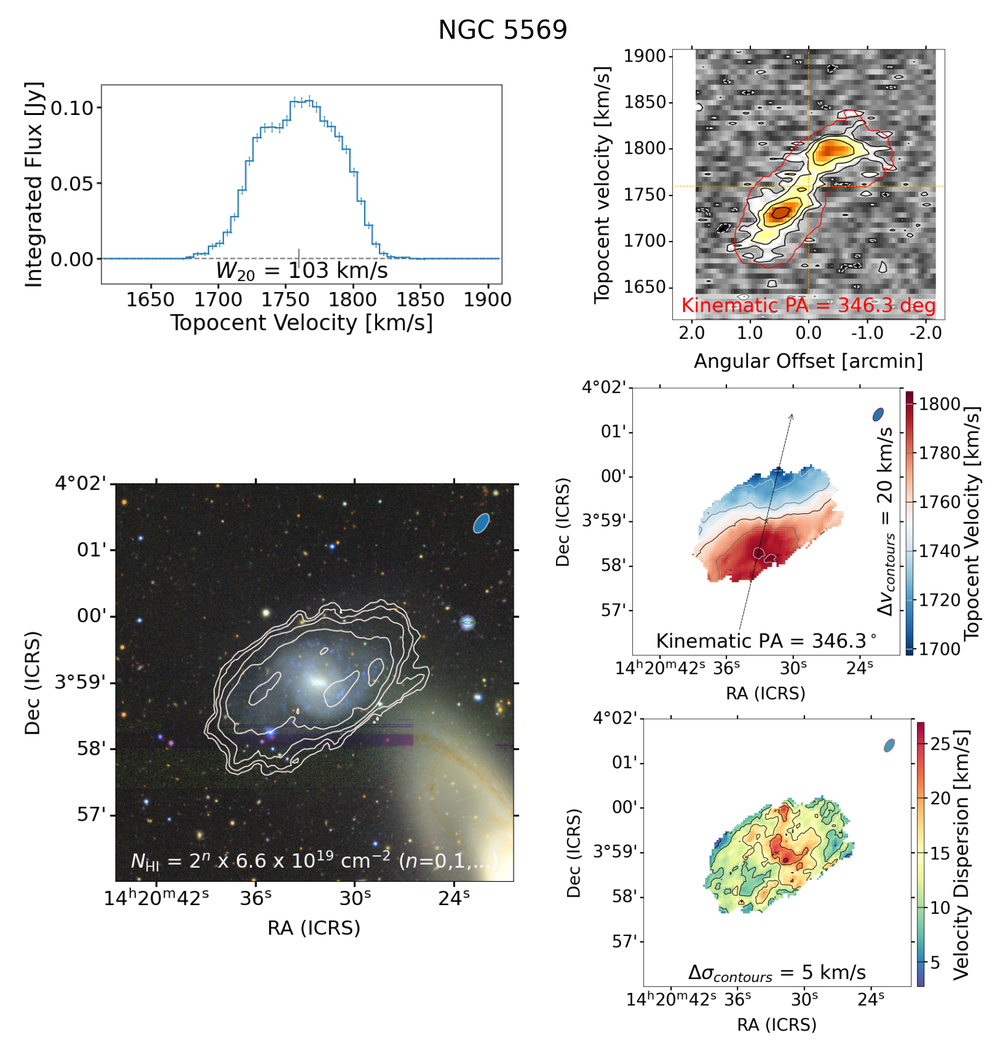}}
    \vspace{0.3cm}
    \resizebox{0.86\hsize}{!}{
    \includegraphics{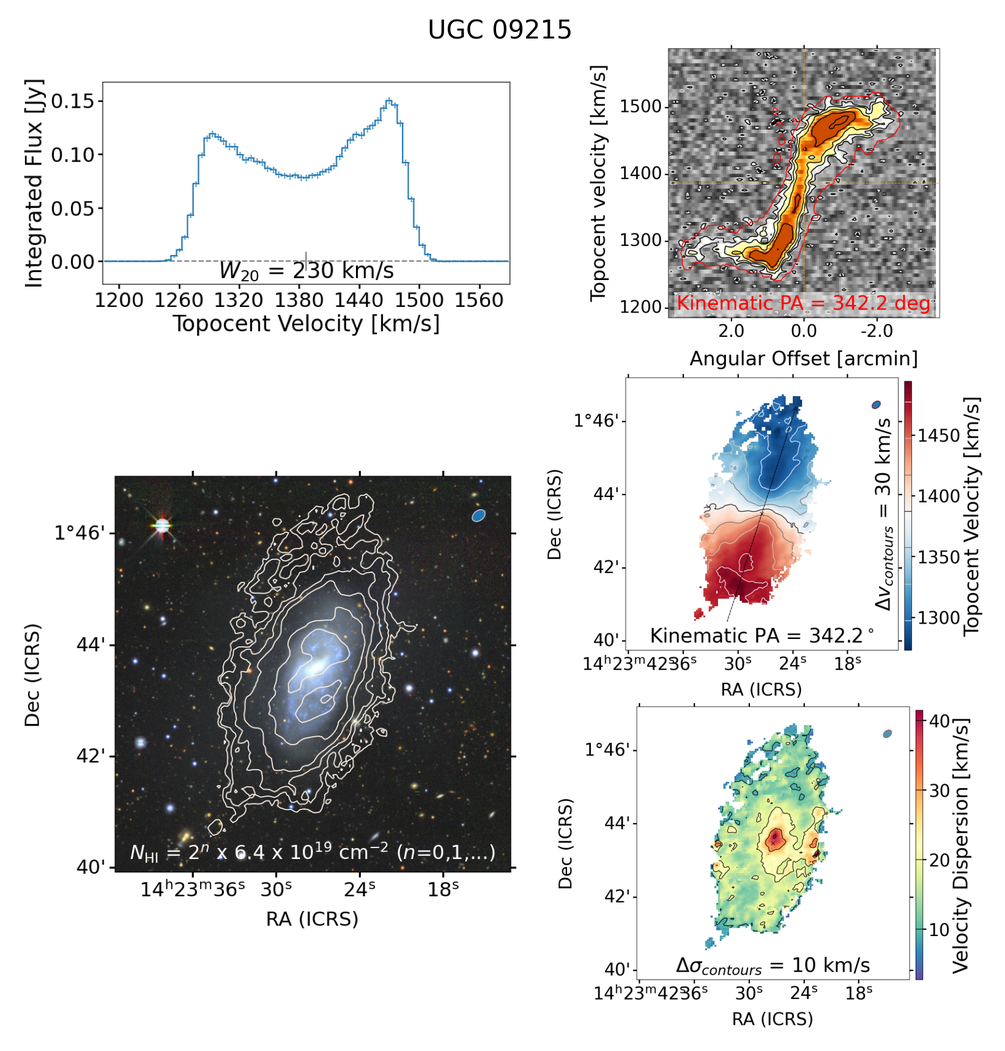}\hspace{0.5cm}
    \includegraphics{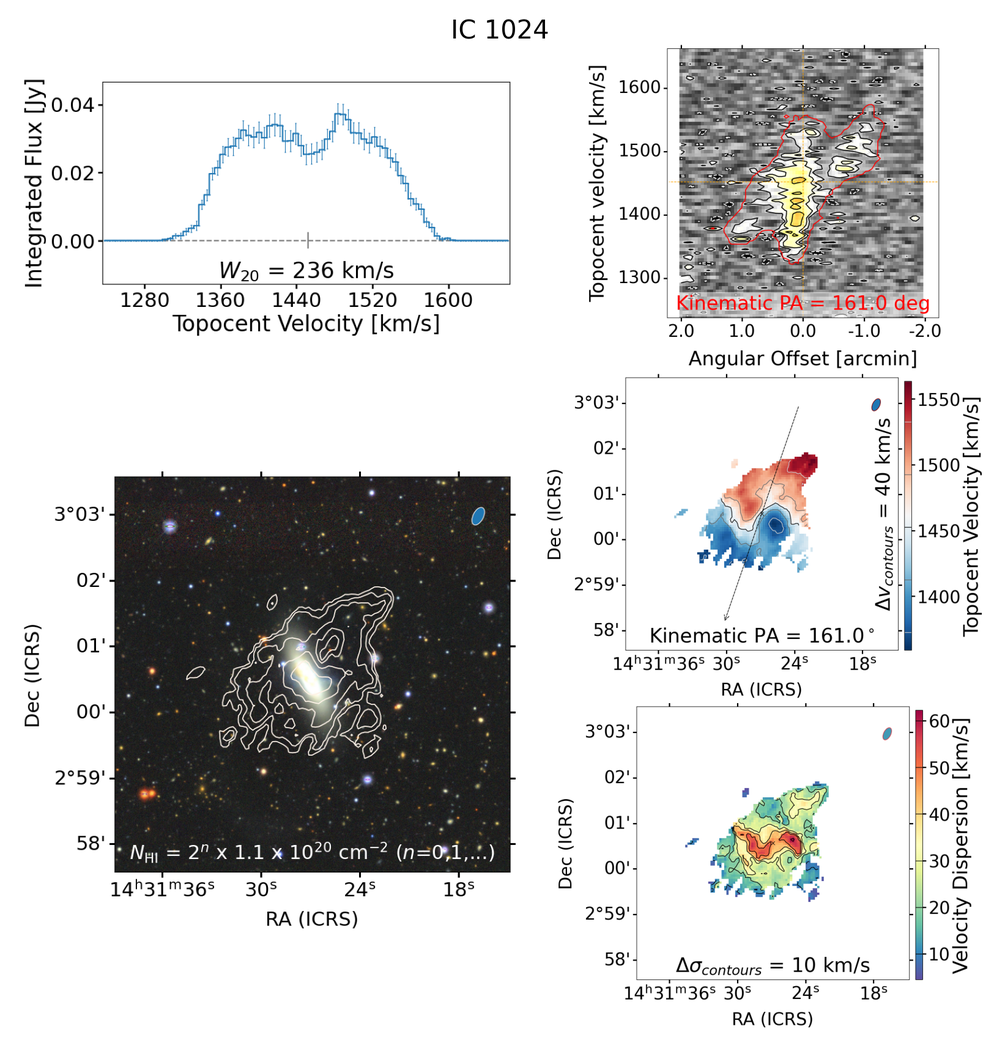}}
    \vspace{0.3cm}
    \resizebox{0.86\hsize}{!}{
    \includegraphics{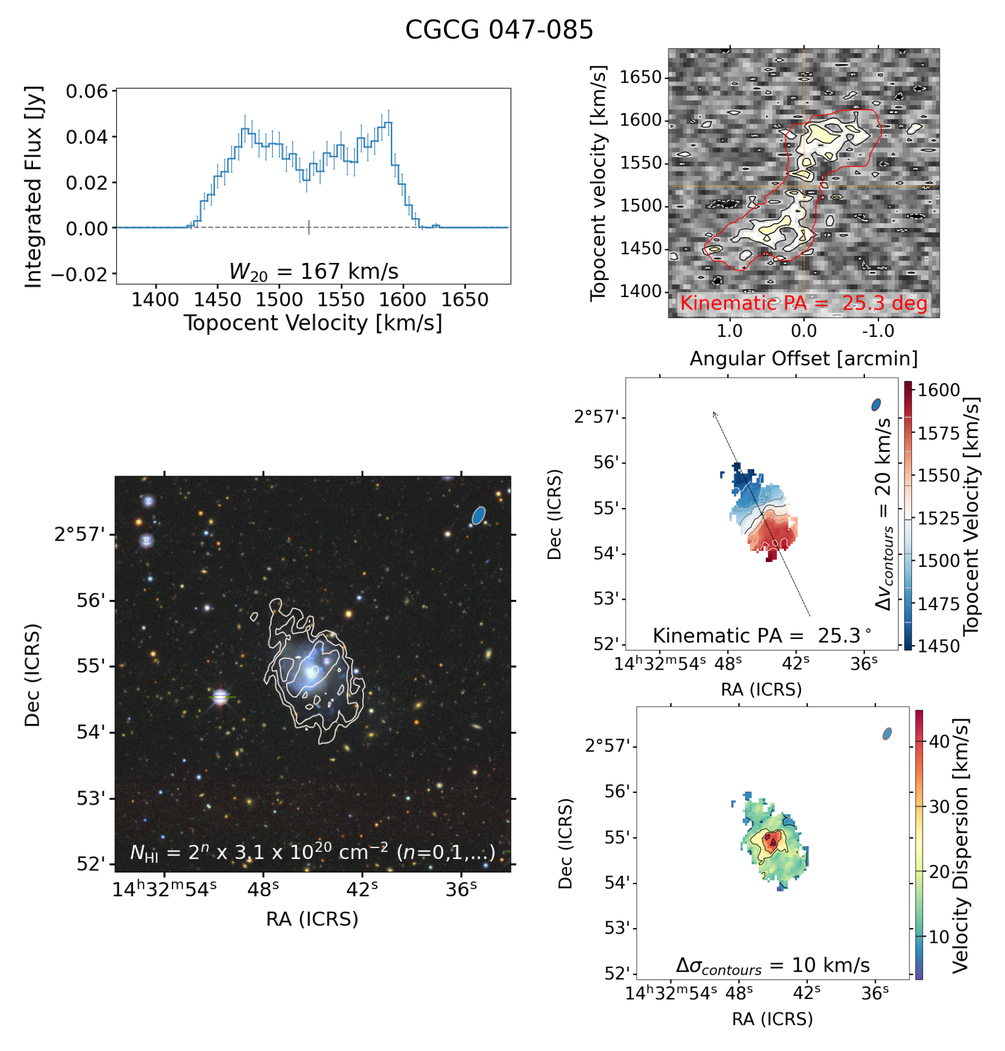}\hspace{0.5cm}
    \includegraphics{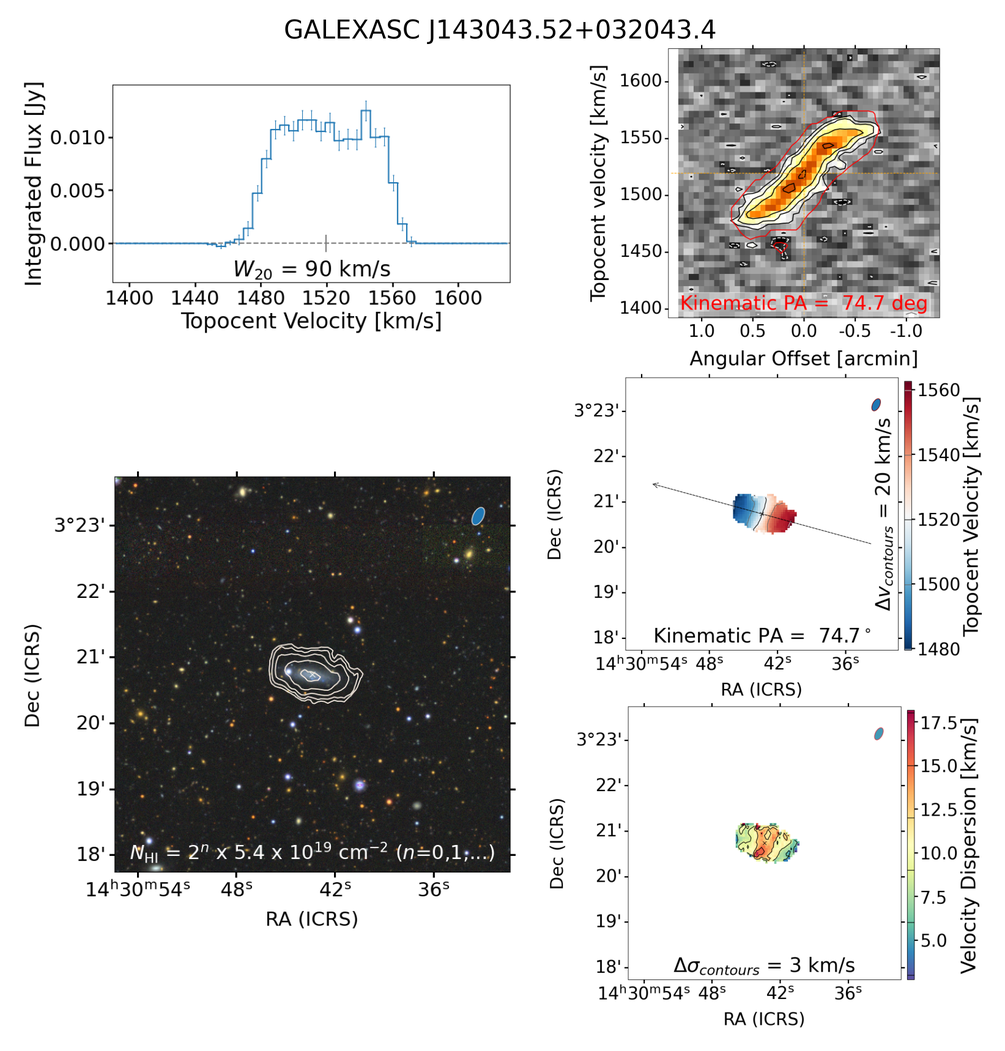}}
    \phantomcaption
\end{figure*}

\clearpage
\begin{figure*}
    \ContinuedFloat
    \centering
    \resizebox{0.86\hsize}{!}{
    \includegraphics{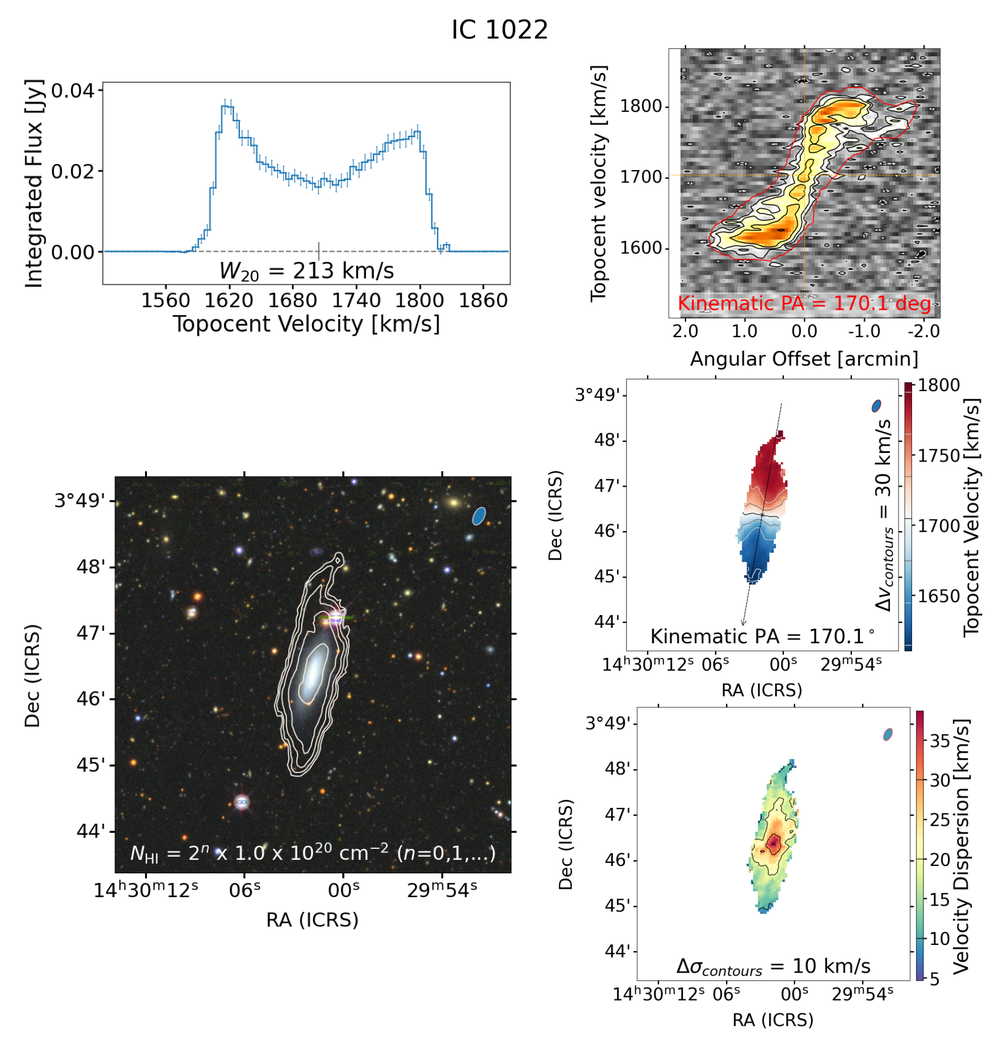}\hspace{0.5cm}
    \includegraphics{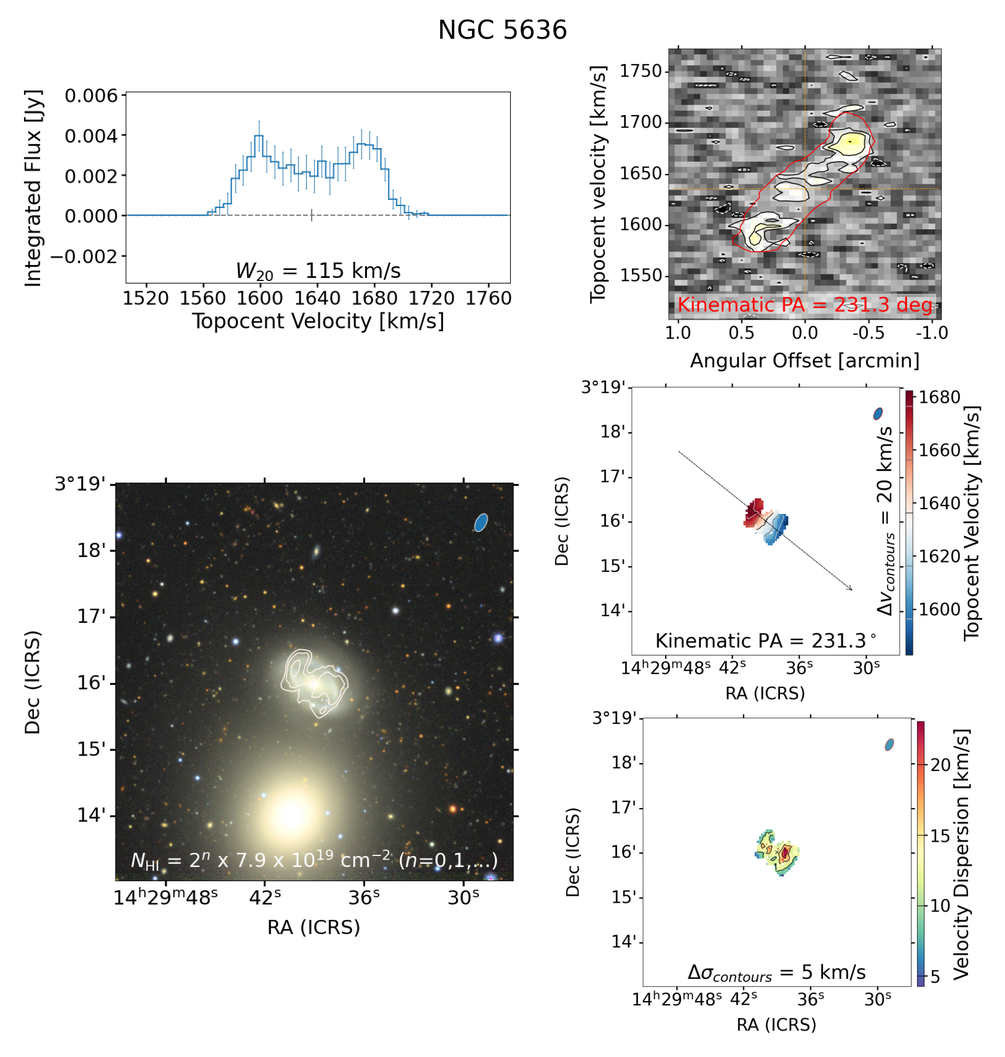}}
    \vspace{0.3cm}
    \resizebox{0.86\hsize}{!}{
    \includegraphics{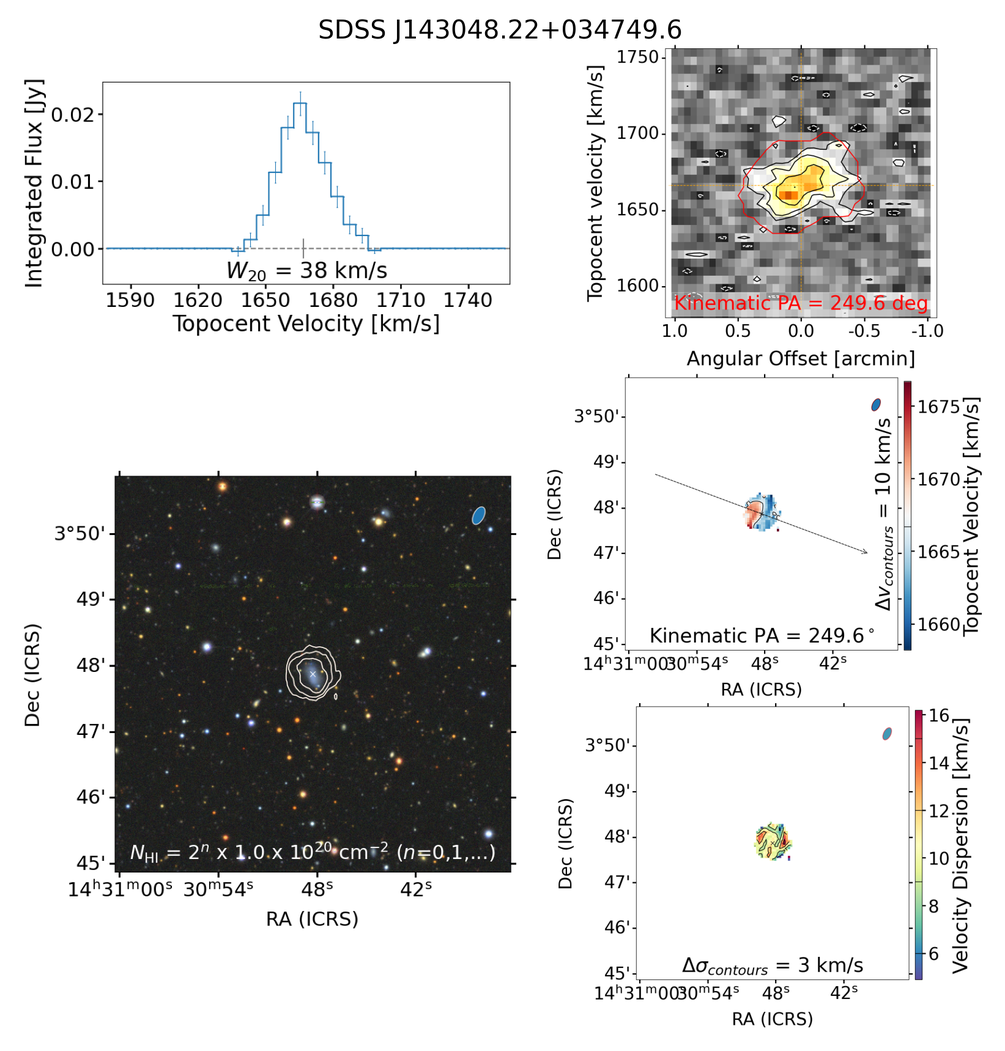}\hspace{0.5cm}
    \includegraphics{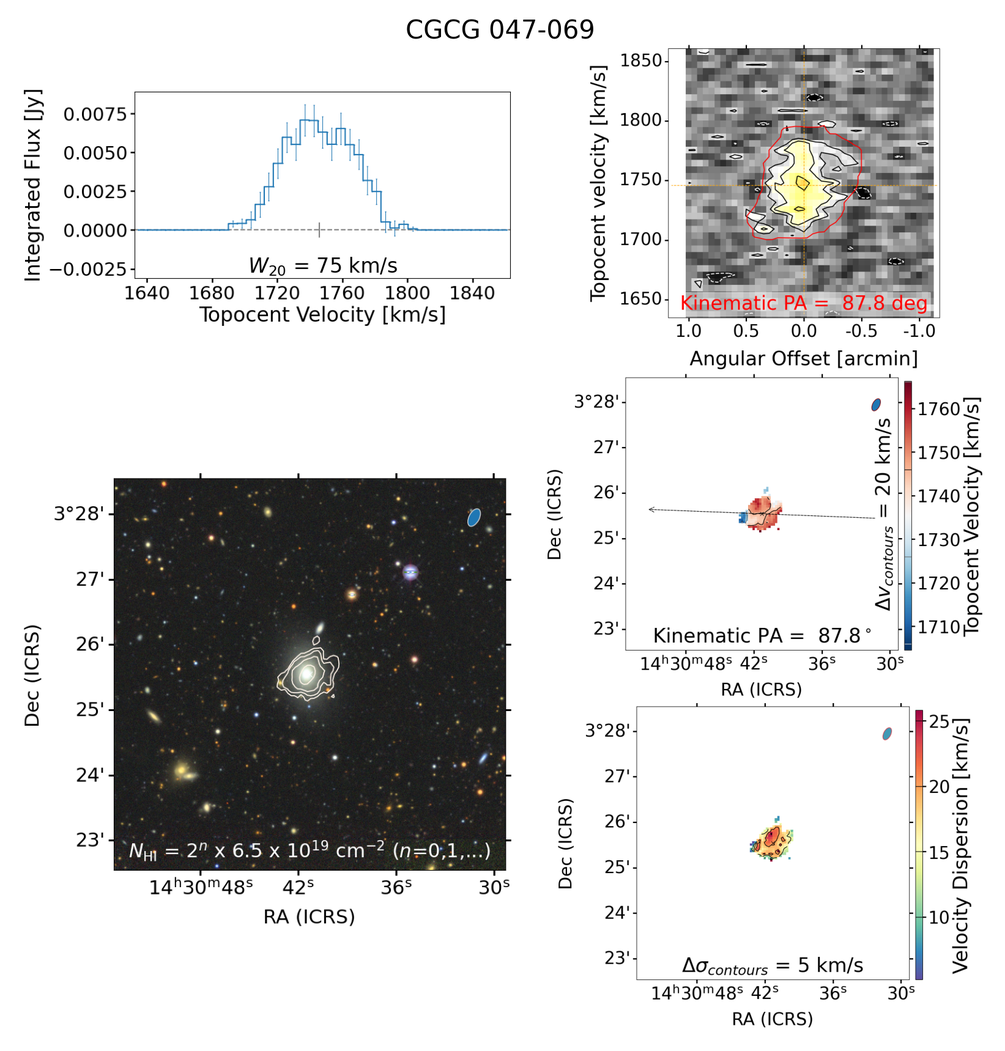}}
    \vspace{0.3cm}
    \resizebox{0.86\hsize}{!}{
    \includegraphics{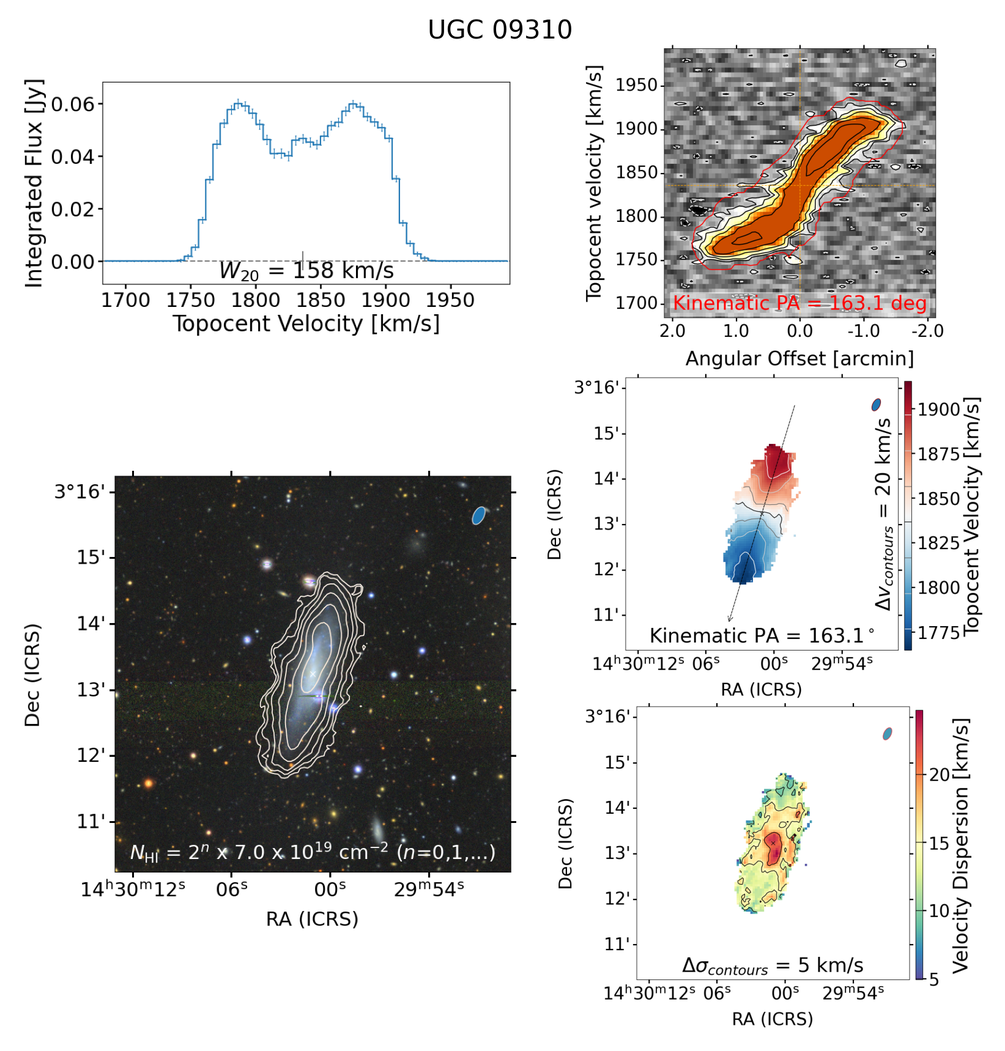}\hspace{0.5cm}
    \includegraphics{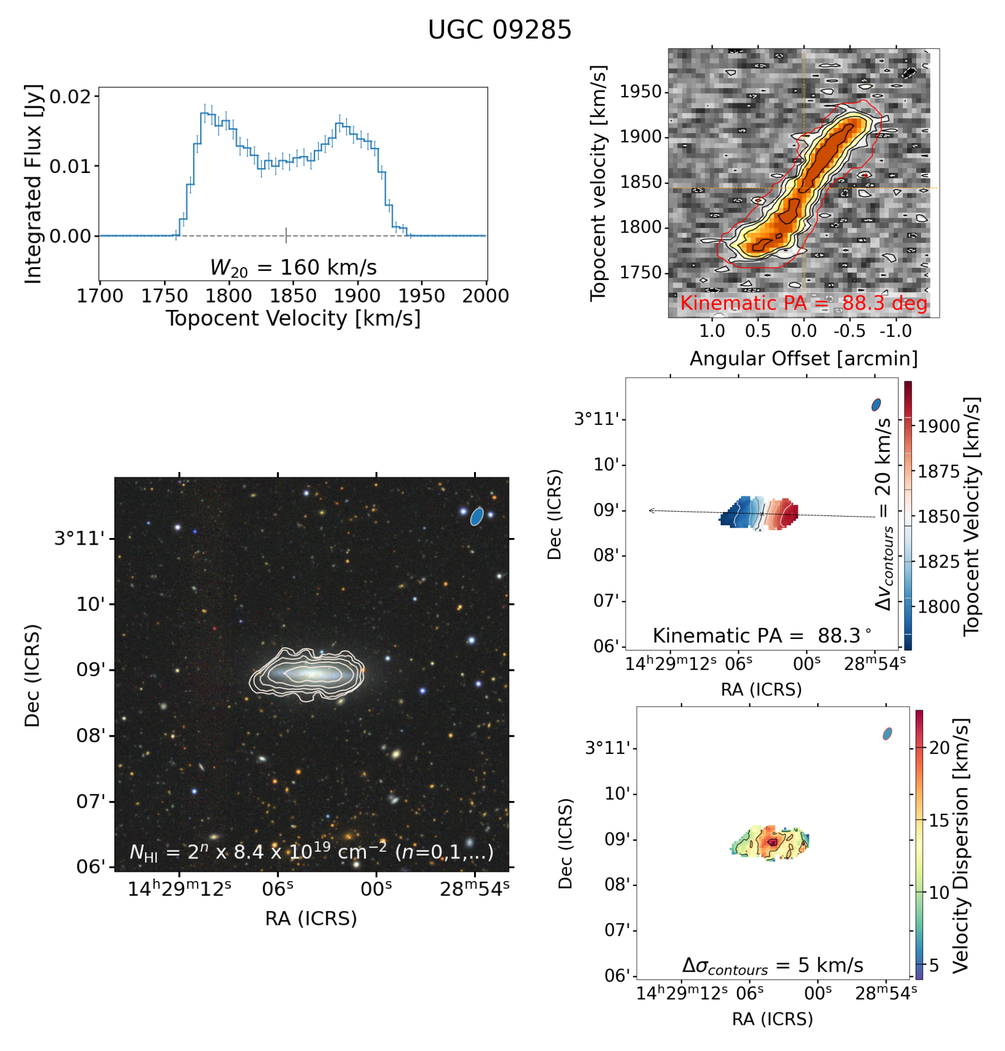}}
    \phantomcaption
\end{figure*}

\clearpage
\begin{figure*}
    \ContinuedFloat
    \centering
    \resizebox{0.86\hsize}{!}{
    \includegraphics{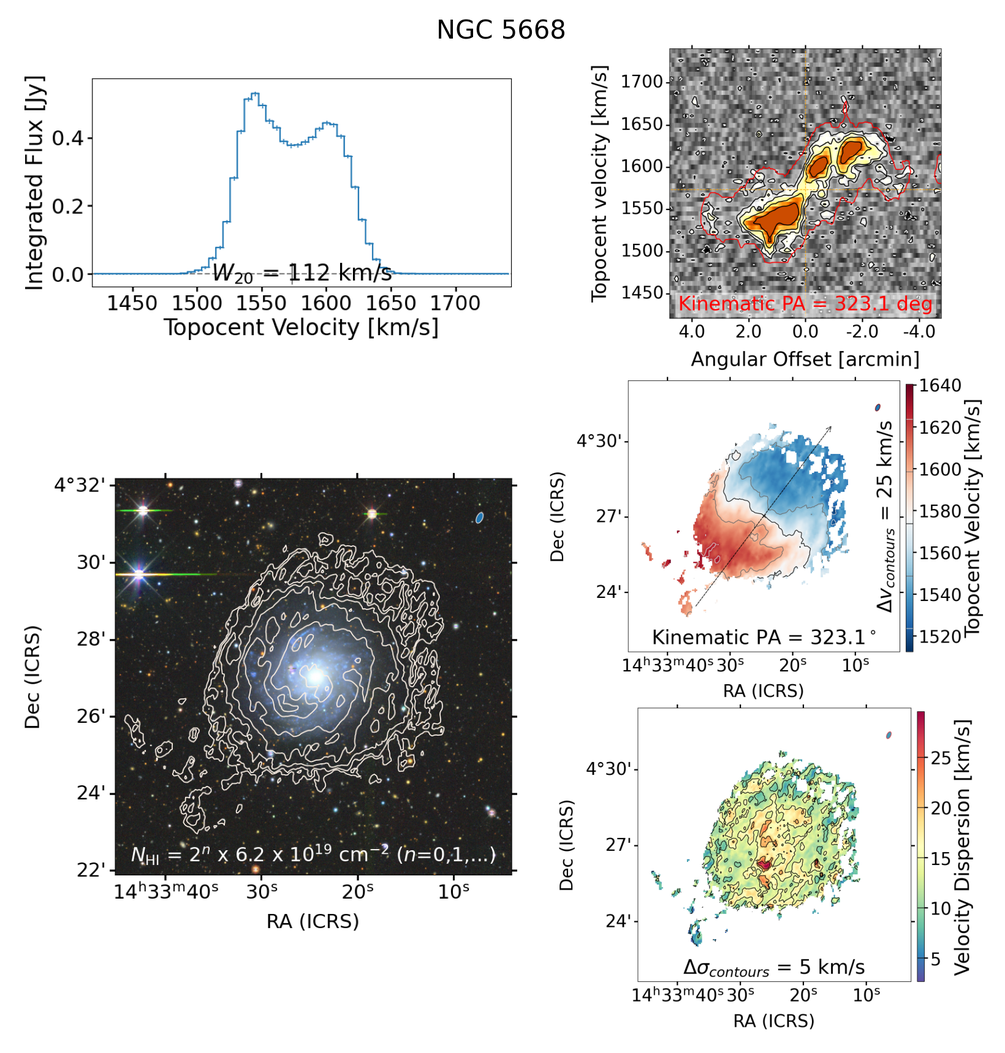}\hspace{0.5cm}
    \includegraphics{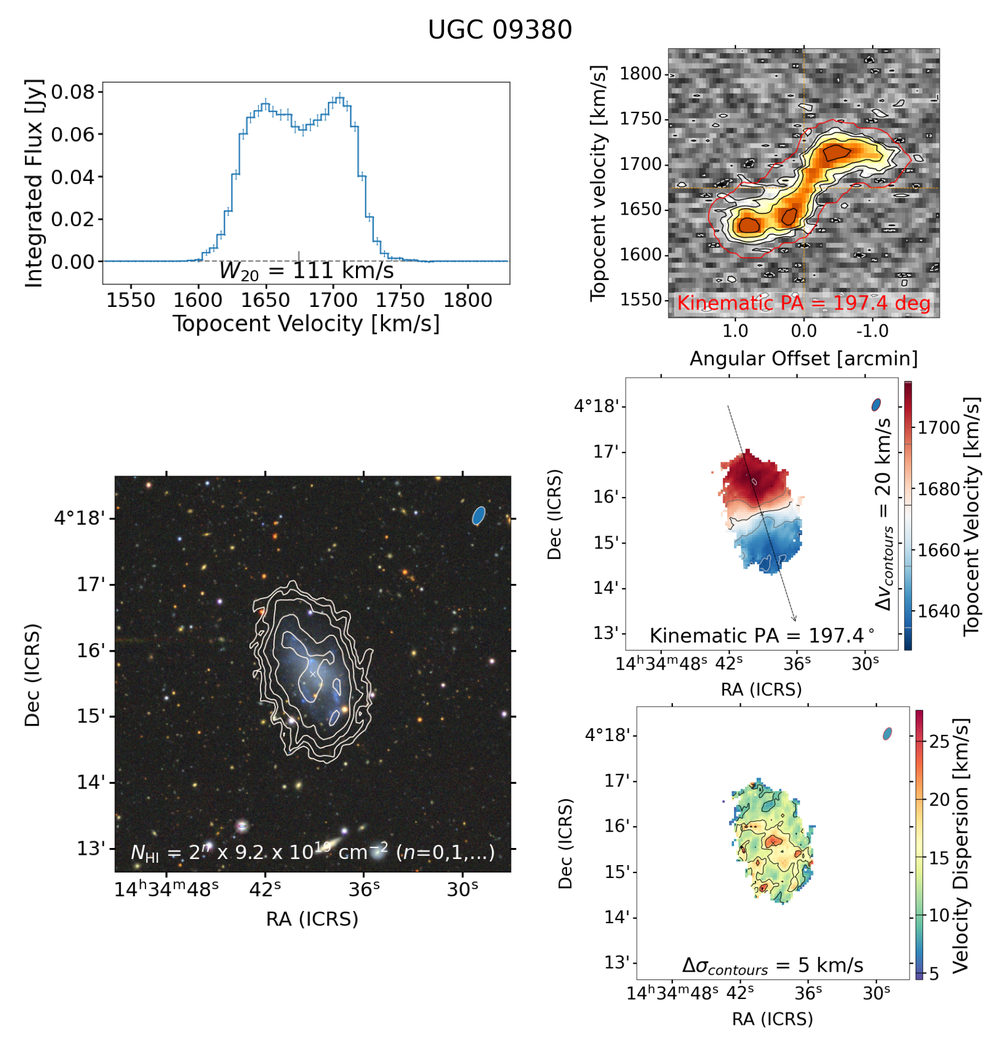}}
    \vspace{0.3cm}
    \resizebox{0.86\hsize}{!}{
    \includegraphics{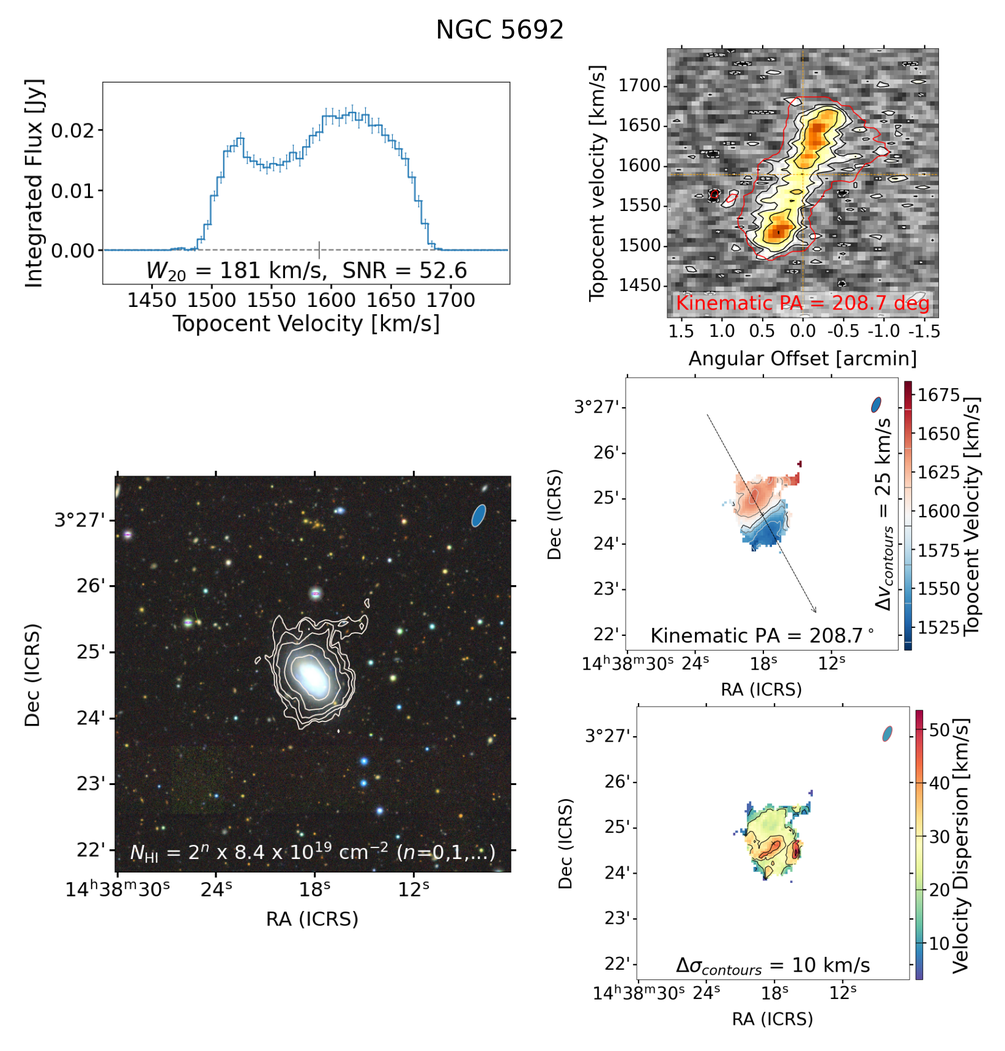}\hspace{0.5cm}
    \includegraphics{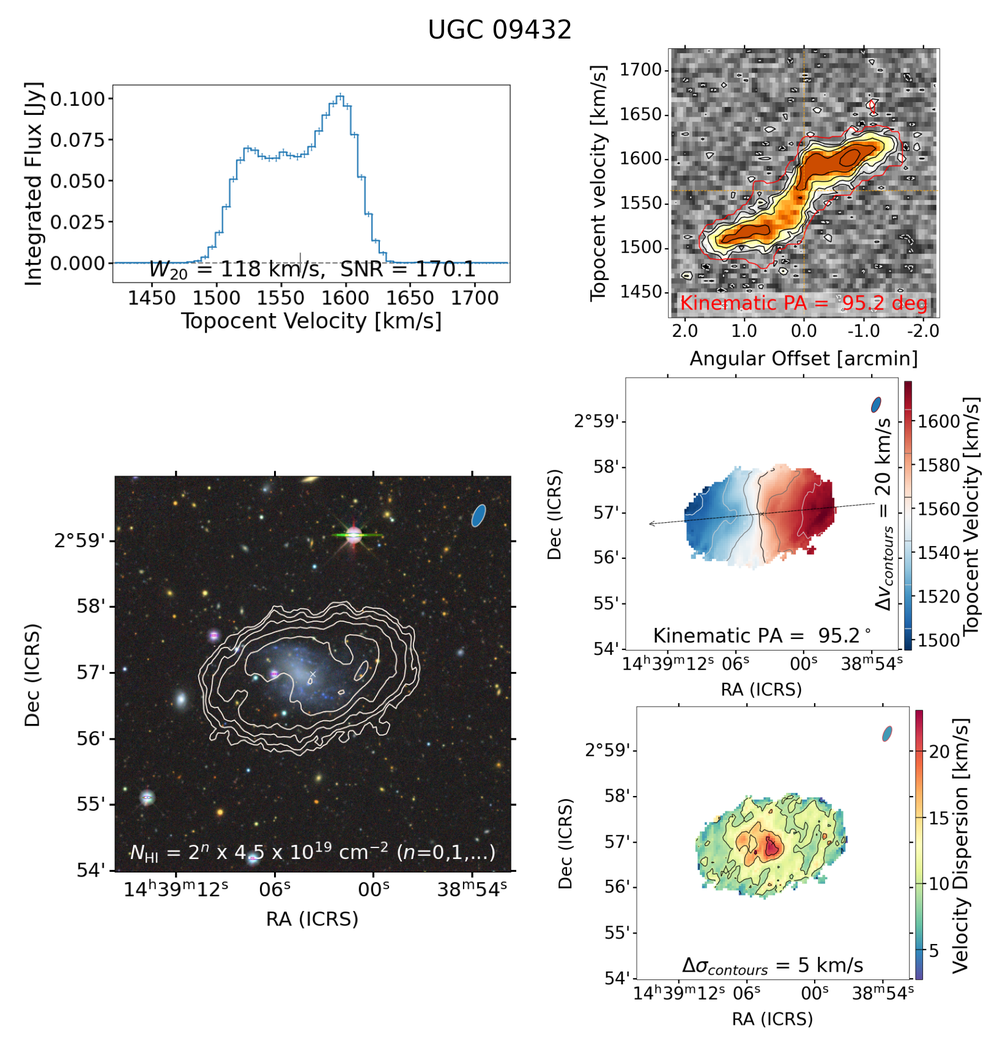}}
    \vspace{0.3cm}
    \resizebox{0.86\hsize}{!}{
    \includegraphics{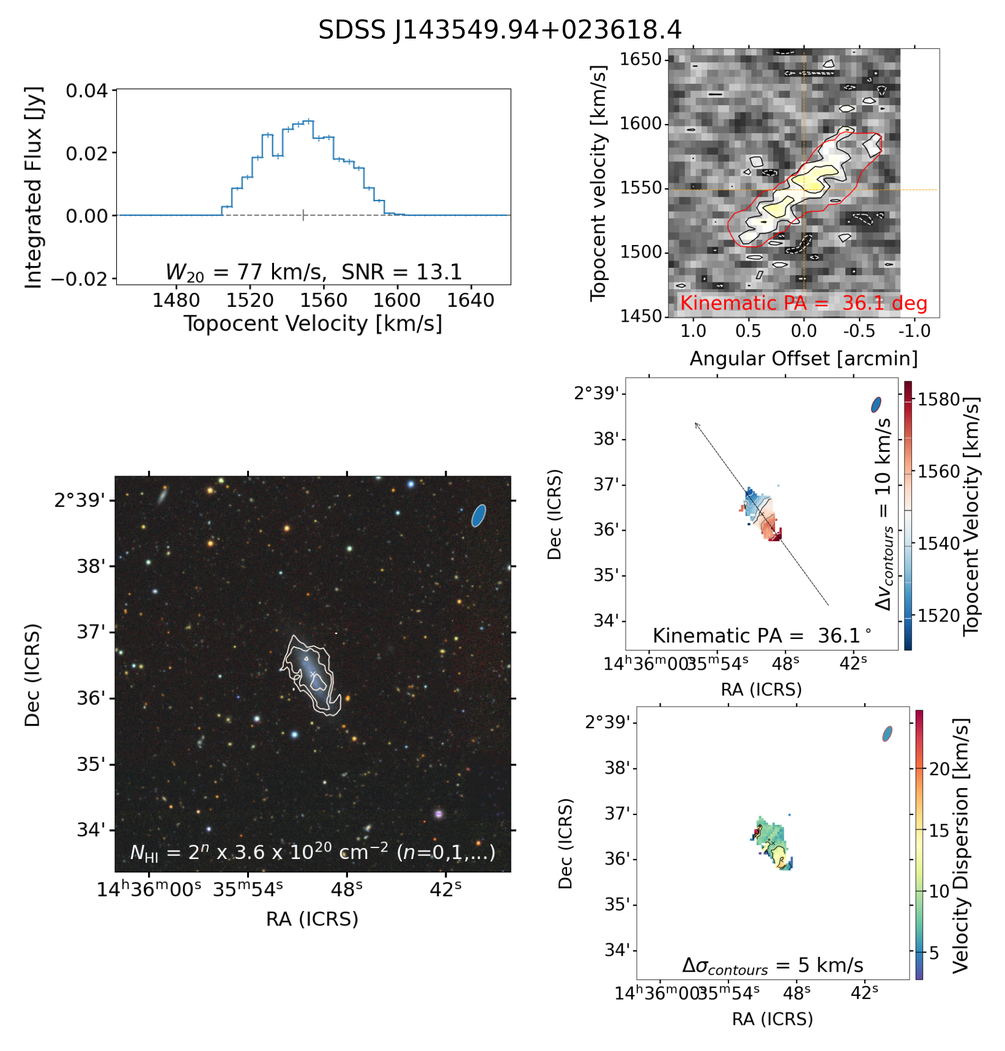}\hspace{0.5cm}
    \includegraphics{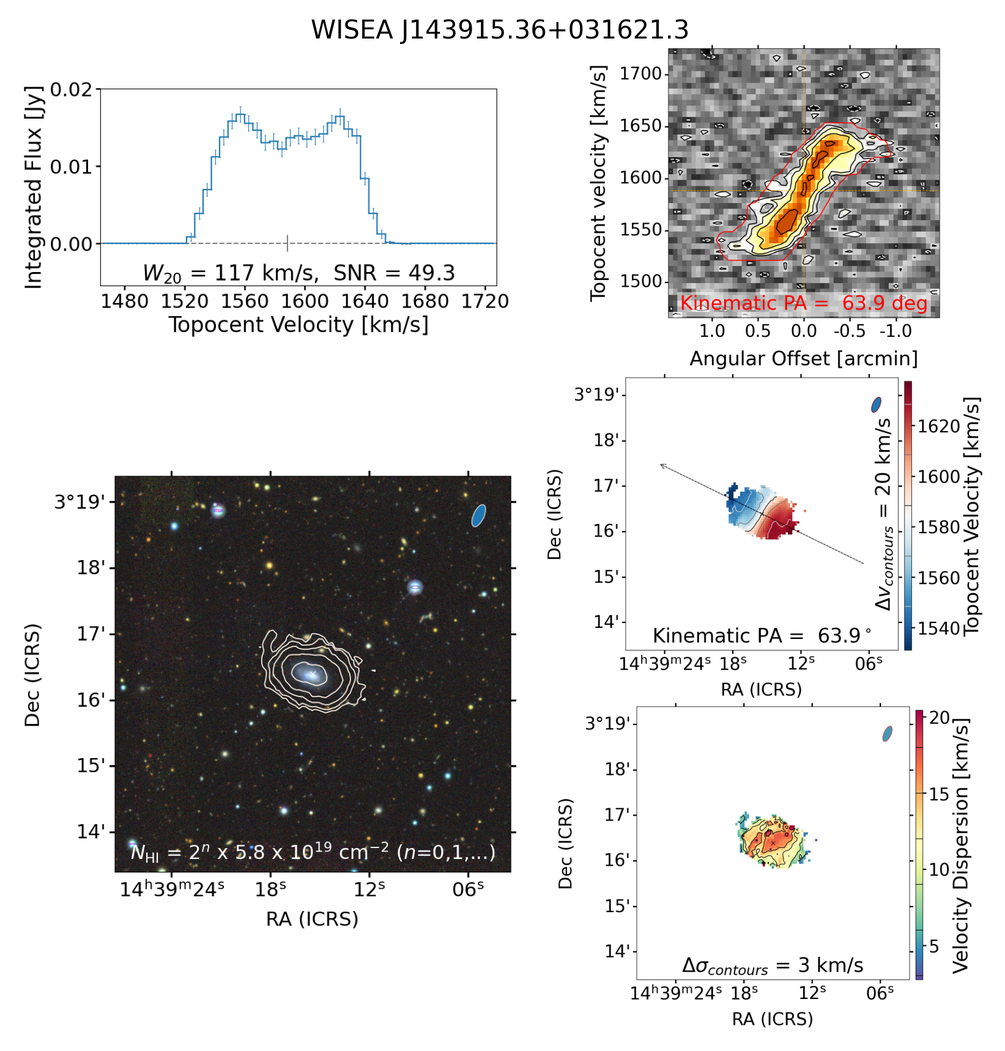}}
    \phantomcaption
\end{figure*}

\clearpage
\begin{figure*}
    \ContinuedFloat
    \centering
    \resizebox{0.86\hsize}{!}{
    \includegraphics{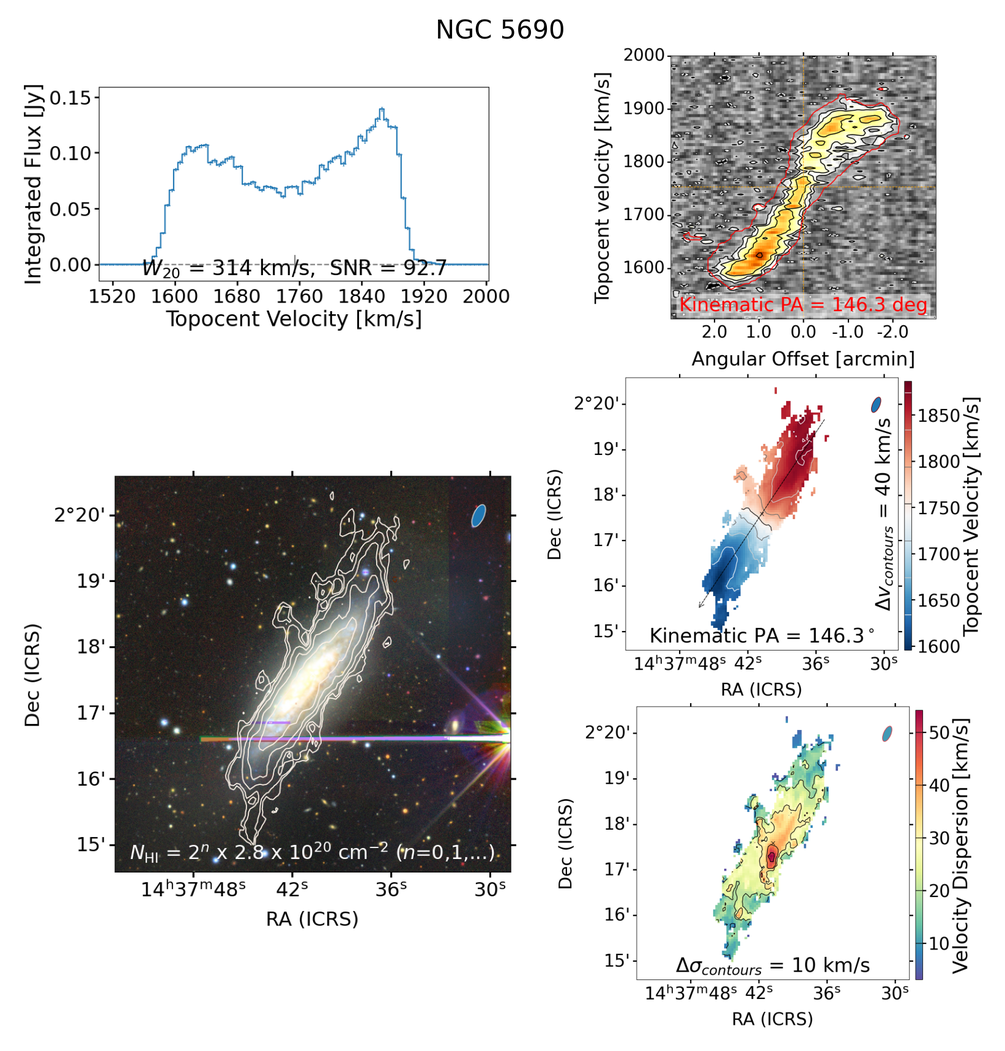}\hspace{0.5cm}
    \includegraphics{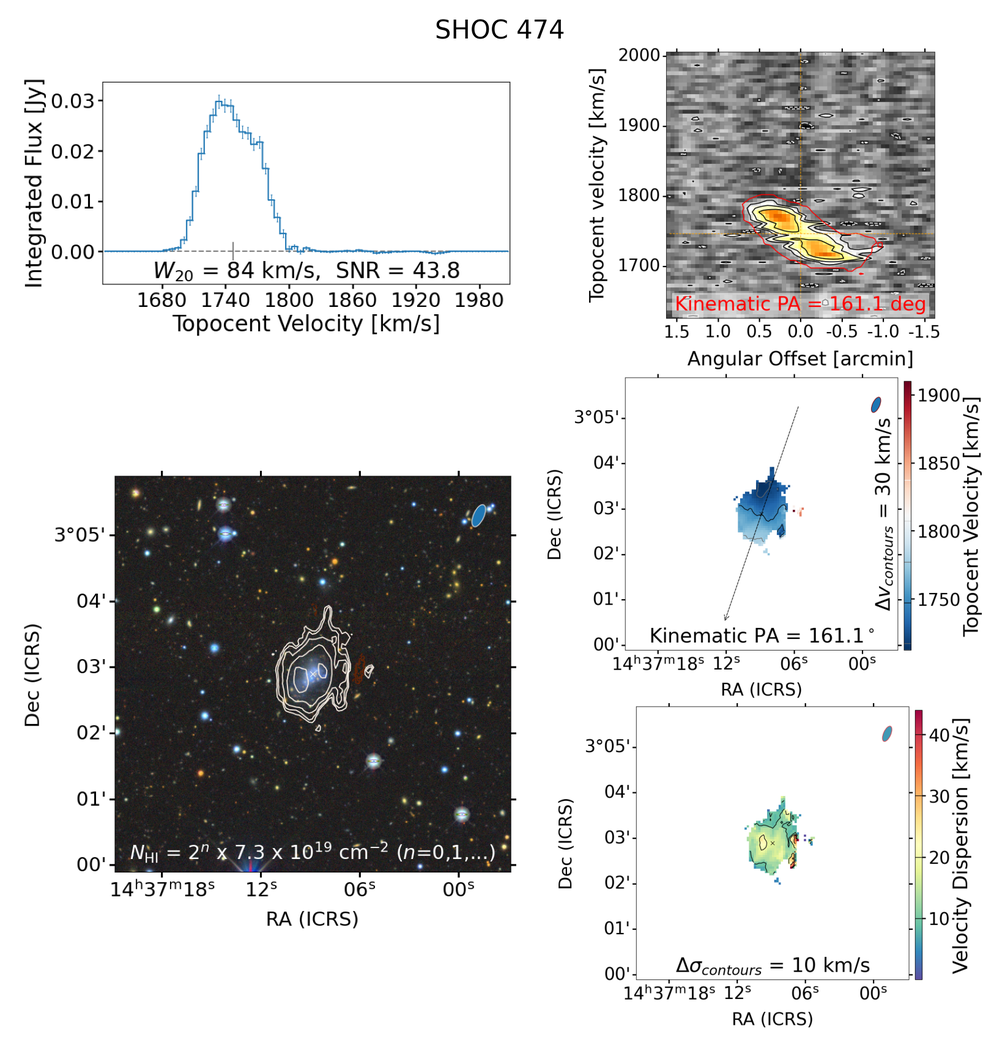}}
    \vspace{0.3cm}
    \resizebox{0.86\hsize}{!}{
    \includegraphics{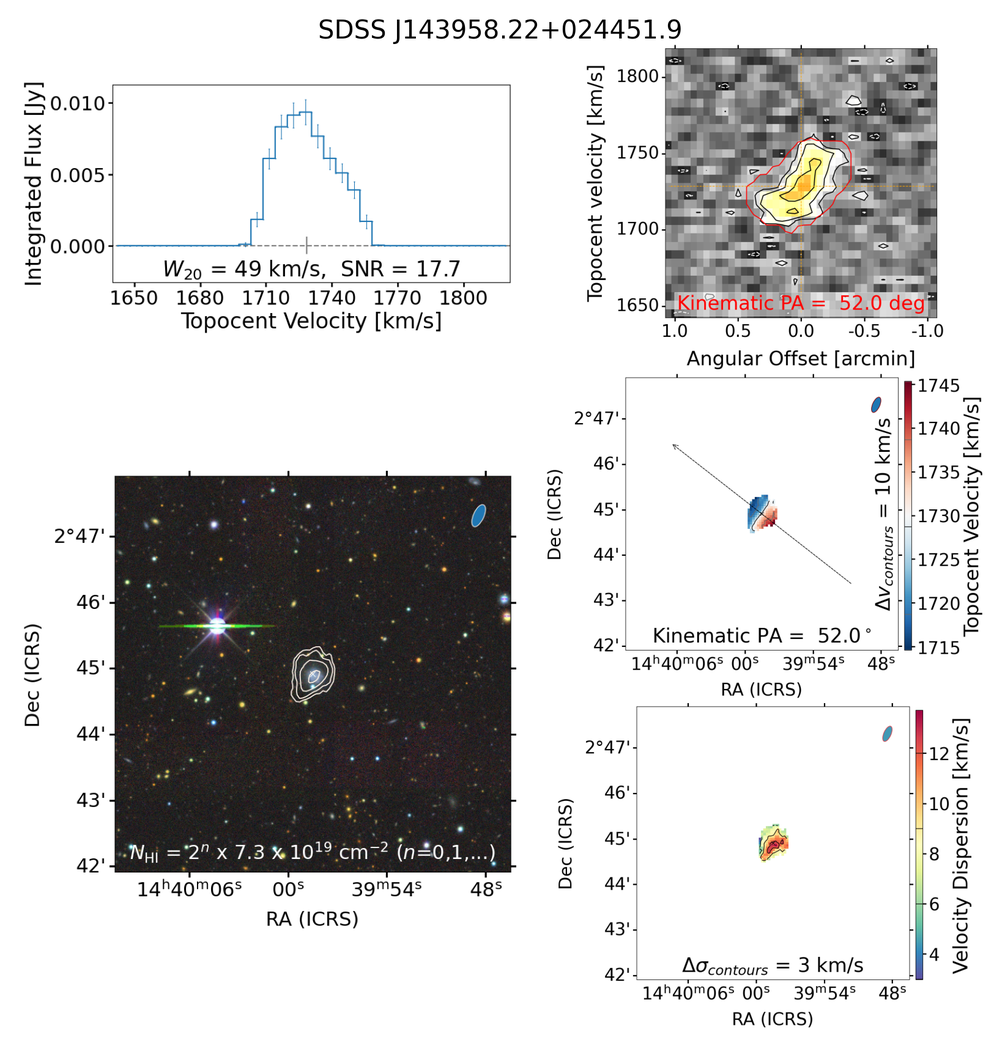}\hspace{0.5cm}
    \includegraphics{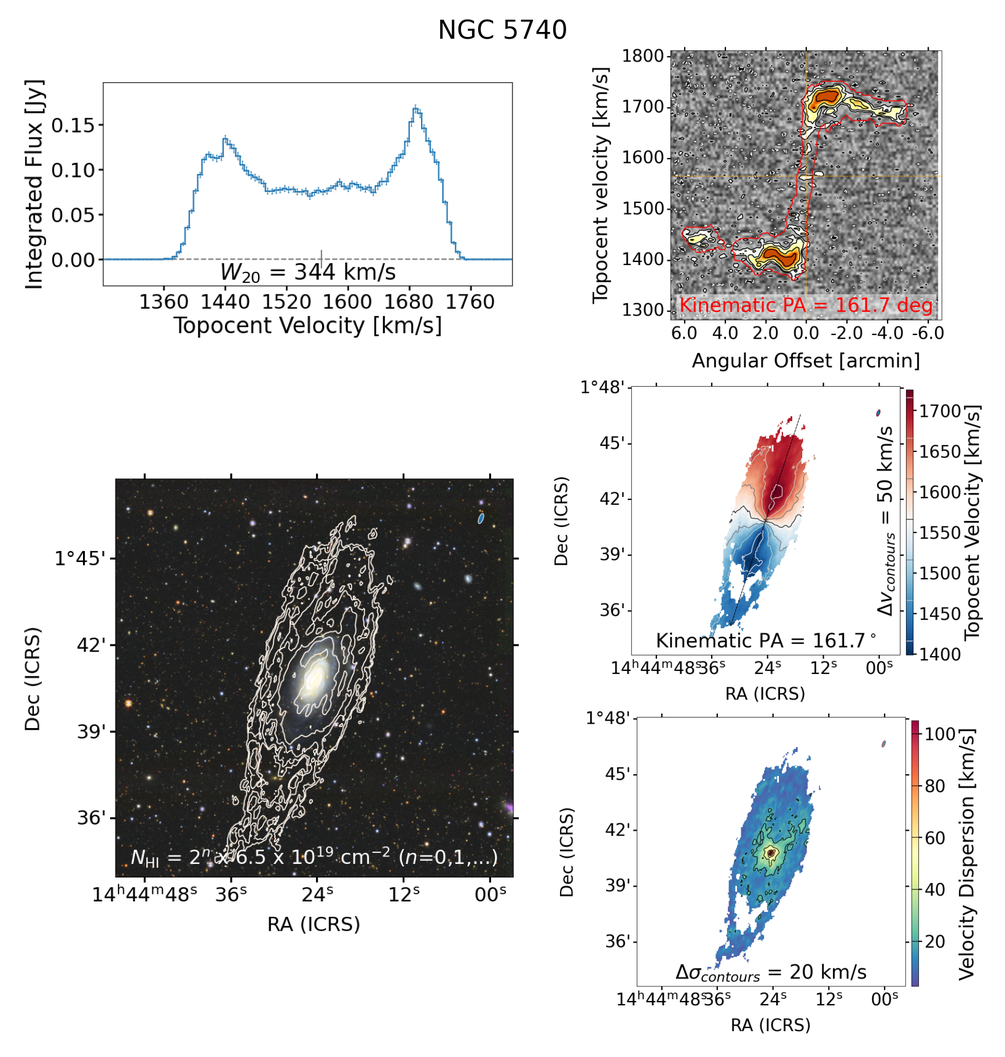}}
    \vspace{0.3cm}
    \resizebox{0.86\hsize}{!}{
    \includegraphics{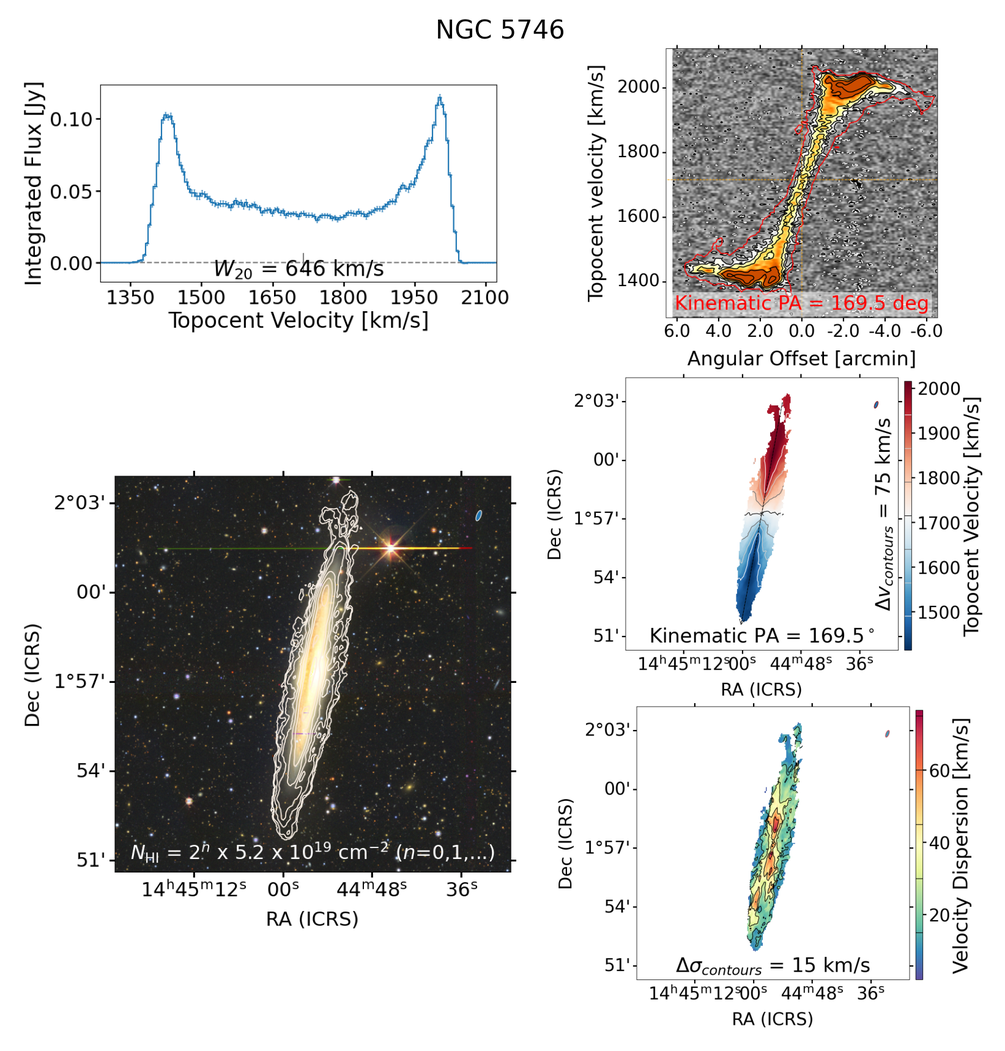}\hspace{0.5cm}
    \includegraphics{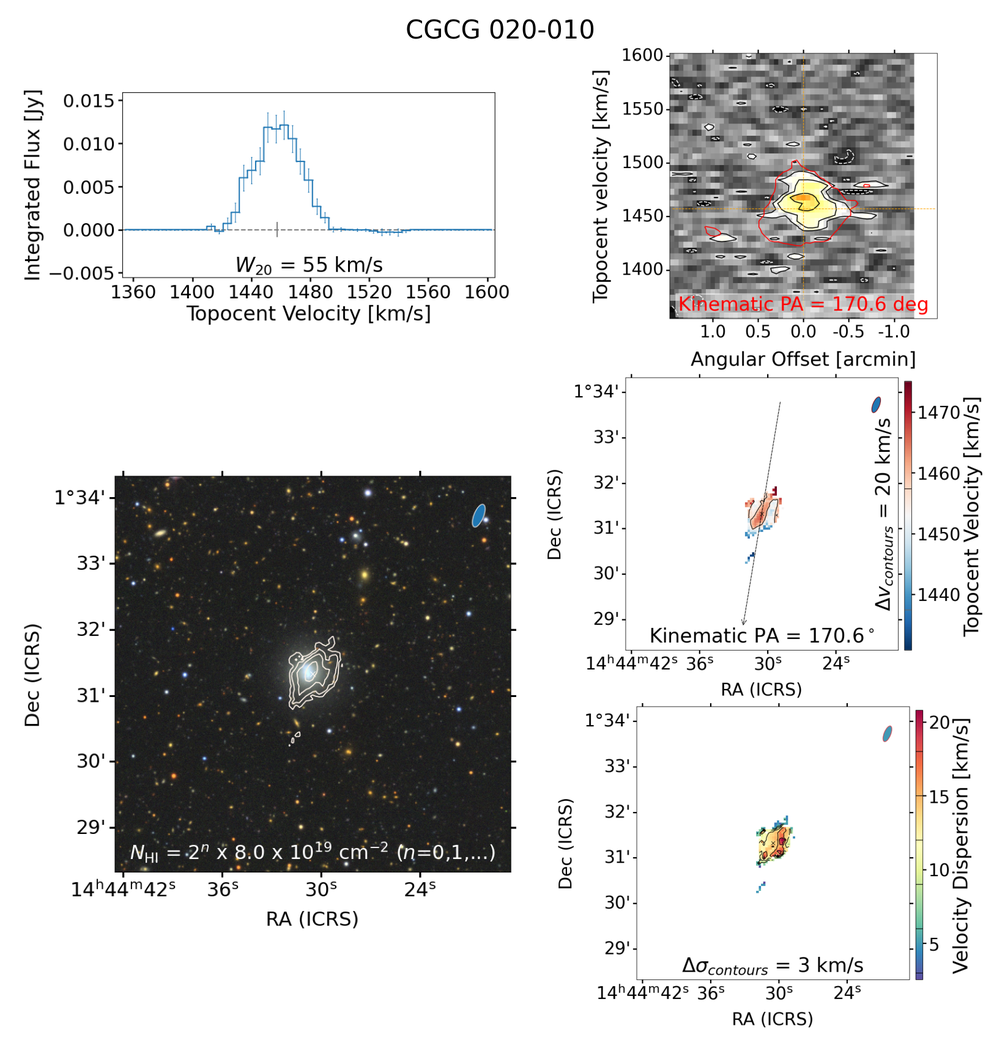}}
    \phantomcaption
\end{figure*}

\clearpage
\begin{figure*}
    \ContinuedFloat
    \centering
    \resizebox{0.86\hsize}{!}{
    \includegraphics{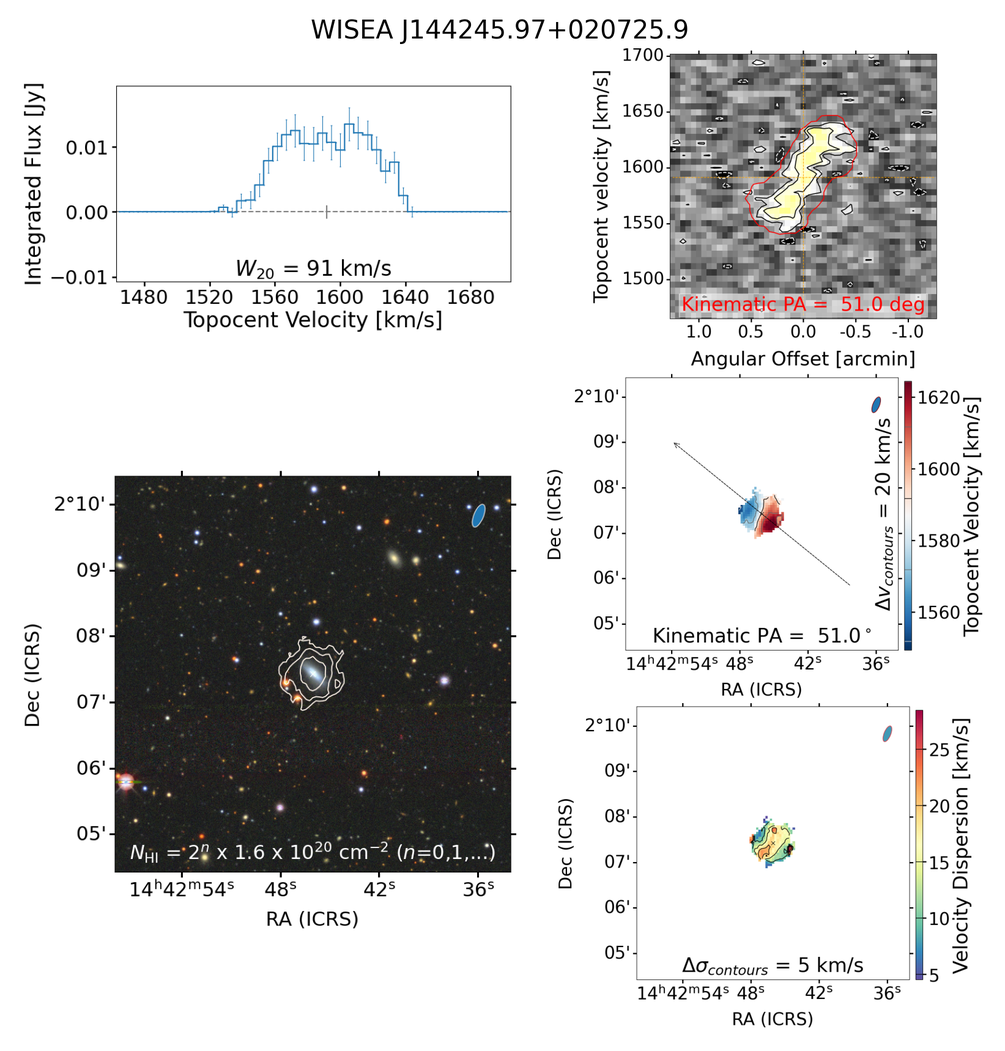}\hspace{0.5cm}
    \includegraphics{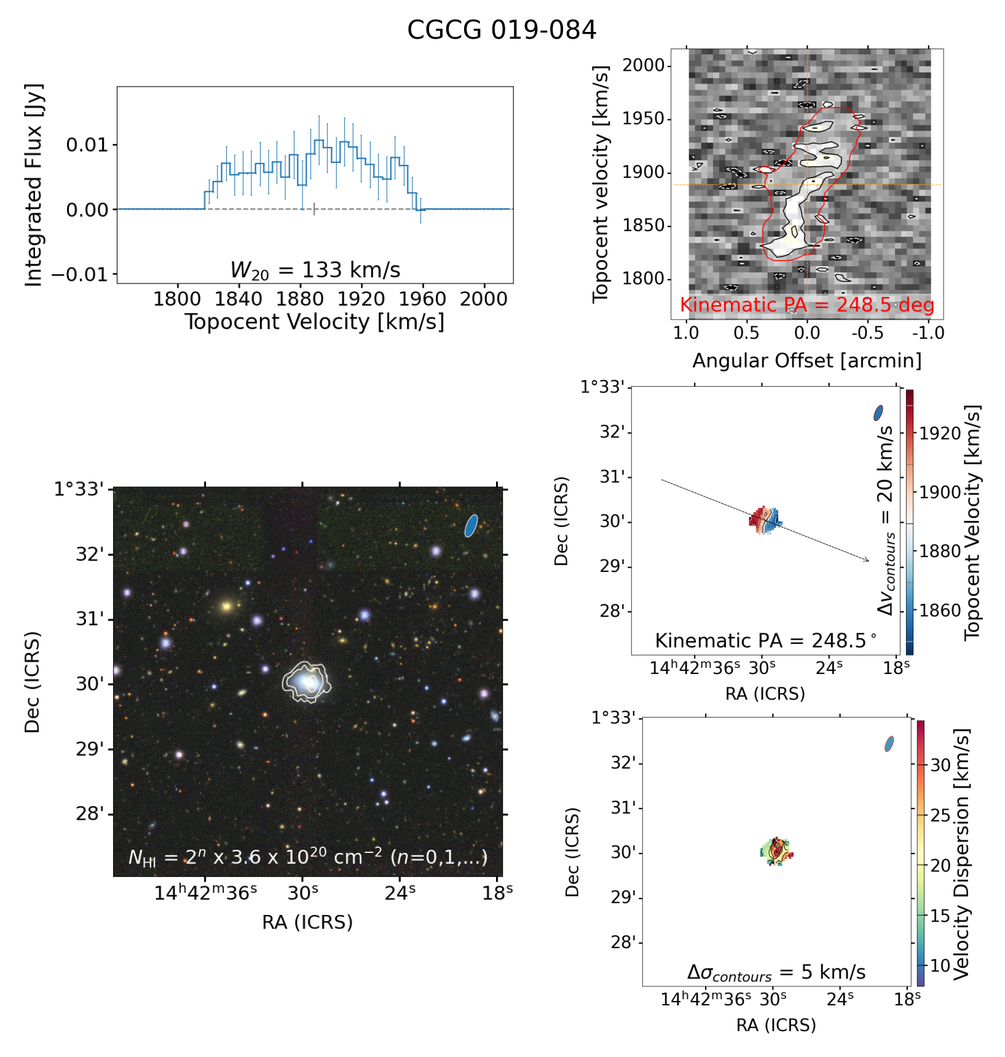}}
    \vspace{0.3cm}
    \resizebox{0.86\hsize}{!}{
    \includegraphics{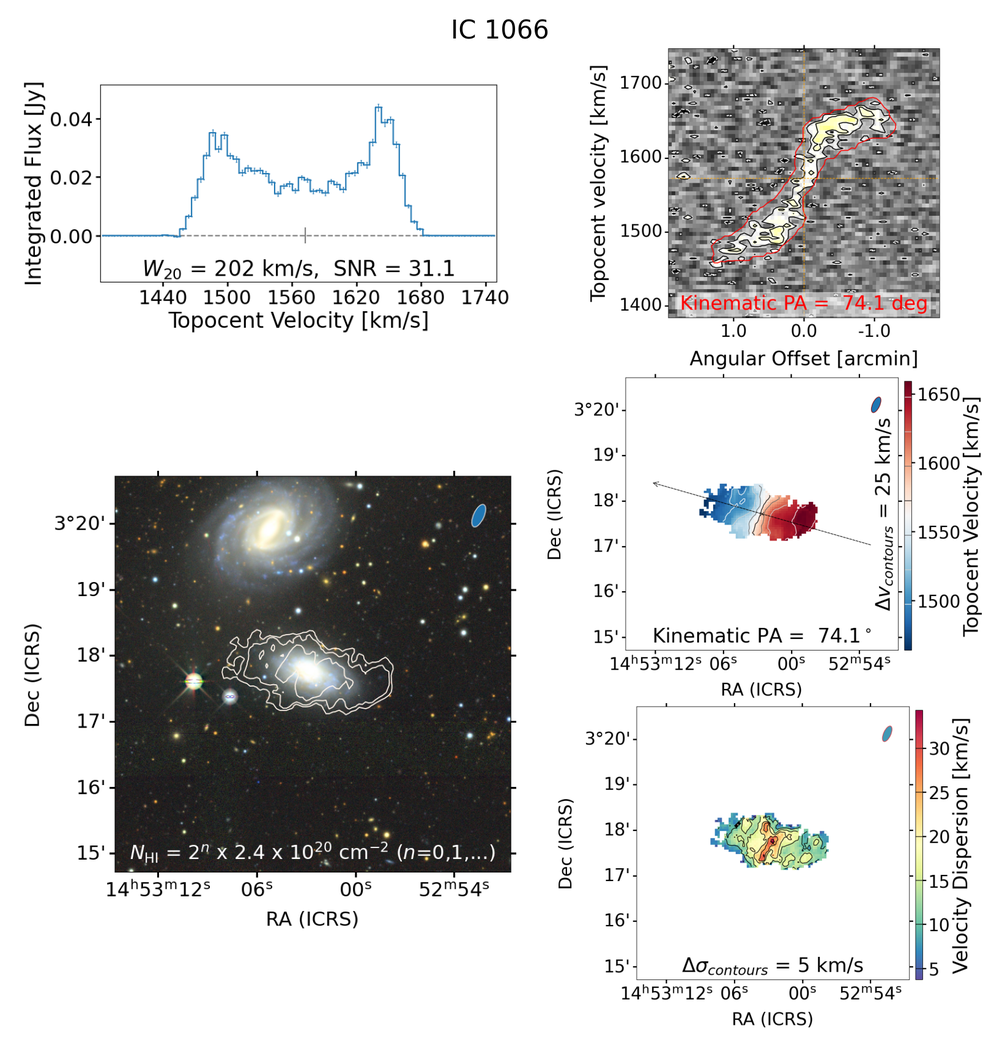}\hspace{0.5cm}
    \includegraphics{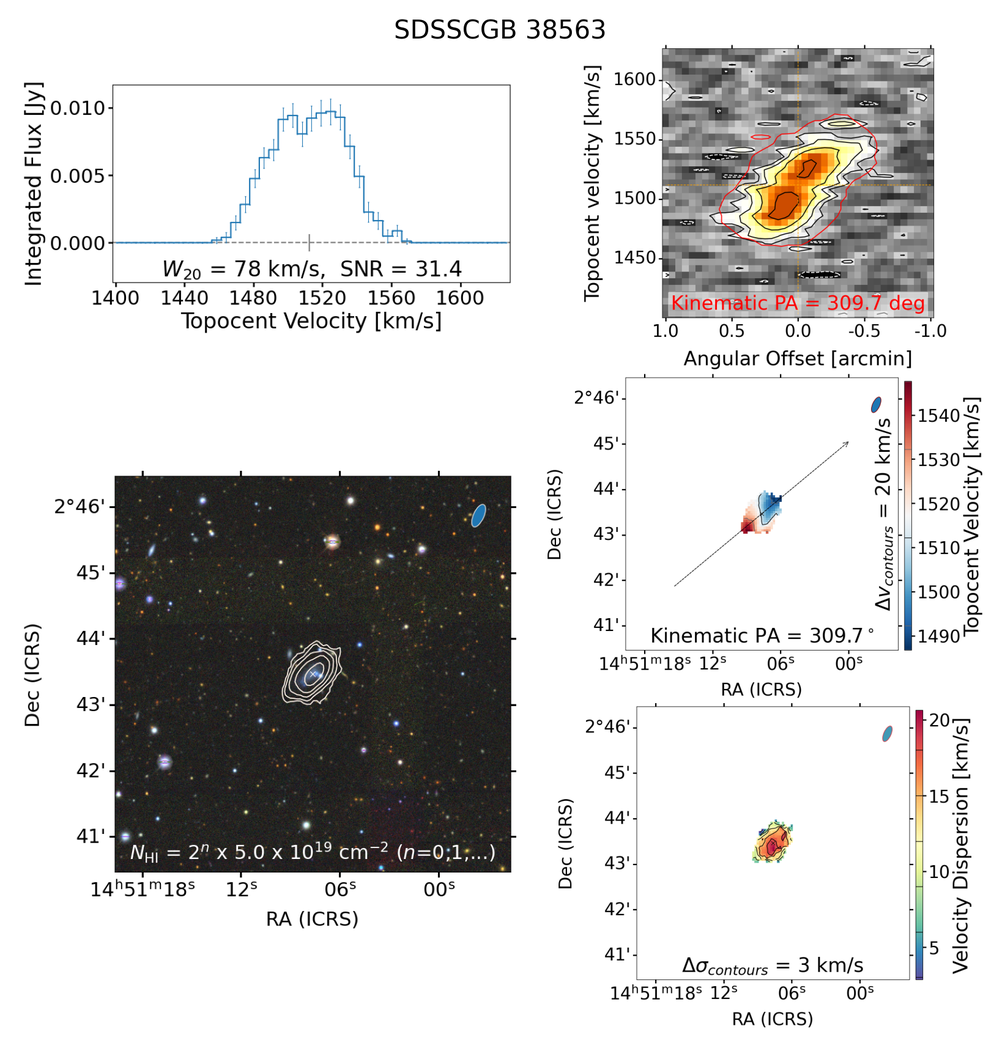}}
    \vspace{0.3cm}
    \resizebox{0.86\hsize}{!}{
    \includegraphics{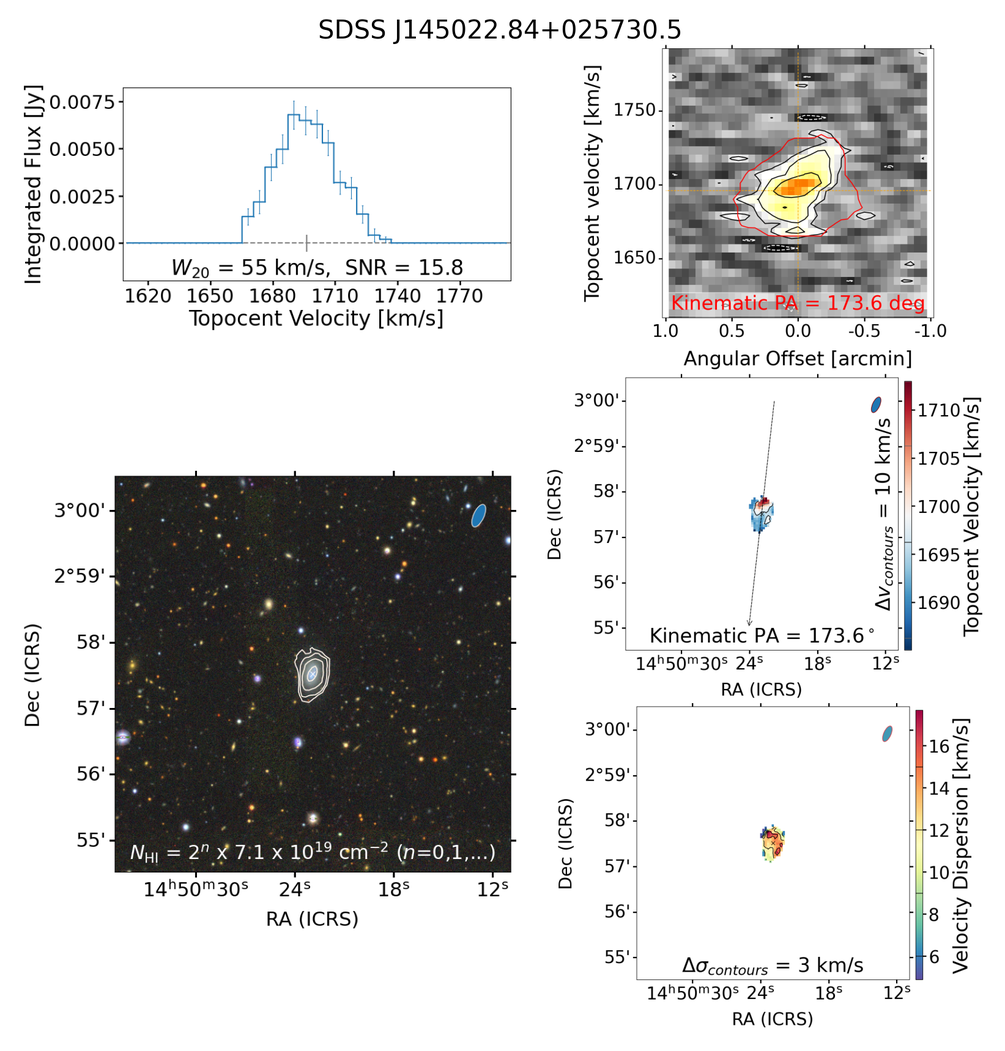}\hspace{0.5cm}
    \includegraphics{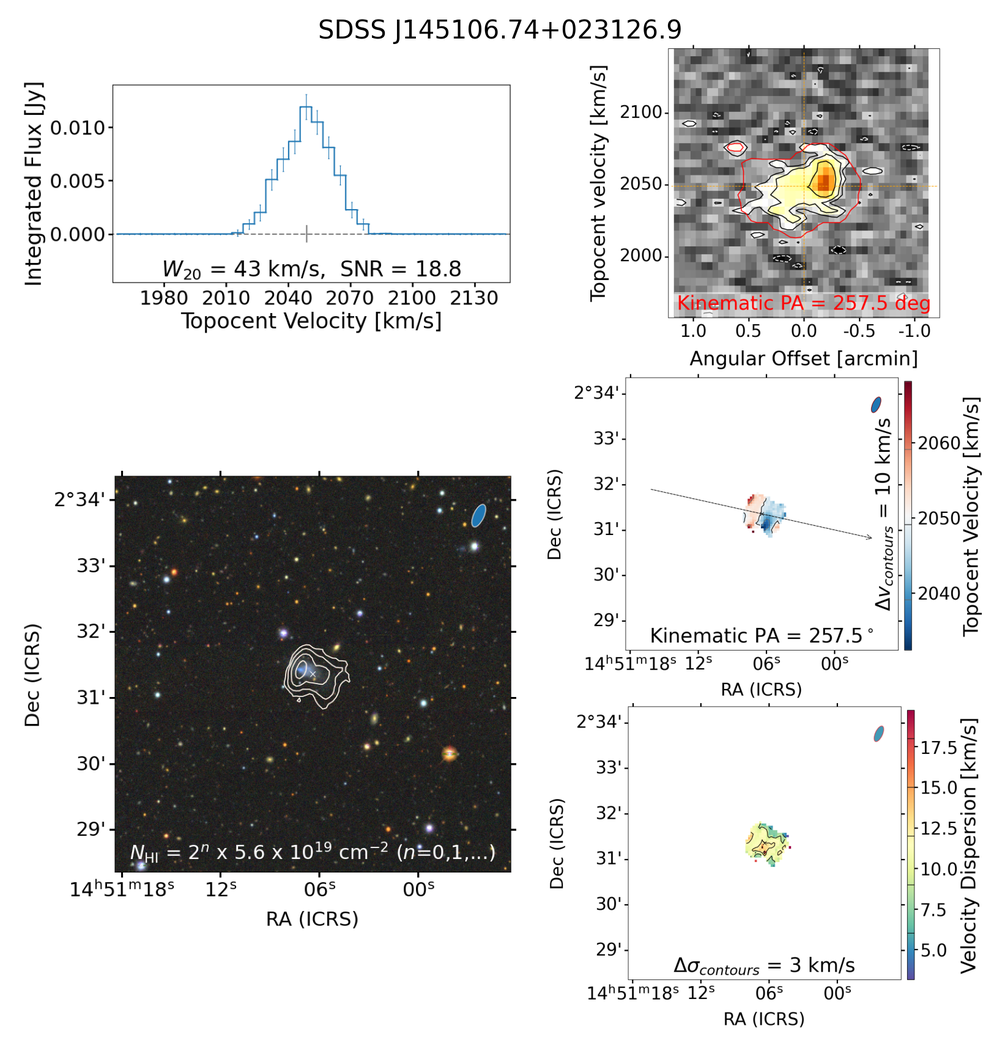}}
    \phantomcaption
\end{figure*}

\clearpage
\begin{figure*}
    \ContinuedFloat
    \centering
    \resizebox{0.86\hsize}{!}{
    \includegraphics{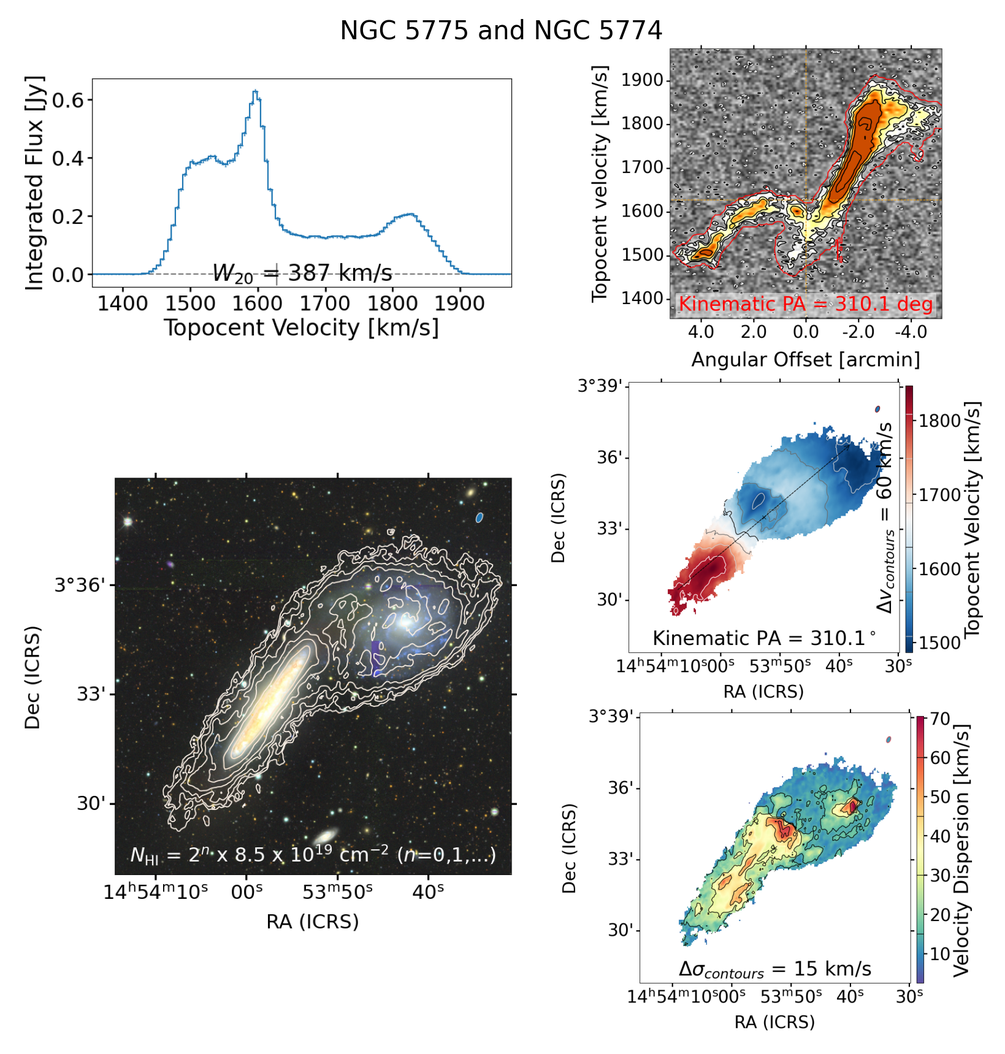}\hspace{0.5cm}
    \includegraphics{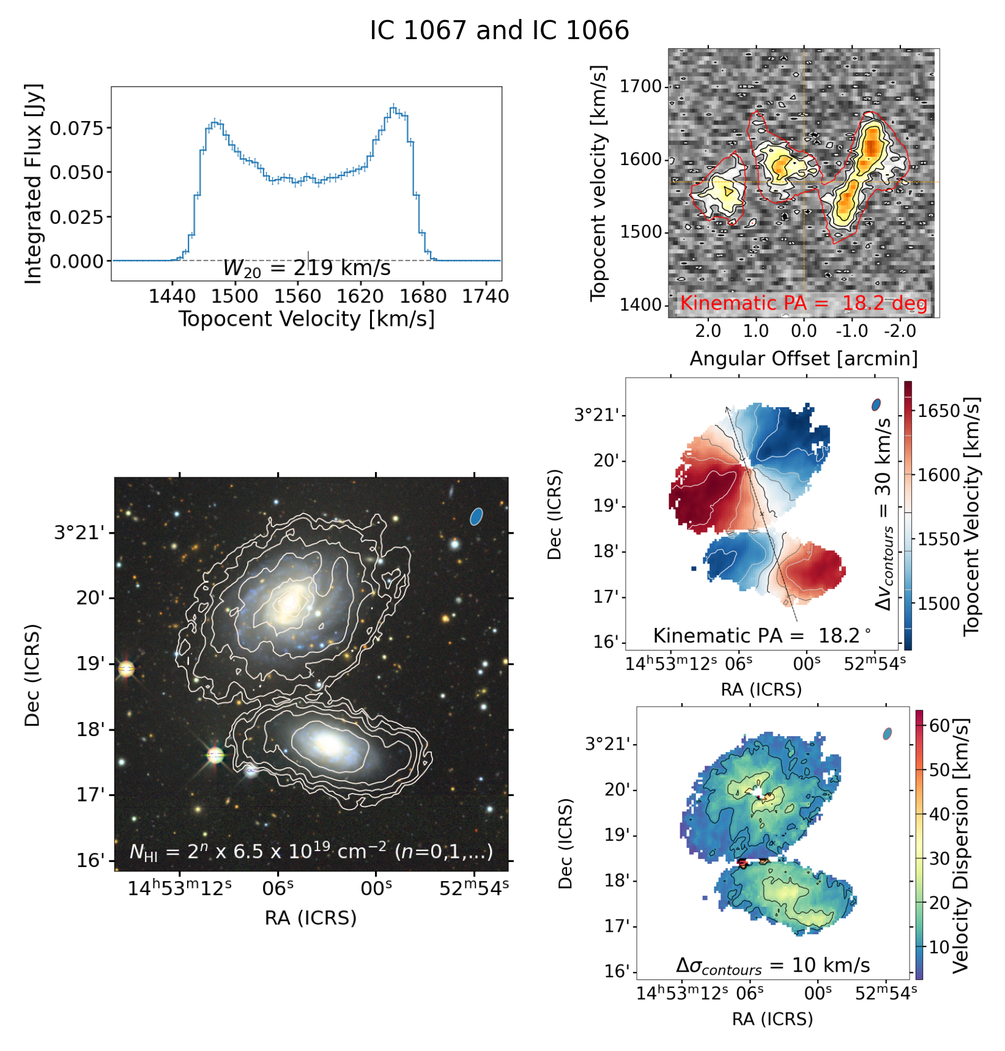}}
    \vspace{0.3cm}
    \resizebox{0.86\hsize}{!}{
    \includegraphics{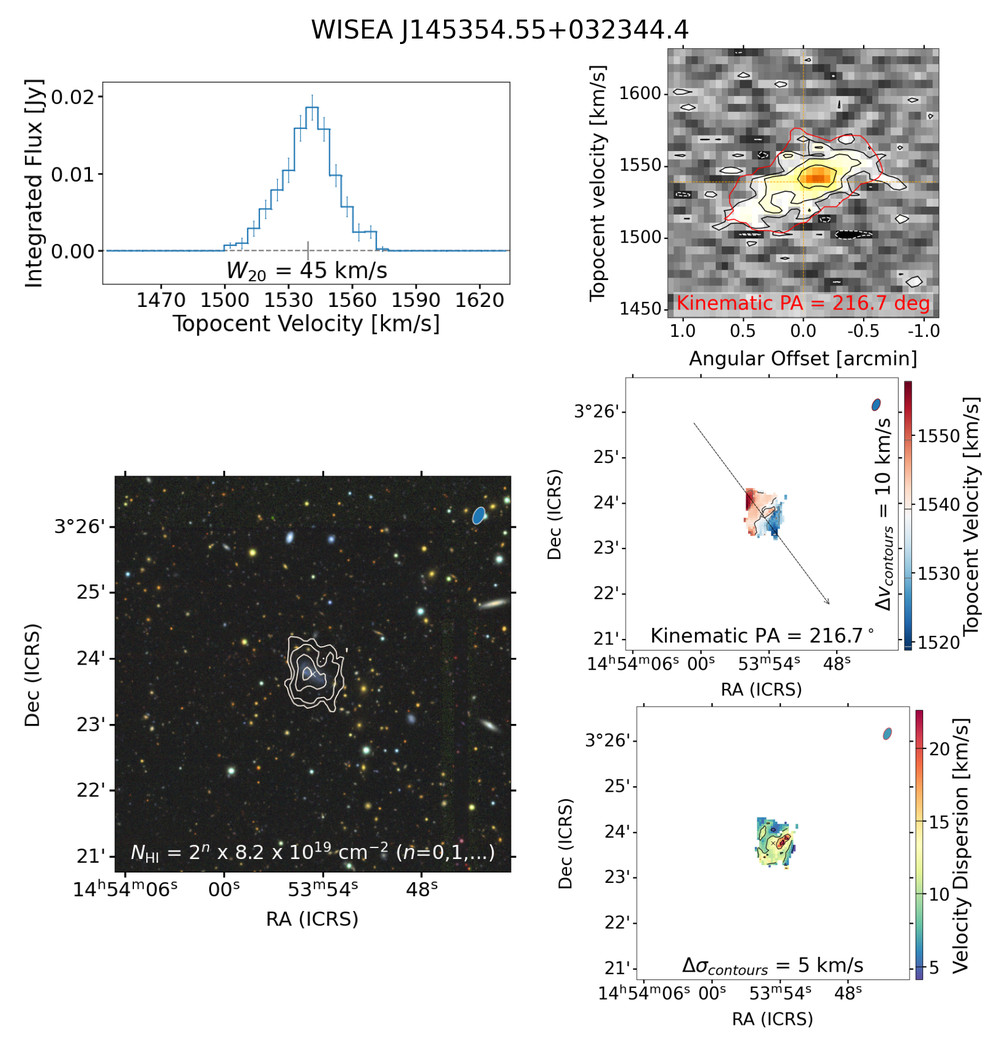}\hspace{0.5cm}
    \includegraphics{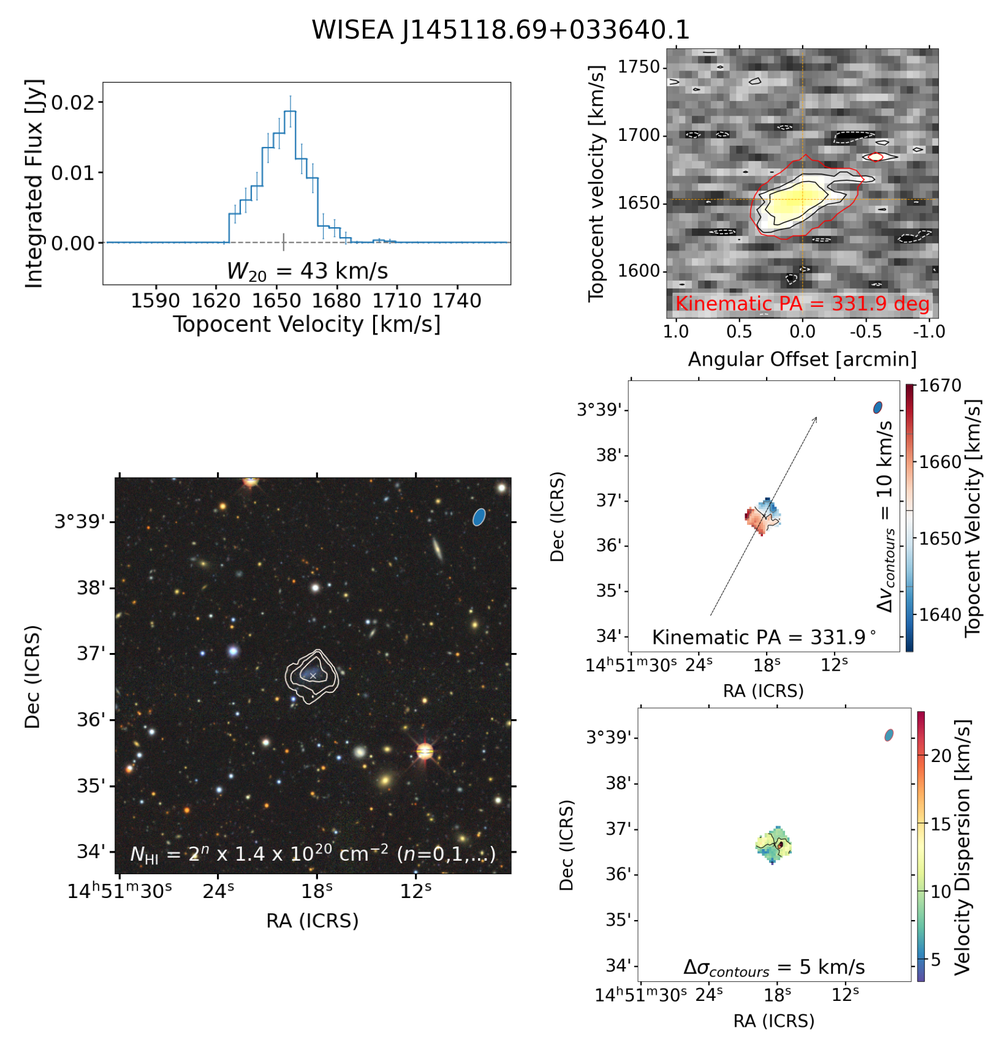}}
    \vspace{0.3cm}
    \resizebox{0.86\hsize}{!}{
    \includegraphics{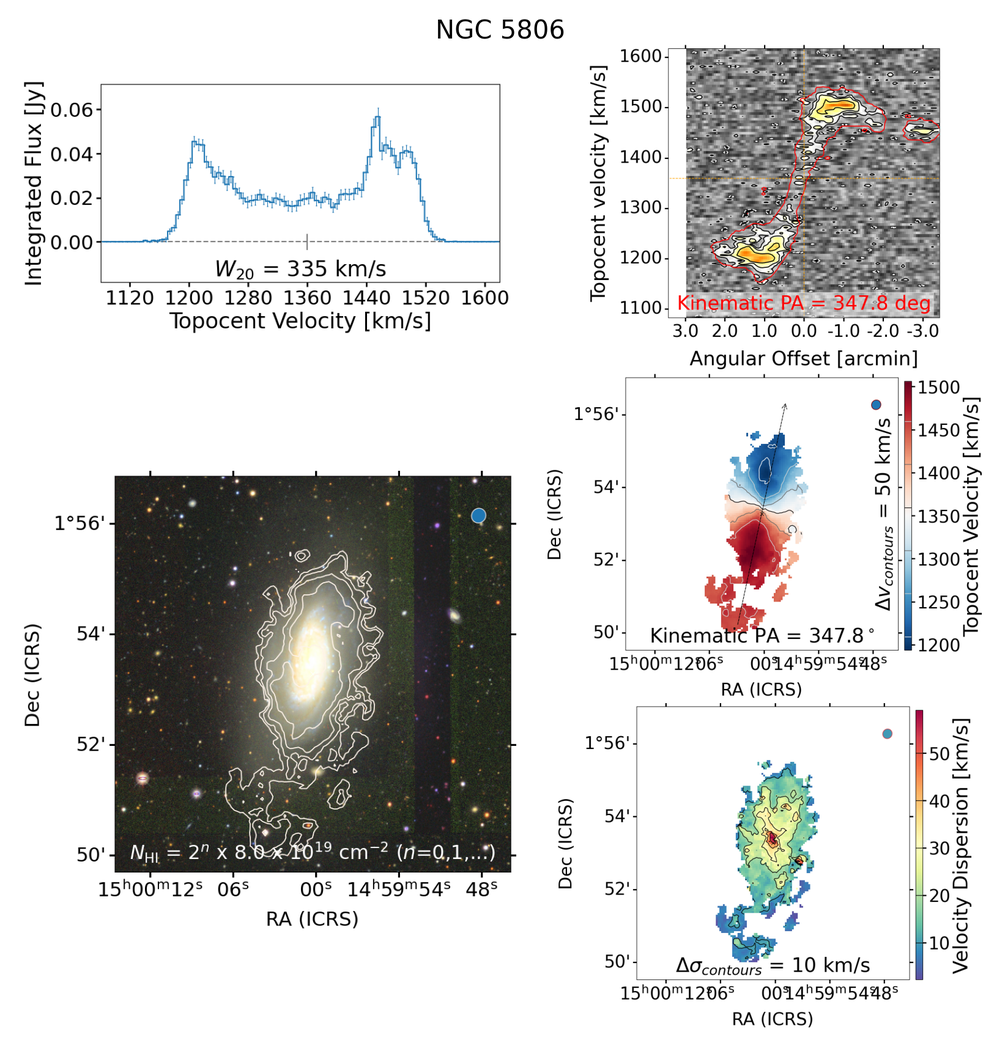}\hspace{0.5cm}
    \includegraphics{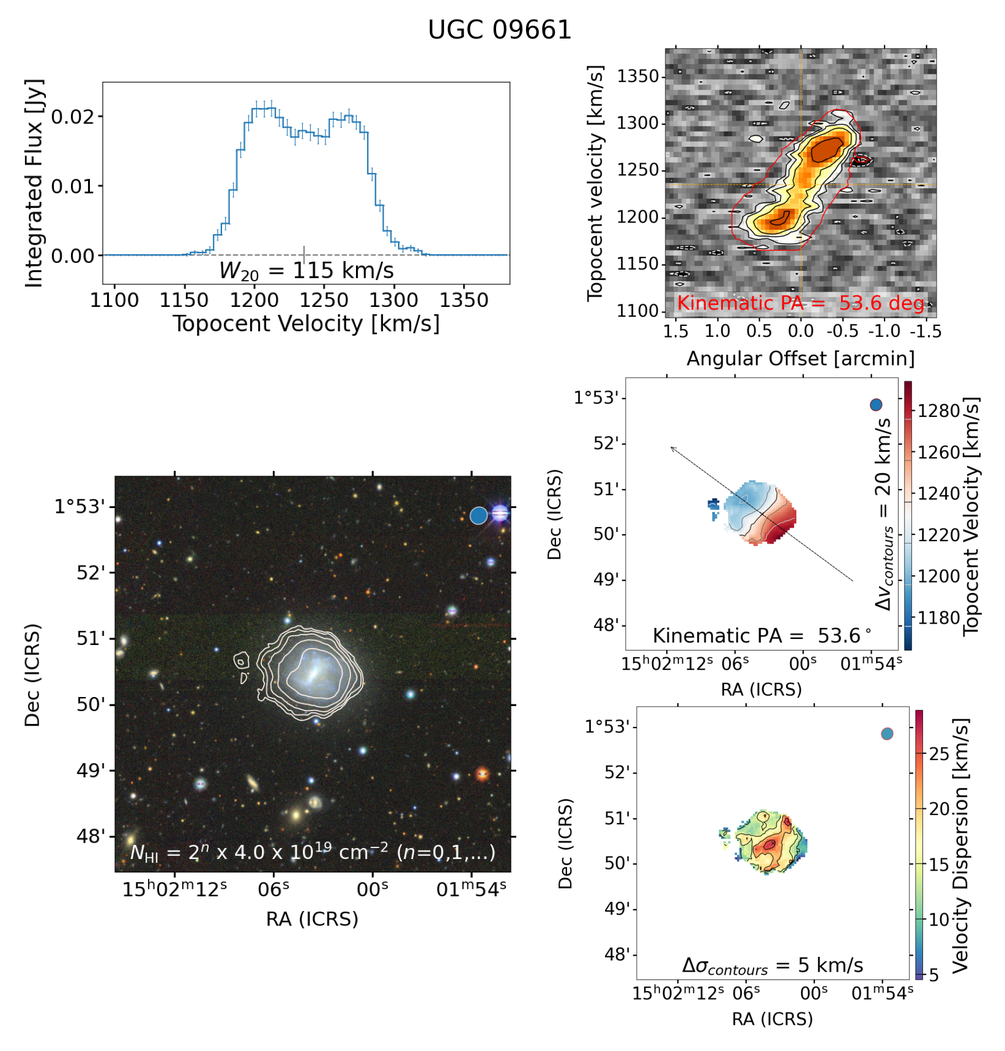}}
    \phantomcaption
\end{figure*}

\clearpage
\begin{figure*}
    \ContinuedFloat
    \centering
    \resizebox{0.86\hsize}{!}{
    \includegraphics{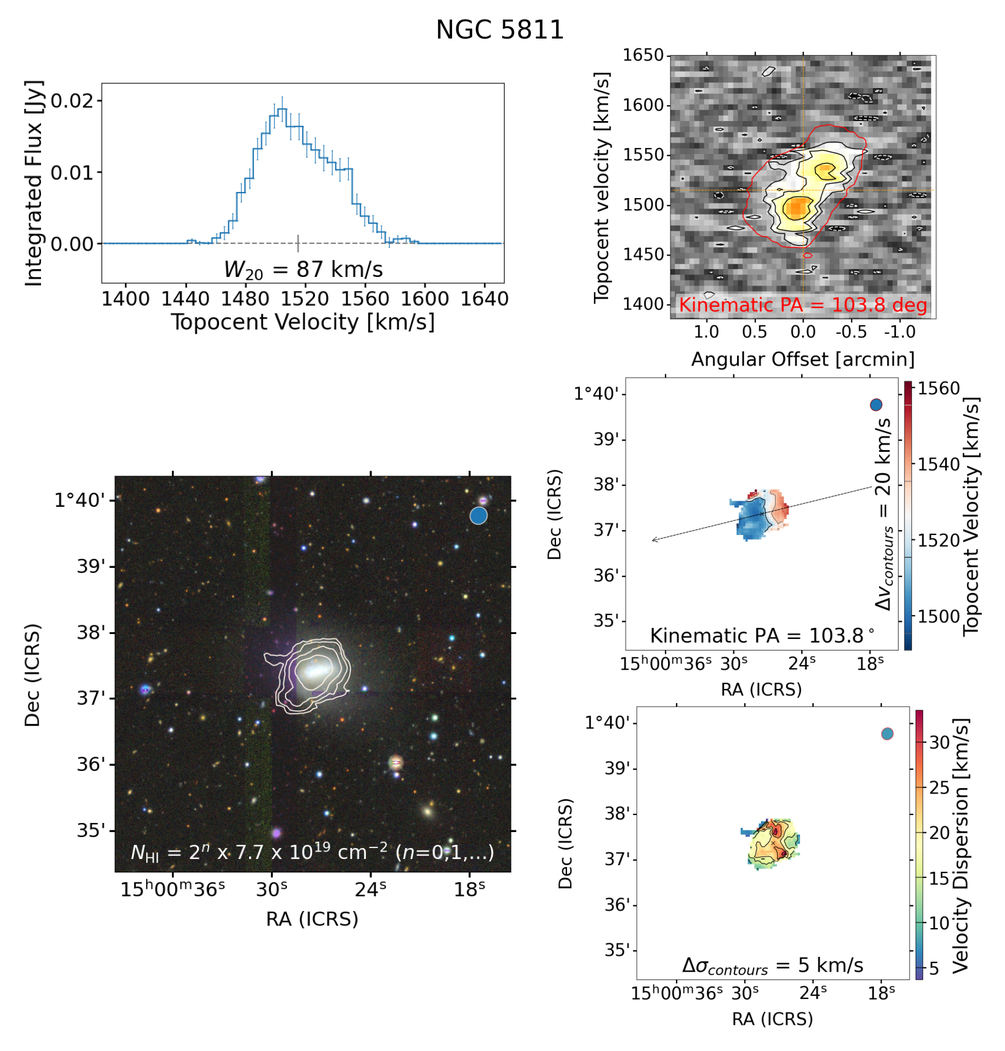}\hspace{0.5cm}
    \includegraphics{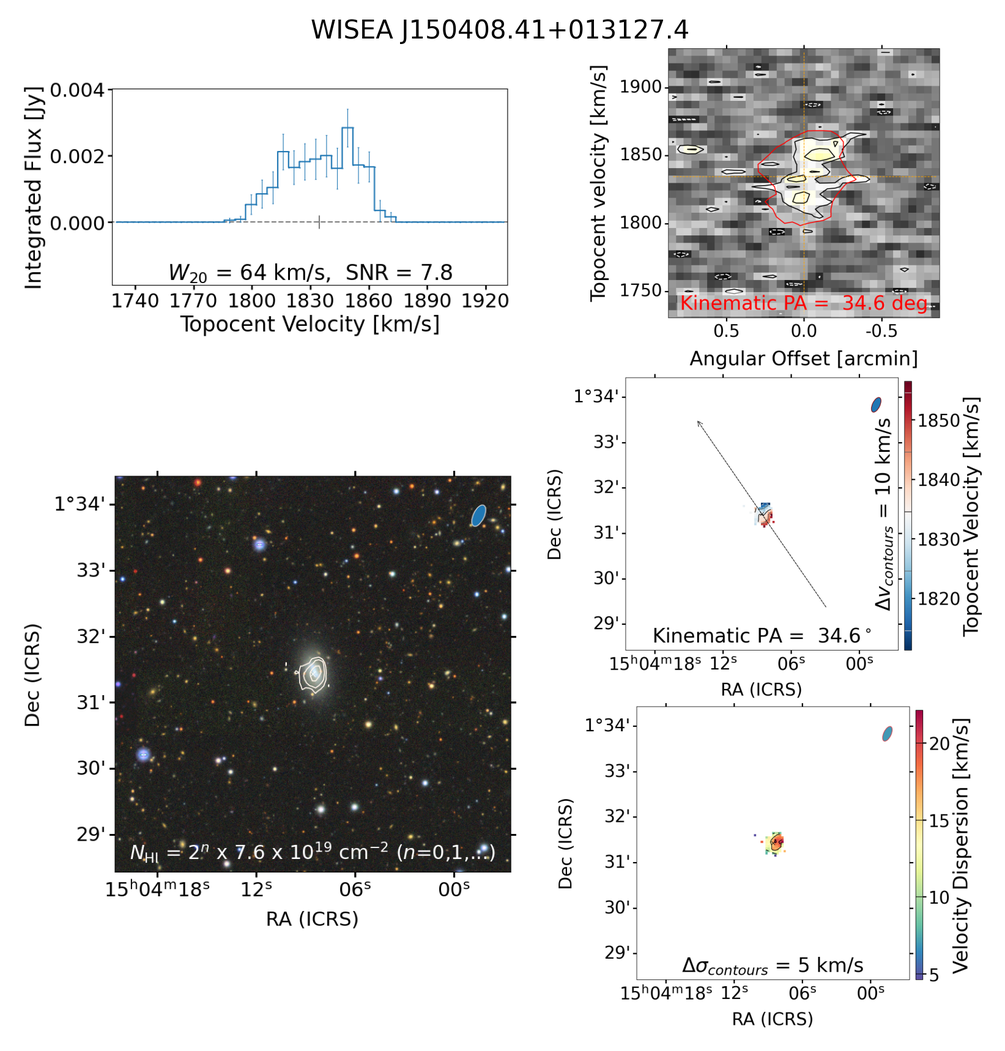}}
    \vspace{0.3cm}
    \resizebox{0.86\hsize}{!}{
    \includegraphics{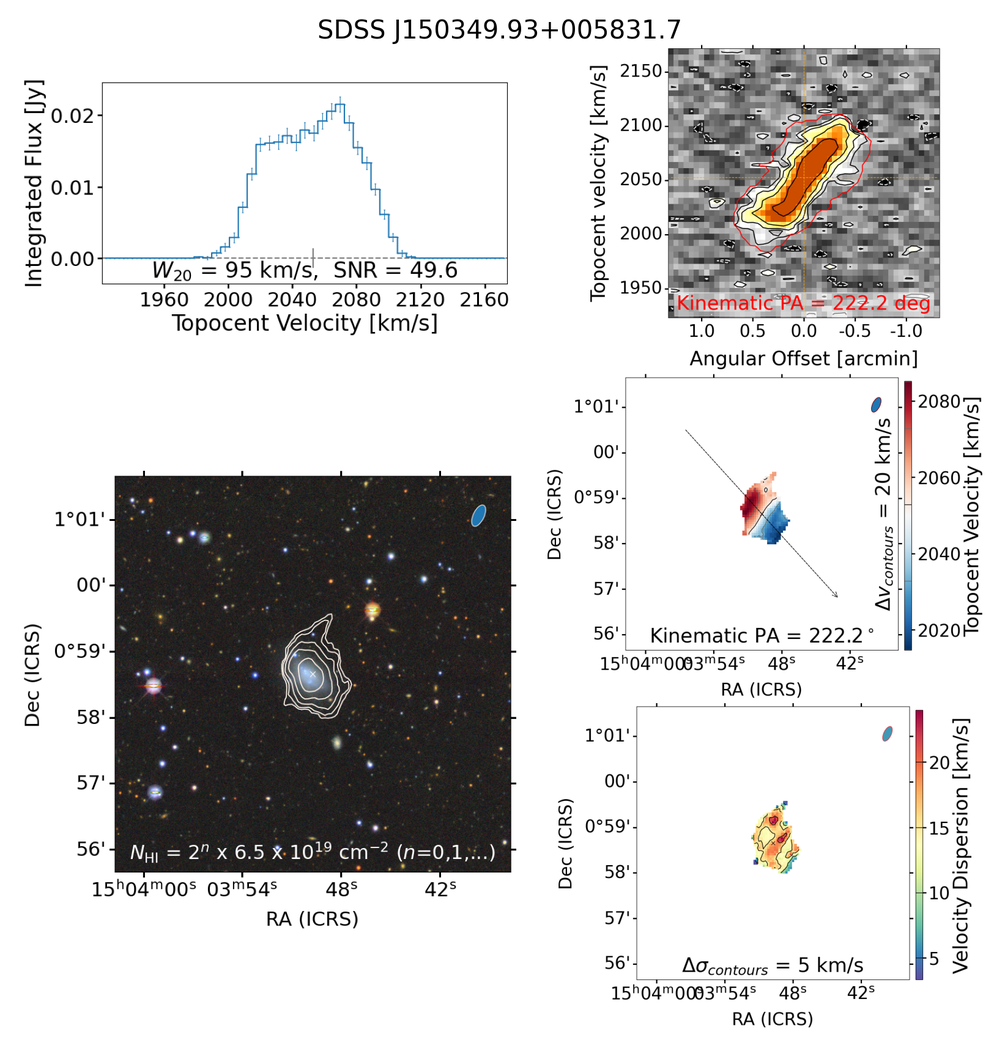}\hspace{0.5cm}
    \includegraphics{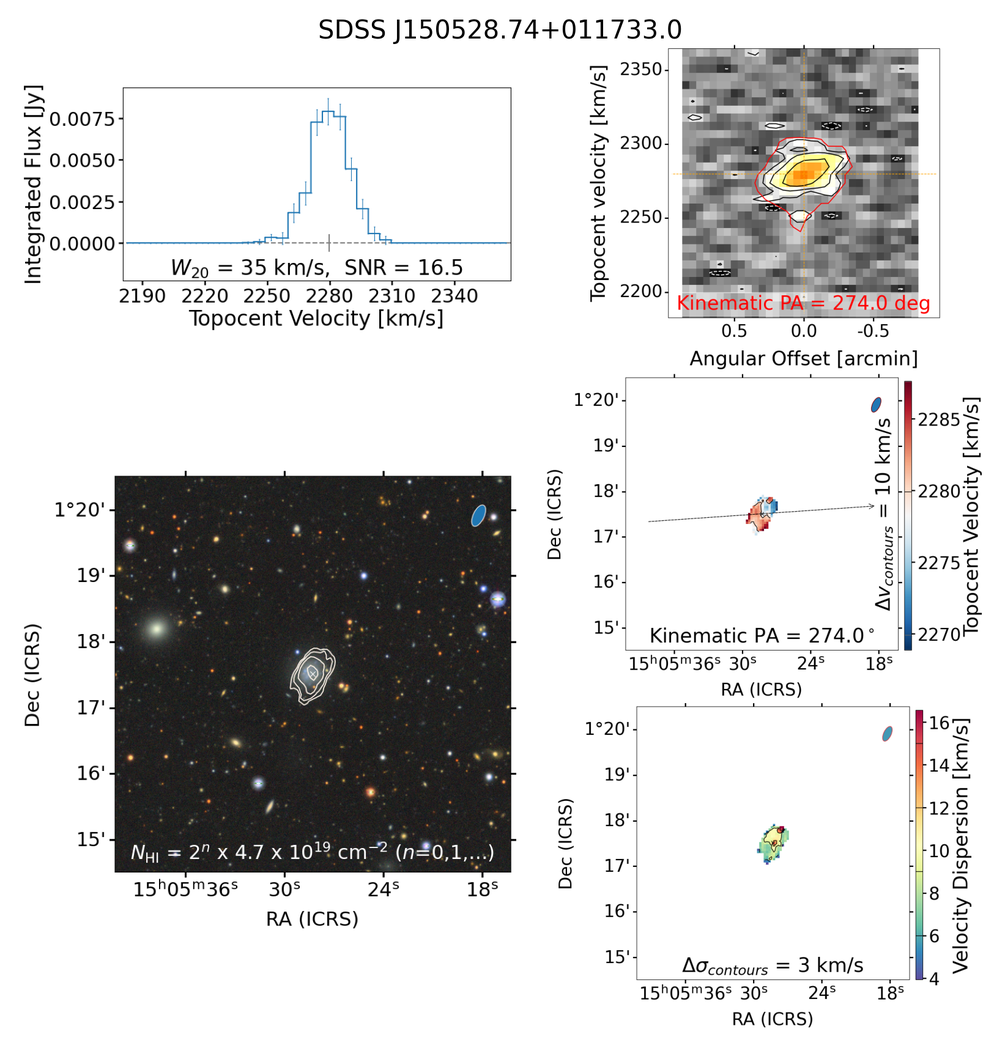}}
    \vspace{0.3cm}
    \resizebox{0.86\hsize}{!}{
    \includegraphics{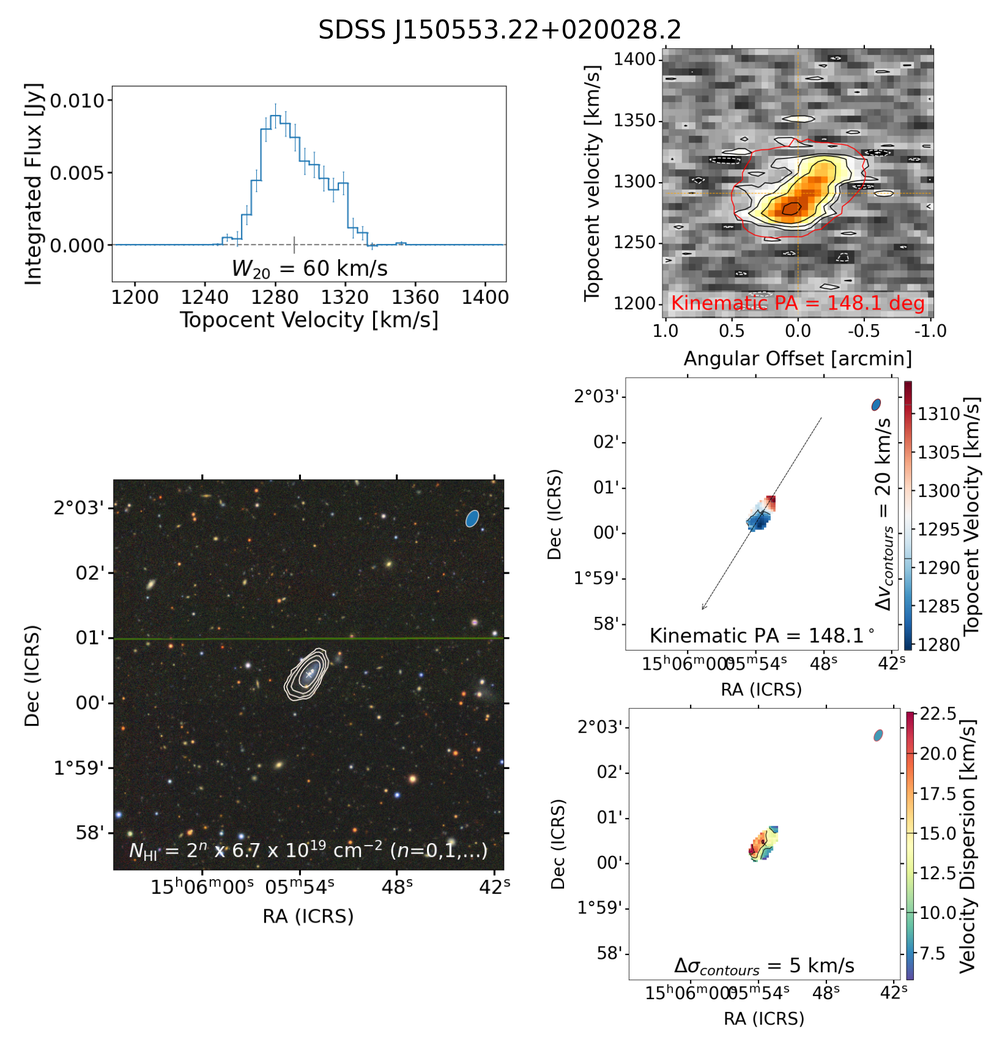}\hspace{0.5cm}
    \includegraphics{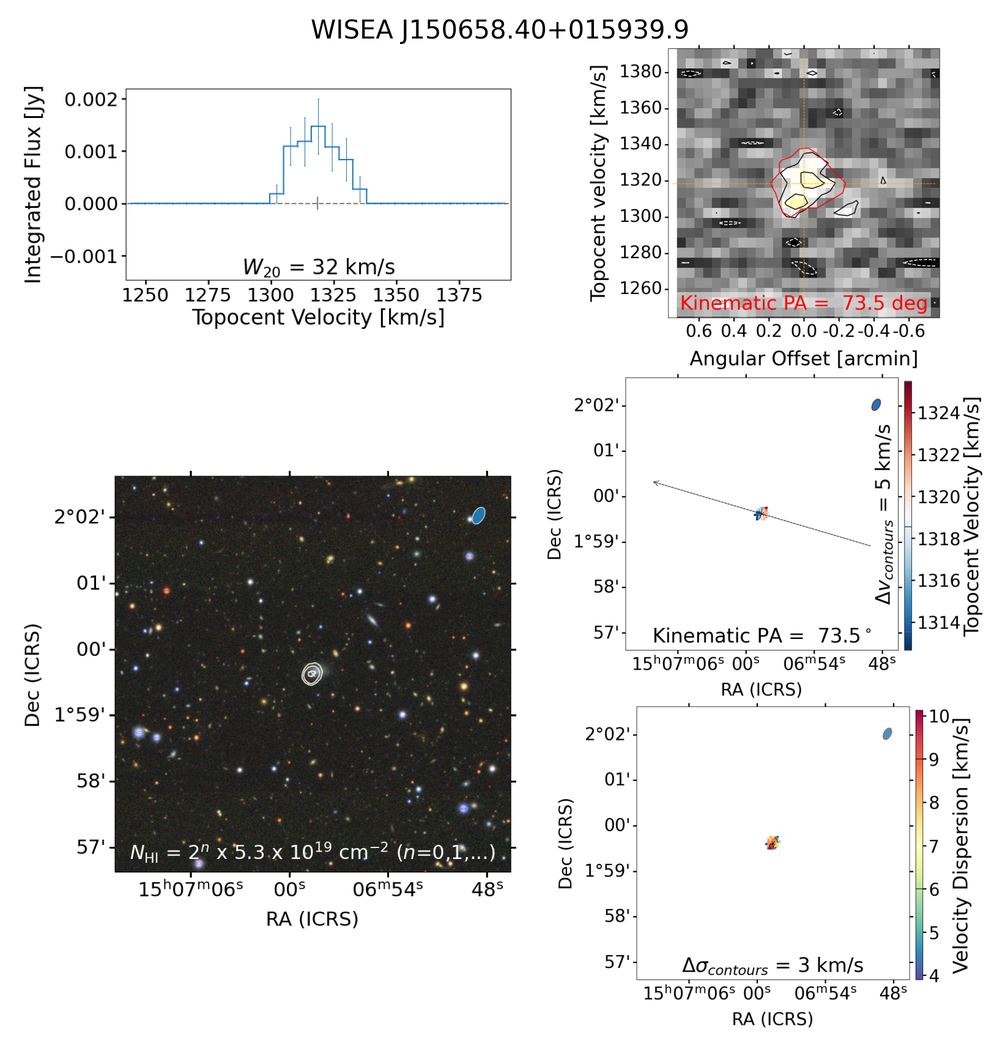}}
    \phantomcaption
\end{figure*}

\clearpage
\begin{figure*}
    \ContinuedFloat
    \centering
    \resizebox{0.86\hsize}{!}{
    \includegraphics{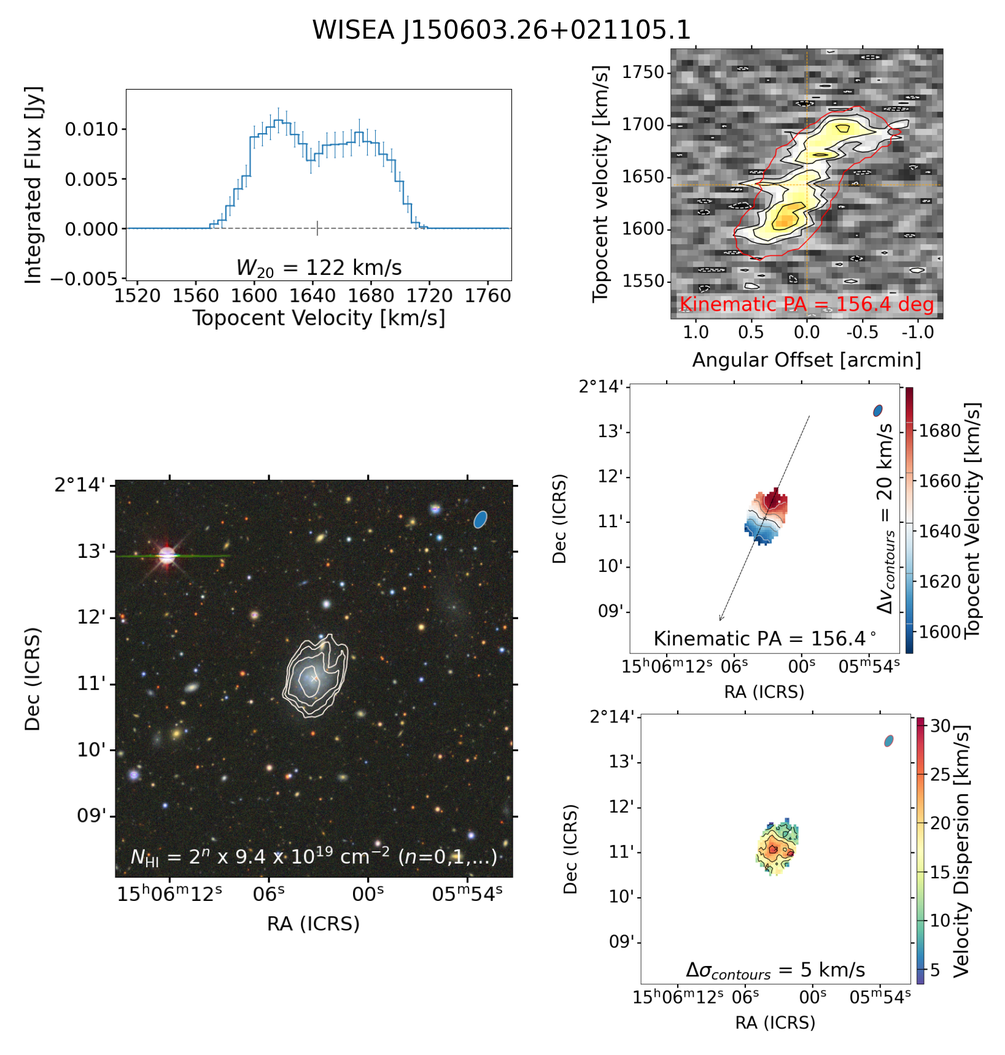}\hspace{0.5cm}
    \includegraphics{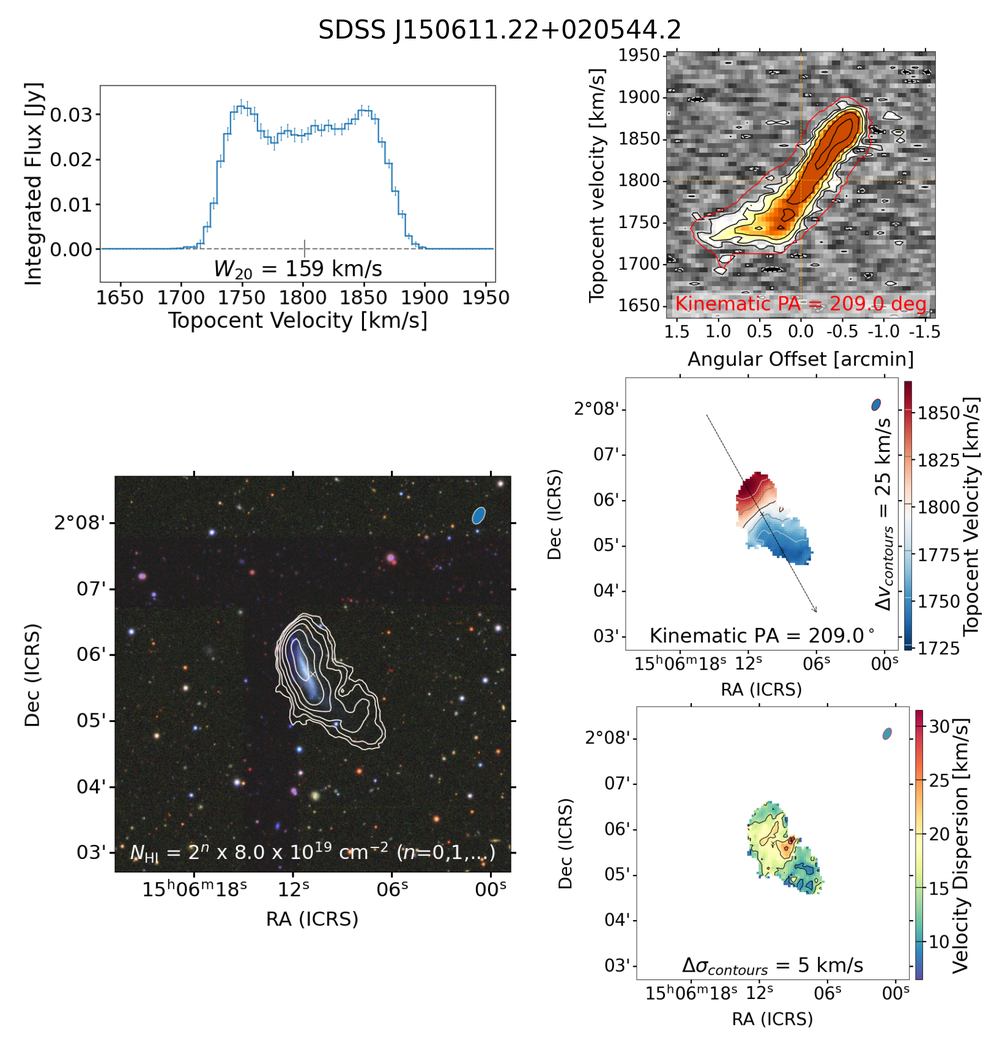}}
    \vspace{0.3cm}
    \resizebox{0.86\hsize}{!}{
    \includegraphics{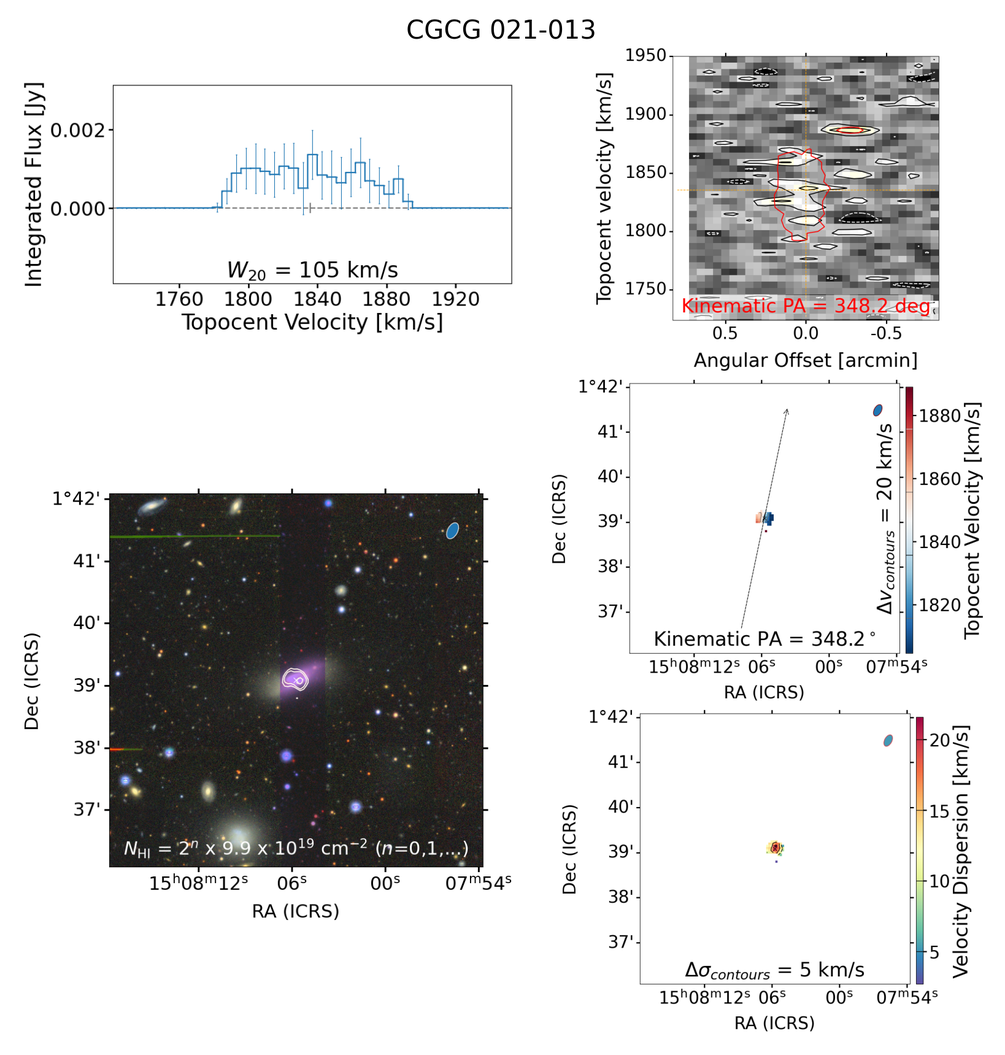}\hspace{0.5cm}
    \includegraphics{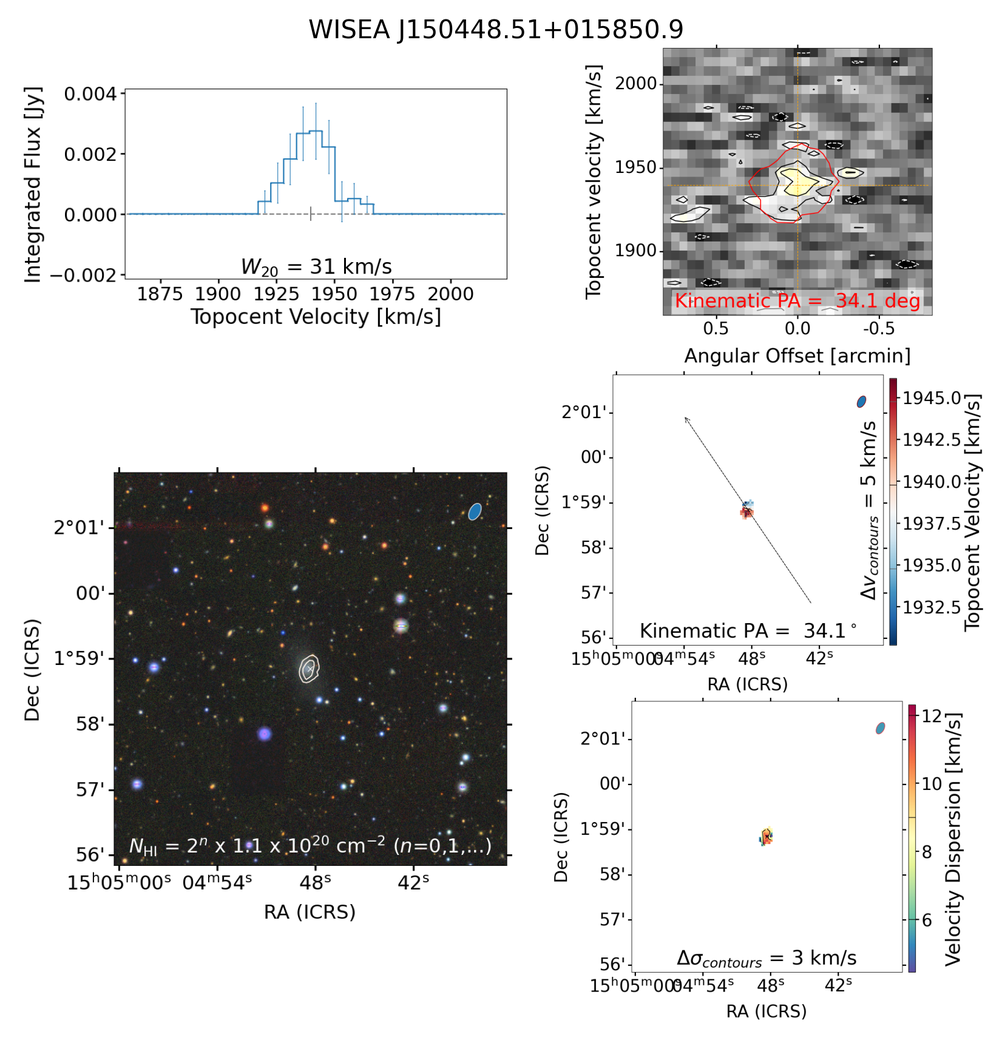}}
    \vspace{0.3cm}
    \resizebox{0.86\hsize}{!}{
    \includegraphics{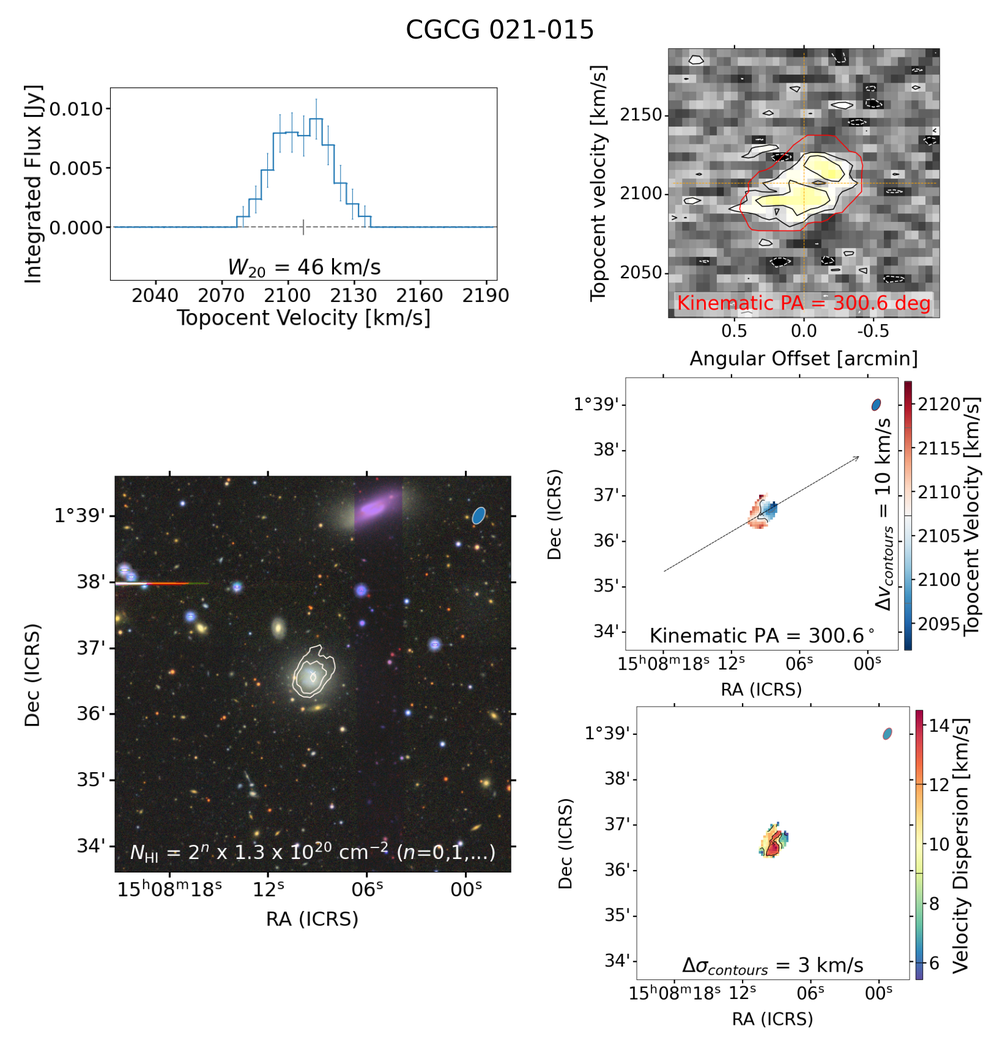}\hspace{0.5cm}
    \includegraphics{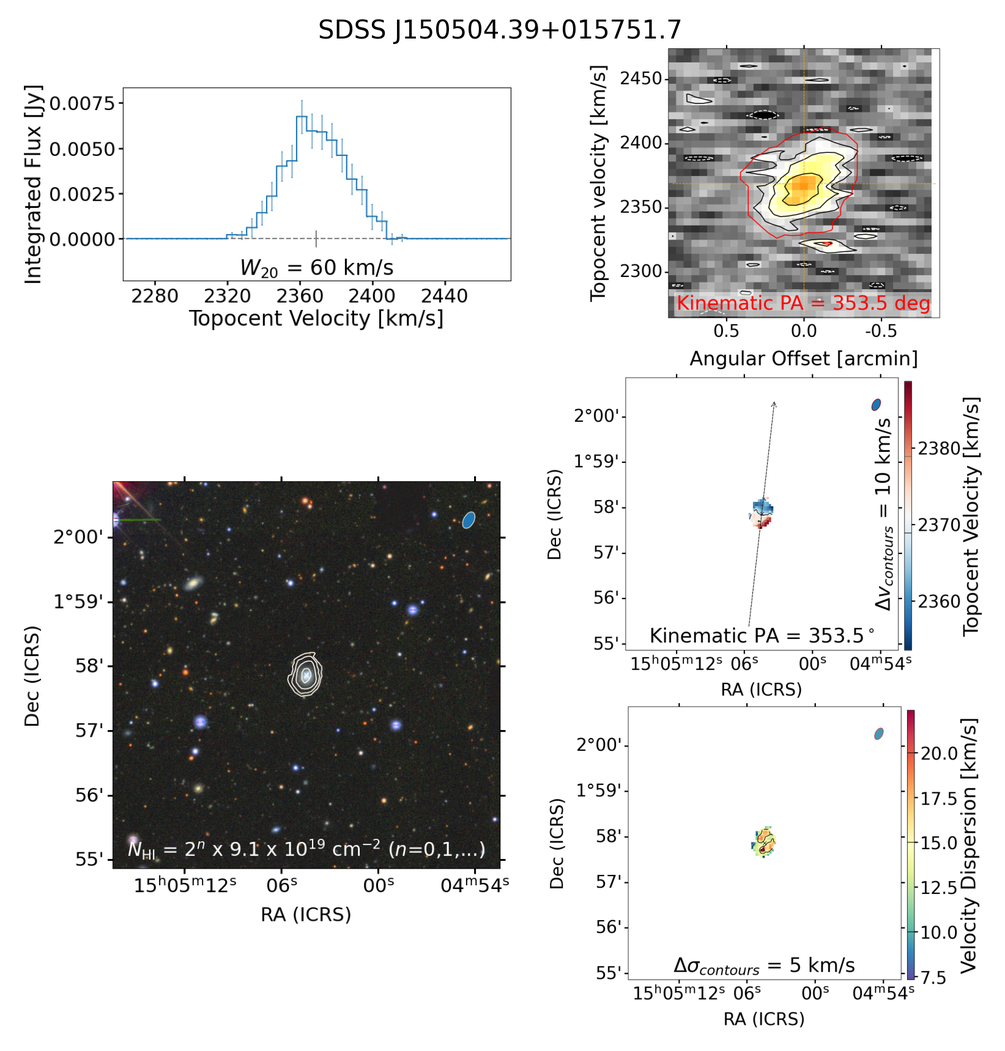}}
    \phantomcaption
\end{figure*}

\clearpage
\begin{figure*}
    \ContinuedFloat
    \centering
    \resizebox{0.86\hsize}{!}{
    \includegraphics{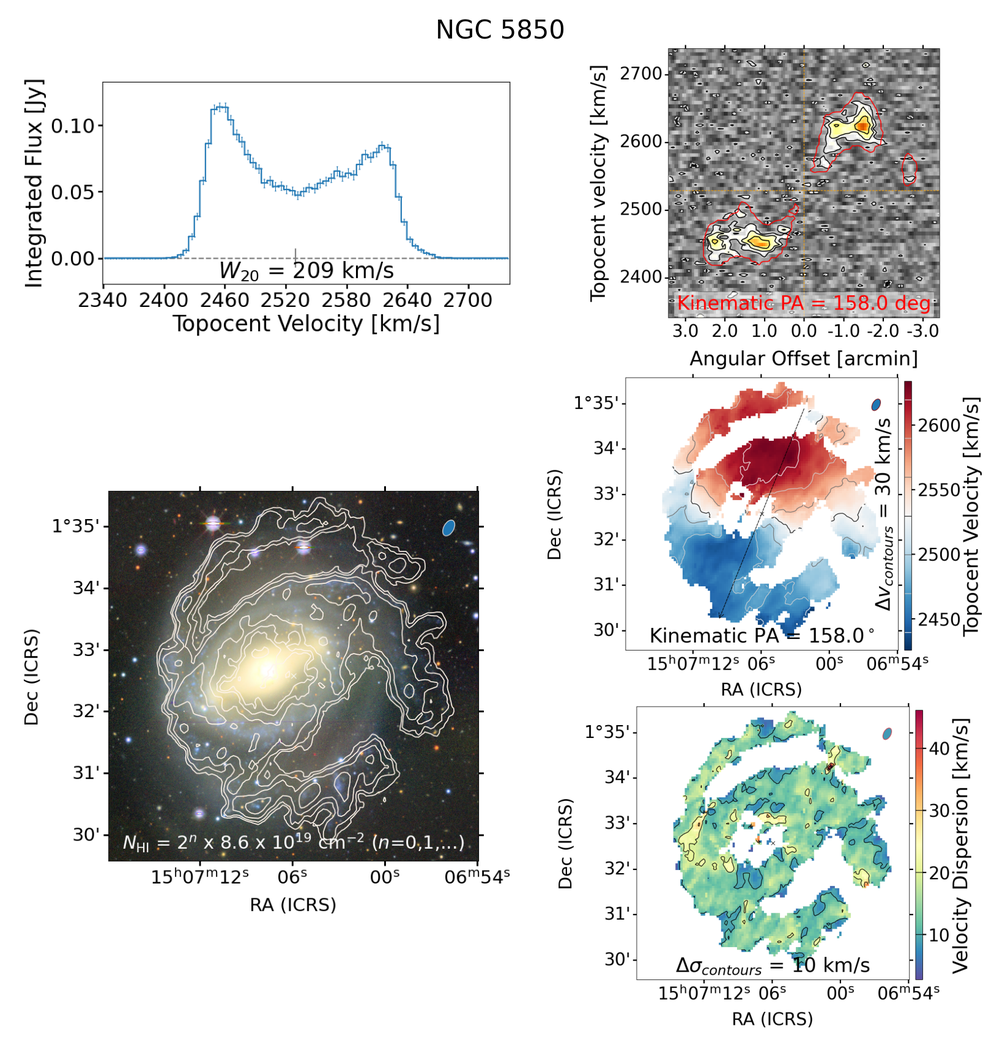}\hspace{0.5cm}
    \includegraphics{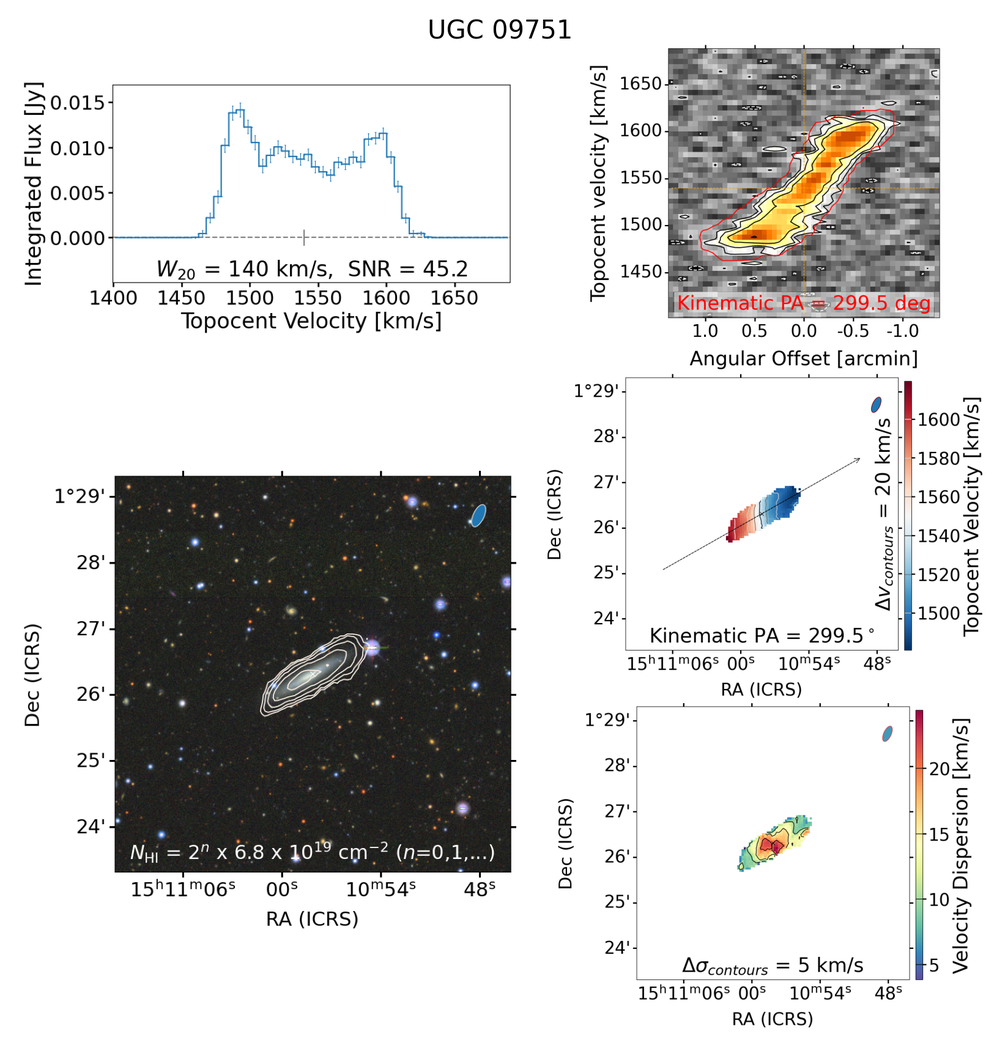}}
    \vspace{0.3cm}
    \resizebox{0.86\hsize}{!}{
    \includegraphics{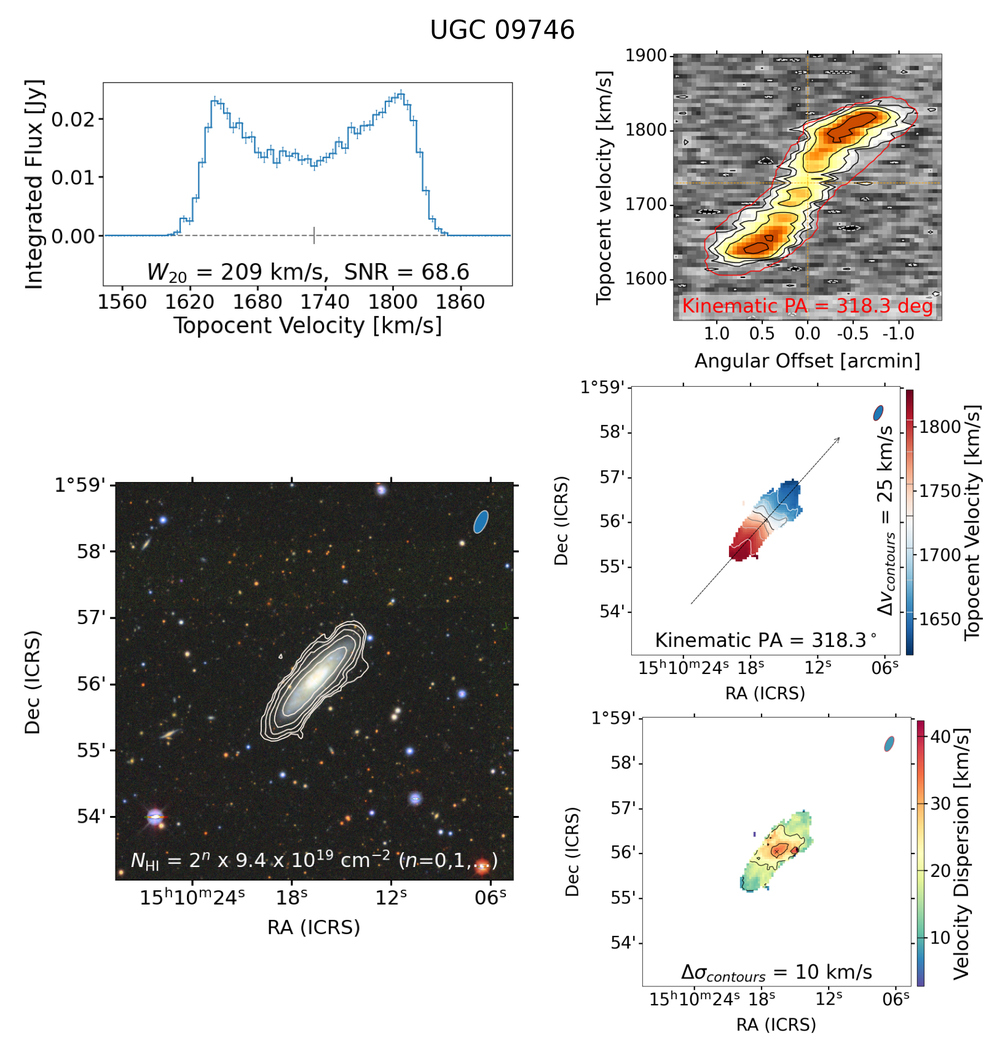}\hspace{0.5cm}
    \includegraphics{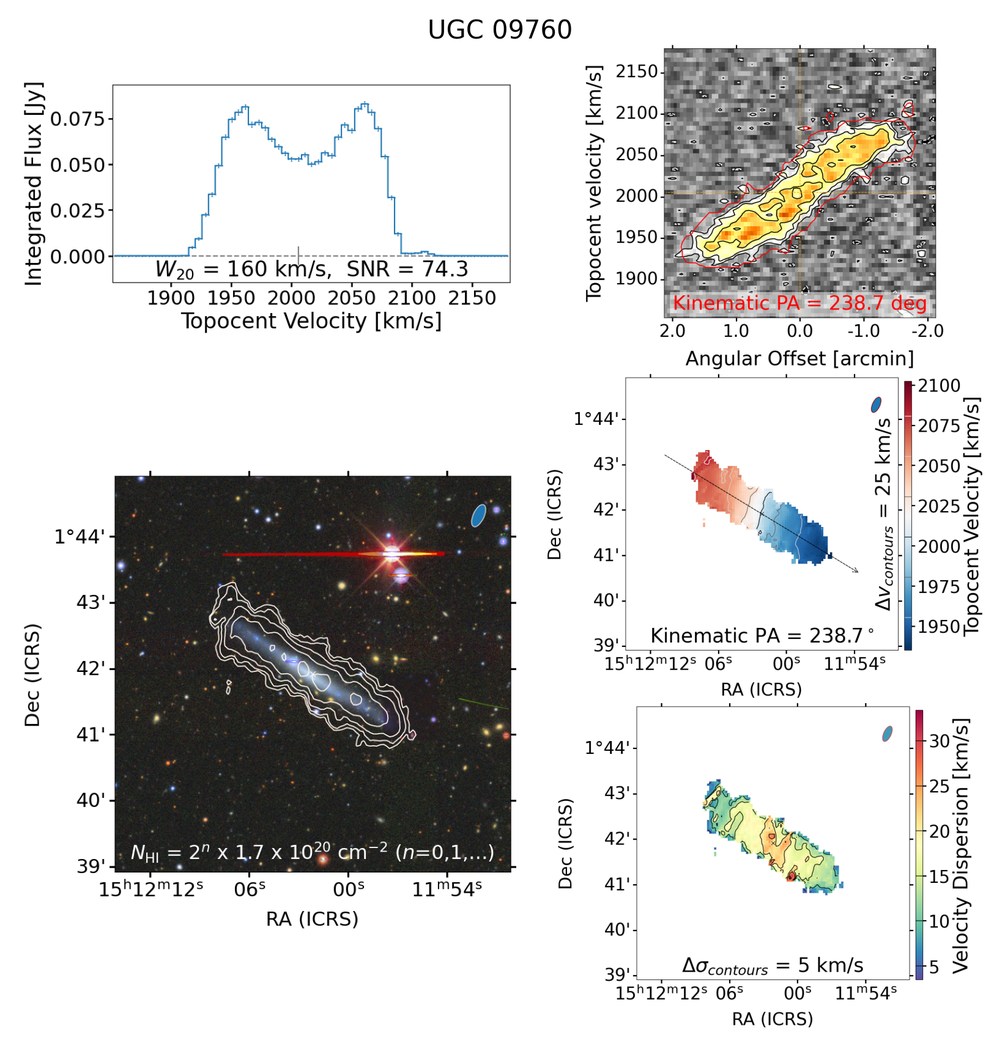}}
    \caption{\HI overlaying optical/near-IR images (bottom left panels of each image), moment-1 (middle right) and moment-2 maps (bottom right), PV diagrams (top right), and spectra (top left) of \HI detections with confirmed optical counterparts. The properties of the \HI maps are diverse and complex, suggesting a plethora of influences. The filament environment is a rich tapestry of baryons, and we will delve more into the scientific implications of this data set in the next papers of this series.}\label{fig:AllGalImage}
\end{figure*}

\end{document}